\documentstyle[bezier]{houches}
\input epsf.tex

\newcommand{\newsection}{\section}
\newcommand{\rf}[1]{(\ref{#1})}
\newcommand{\beq}{\begin{equation}}
\newcommand{\eeq}{\end{equation}}
\newcommand{\beqn}{\begin{displaymath}}
\newcommand{\eeqn}{\end{displaymath}}
\newcommand{\bea}{\begin{eqnarray}}
\newcommand{\eea}{\end{eqnarray}}
\newcommand{\g}{\gamma}
\renewcommand{\l}{\lambda}
\renewcommand{\L}{\Lambda}

\renewcommand{\b}{\beta}
\renewcommand{\a}{\alpha}
\newcommand{\n}{\nu}
\newcommand{\m}{\mu}
\newcommand{\th}{\theta}
\newcommand{\ep}{\varepsilon}
\newcommand{\om}{\omega}
\newcommand{\sg}{\sigma}
\newcommand{\k}{\kappa}
\newcommand{\del}{\delta}

\newcommand{\Dl}{\Delta}
\newcommand{\oh}{\frac{1}{2}}
\newcommand{\oq}{\frac{1}{4}}
\newcommand{\dg}{\dagger}
\newcommand{\Tr}{{\rm Tr}\;}
\newcommand{\cH}{{\cal H}}
\newcommand{\cL}{{\cal L}}
\newcommand{\cD}{{\cal D}}

\newcommand{\cM}{{\cal M}}
\newcommand{\cT}{{\cal T}}
\newcommand{\cN}{{\cal N}}
\newcommand{\cP}{{\cal P}}
\newcommand{\prt}{\partial}
\newcommand{\dcmp}{\frac{\d\om}{2\pi i}}
\newcommand{\dcmpV}{\dcmp\,\frac{V'(\om)}{z-\om}}
\newcommand{\dcmpx}{\frac{\d\om}{4\pi i}}
\newcommand{\dcmpxV}{\dcmpx\,\frac{\om V'(\om)}{z^2-\om^2}}
\newcommand{\pdp}{\phi^\dg\phi}
\newcommand{\ra}{\rangle}
\newcommand{\la}{\langle}
\newcommand{\ket}{\right\ra}
\newcommand{\bra}{\left\la}

\newcommand{\he}{\hat{e}}
\newcommand{\hK}{\hat{K}}
\newcommand{\hKl}{\hK_\l}
\newcommand{\hG}{\hat{G}}
\newcommand{\Dm}{\Delta \m}
\newcommand{\bchi}{\bar{\chi}}
\newcommand{\bm}{\bar{\mu}}
\newcommand{\bG}{\bar{G}}
\newcommand{\bT}{\bar{T}}
\newcommand{\nn}{\nonumber}

\begin{document}

\author[J. Ambj\o rn]{JAN AMBJ\O RN}
\address{ The Niels Bohr Institute \\
Blegdamsvej 17, DK-2100 Copenhagen \O \\
Denmark}

\chapter{Quantization of Geometry}

\newsection{Introduction}

The unification of the theory of critical phenomena and Euclidean
quantum field theory has been immensely yielding for both areas.
The theory of critical phenomena improved  in a very important way
our understanding of renormalization. The concepts of universality
classes, critical points, marginal operators etc.
have made many aspects of the renormalization procedure
and the renormalization group equations much more intuitive
for the field theorists, while the full machinery of
Green functions and Feynman integrals have been a most important
technical ingredient, which has influenced the theory of
critical phenomena.

There is a special area of field theory which is most intimately
linked to statistical mechanics, but where it nevertheless was
felt that one could not use the methods from statistical mechanics:
the quantization of geometrical objects, i.e.  the
first quantization of the free relativistic particle, the relativistic
string and the quantization of gravity. In all cases the classical
action is defined entirely in terms of simple geometrical
expressions which, however, when written in terms of an explicit
parametrization, become ugly and difficult to
treat by analytical methods: the length  of a curve will involve the
square root of the coordinates, just to take an example. Having chosen
a parametrization one gets the additional problem that the final
result should not depend on this parametrization.
Rather absurdly, this is often viewed as a virtue: the theory has
a large invariance, namely invariance under diffeomorphism.
In reality it reflects our inability to quantize the
physical, geometric degrees of freedom. Much is lost
compared to the beauty present if we simply view the system
as a statistical ensemble of geometrical objects, the partition function
being defined as the integral over all such objects, the weight being the
exponential of the classical action which itself is defined 
entirely in terms of the geometry.

When the link is made between statistical systems
and Euclidean quantum field theory, the discretization of space 
is usually a key ingredient. By discretizing ordinary space and 
restricting the volume to be finite we approximate the field theoretical
problem by a finite dimensional problem, which can often be viewed as 
a generalized lattice spin problem. In the infinite volume limit 
we can look for phase transitions of the spin system, and if these 
are characterized by a divergent correlation length, it is possible
to forget the underlying lattice structure: we can take the lattice
spacing to zero compared with the correlation length, and in this 
limit we might recover a continuum field theory. Usually masses
and coupling constants are defined not {\it at} the critical
point but by {\it the approach to} the critical point. 
In this process space is just playing a spectator role. We do not demand 
any {\it local} invariance maintained and this is why we can 
discretized space without any problems. Lattice operators which 
break explicitly Euclidean invariance will be suppressed in
the scaling limit. Local invariance usually mixes
high an low frequencies and is much more difficult to
discretize in a simple way. This is often used as an 
argument against attempts to discretize geometrical theories
and it is presumably correct that it is rather fruitless to 
attempt a discretization of a given parametrization of the 
geometry. If we, on the contrary, choose to discretize the geometry
directly there will be no problems at all. The main theme
of these lectures will be that a discretization of geometry 
is natural, it fits perfectly with a statistical mechanics
interpretation of the theories and the whole machinery
of critical phenomena and scaling limits can be applied in a very powerful
way to geometrical theories. In addition we  will get a pleasant
surprise: some of the theories are easier to solve directly
at the discretized level than in the continuum, and
this fact allows us to study the scaling limit in considerable detail.

What is meant by ``geometrical theories''? These are theories
which describe the propagation of ($d$-$1$)-dimensional manifolds
by summing over an appropriate class of $d$-dimensional manifolds
of which they are boundaries. The action might depend only on the
intrinsic geometry of the $d$-dimensional manifolds. If the manifolds
are embedded in $R^D$ the action  might in addition depend
on the extrinsic geometry.

The simplest of such theories describes the propagation of
point particles, i.e. we consider the theory of paths $P(x,y)$ between
two space points $x$ and $y$ which belong to $R^D$.
The simplest action for the paths will be
\beq
S[P(x,y)] = m \int_{P(x,y)} \d l + \l\int_{P(x,y)}
\d l\; |k| + \cdots.
\label{1.1}
\eeq
The first term only refers to an ``intrinsic'' property of the
path,  its length, which can be defined without reference to
the target space where it is embedded. The second term, a curvature
term, refers explicitly to the embedding. In principle we can add
higher powers of the curvature and also torsion terms to the action.

If we move up one dimension we get a theory which describes the
propagation of one dimensional objects (strings), i.e. we consider the
theory of surfaces $S(l_i)$ spanned between boundary strings (or loops)
$l_i$, $i=1,..,n$. The simplest action for such surfaces will be:
\beq
S[M(l_i)] = \m \int_{M(l_i)} \d A(M)+\l \int_{M(l_i)} \d A(M) H^2 \cdot
\label{1.2}
\eeq
where $M(l_i)$ denotes a manifold with boundaries $l_i$ and
$\d A(M)$ the area element of the manifold. Here  a
number of interpretations are possible.
We can choose to view the area as the  induced
area of a surface embedded in, say, $R^D$. This is analogous to the
interpretation given in \rf{1.1}. Then it is possible to add further
terms depending on the extrinsic geometry as indicated in
eq. \rf{1.2} where $H$ denotes the extrinsic curvature term.
In two dimensions
we have in addition the possibility to formulate the theory without reference
to a target space  (we have the same possibility in one dimension,
but the theory will be trivial unless matter fields are added, as
will be discussed later). By viewing the
variables as intrinsic we consider two-dimensional quantum gravity
and the action can be written as
\beq
S[M(l_i)] =  \int_{M(l_i)}\d^2\xi \sqrt{g}
\left(\m -\l_1 R + \l_2 R^2 \cdots \right).
\label{1.4}
\eeq
In this formula $g_{ab}$ denotes an internal metric. The first term
is still the area term now written in terms of the internal metric.
$R$ denotes the intrinsic or Gaussian curvature of the manifold.

When we finally move to higher dimensions we usually have no
interest in referring to some embedding space and we will
discuss the propagation of ($d-1$)-dimensional manifolds $b_i$
via $d$-dimensional manifolds which have the $b_i$'s as their boundary
entirely in terms of intrinsic variables.
The action will be
\beq
S[M(b_i)] = \int_{M(b_i)}
\d^d\xi \sqrt{g} \left(\m -\l_1 R + \l_2 R^2 \cdots \right)+{\rm b.t.},
\label{1.5}
\eeq
where {\em b.t.} denotes  boundary terms.
In a formula like \rf{1.5}
the natural variables to consider are equivalence classes $[g_{ab}]$
of metrics. In a continuum formulation it is quite difficult
to work directly with such variables. One almost inevitably ends
up with $g_{ab}$ themselves. Somewhat surprisingly, the regularized
(discretized) quantum theory offers the possibility to work
directly with equivalence classes.

The natural way to define the quantum theory corresponding
to the above classical actions is to use the path integral.
(First) quantization tells us that the propagator $G(b_1,...,b_n)$
between $d-1$ dimensional boundaries will be
\beq
G(b_i) = \int \d M{(b_i)} \; \e^{-S [M{(b_i)}]},
\label{1.6}
\eeq
where the summation is over $d$-dimensional geometries.
From this point of view eq. \rf{1.6} defines the theory
of fluctuating geometries.
We have already defined the action. One has to contribute
a meaning to the integration over geometries. A key ingredient
in doing so will be to approximate in a natural way the smooth structures
by piecewise linear structures. In this way \rf{1.6} will
be an ordinary statistical system. It is possible to discuss the
critical properties of this system and they agree with 
the results obtained by continuum methods whenever they are known.
In addition \rf{1.6} will provide us with a nonperturbative definition
of the theory in cases were continuum methods seem powerless.

In these lectures the intrinsic properties of our geometric objects
will be described by the metric. However, as will be clear, this
description can be replaced by any other, using more appropriate
variables, if needed. The only requirement seems to be that
these variables have a natural description on piecewise linear
structures.

In the following I will try systematically to develop the quantum theory
of geometric objects starting  from the simplest one-dimensional
objects in \rf{1.1} and ending with the  four-dimensional objects in eq. 
\rf{1.6} relevant to quantum gravity.
\newsection{Bosonic propagators and random paths}

\subsection{Quantization}

The classical action of the free relativistic particle in $R^D$ moving
from $x$ to $y$ is, as already mentioned, expressed by:
\beq
S[P(x,y)] = m_0 \int_{P(x,y)} \; \d l ,
\label{2.1}
\eeq

The classical equations of motion are derived by
choosing a parametrization of eq. \rf{2.1} :
\beq
x(\xi): [0,1] \to R^D,~~~~x(0) = x,~~x(1) = y\,.
\label{2.2}
\eeq
With this parametrization one gets:
\beq
S= m_0 \int^1_0 \d\xi \;\sqrt{\left(\dot{x}^\m\right)^2}\, ,~~~~~
\label{2.2a}
\eeq
\beq
\frac{\del S}{\del x^\m (\xi)}=
\frac{\d}{\d\xi} \left( \dot{x}^\m/|\dot{x}|\right) =0,
\eeq
where $\dot{x} \equiv \d x/\d\xi$.
The obvious solution to the classical equation of motion is
$\dot{x} = const.$, i.e. the straight line between $x$ and $y$,
but any reparametrization of the straight line:
\beq
x^\m(\xi) \to x^\m(f(\xi)),~~~~f(0)=0,~f(1)=1,~~ \dot{f} >0, \label{2.3}
\eeq
is a solution too. This is a reflection of the reparametrization invariance
of the  geometrical action \rf{2.1}

First quantization of the system is implemented via the path integral.
We get the propagator $G(x,y)$ by summing over {\it all} paths connecting
$x$ and $y$ and weighted by $\e^{-S[P]}$. The quantum aspect comes
precisely from the fact that not only the path which solves the
classical equations of motion contributes to the propagation:
\beq
G(x,y;m_0) = \int \cD P(x,y) \; \e^{-S[P(x,y)]}. \label{2.10}
\eeq
Each path should be counted only once in eq. \rf{2.10}. Reparametrizations
like \rf{2.3} should not be counted as different paths.

In order to contribute a meaning to $\int \cD P$ we have to introduce
a cut-off. Note that while we usually think about smooth paths, the
action is in fact defined on a larger class of paths, the ones
which are only piecewise smooth. Let us introduce a cut-off by
considering only piecewise linear paths where each step on the path
is of length $a$. This implies that we refrain from discussing
structures smaller than  $a$. Note that the {\it cut-off introduced
in this way by definition is reparametrization invariant}, since it
refers directly to a length in $R^D$. The possible length of the
paths will now be a multiple of $a$ : $l=n a$ and the action
of such a path will be $S = m_0 l$. For each piecewise linear path
between $x$ and $y$ consisting of $n$ pieces of length $a$
we have to integrate
over the possible positions of the $n-1$ interior points compatible
with the length assignment $a$. If we denote the vectors of the
$n$ linear parts by $ a \hat{e}_i$, $\hat{e}_i$ being a unit vector
in $R^D$, we have:
\beq
\int \cD P(x,y) \to \sum_{n=1}^\infty
\int \prod_{i=0}^n \d\hat{e}_i \; \del(a \sum \hat{e}_i -
(y-x)), \label{2.11}
\eeq
\beq
G_a(x,y;m_0) = \sum_{n=1}^\infty \e^{-m_0a n}
\int \prod_{i=0}^n \d\hat{e}_i \; \del(a \sum \hat{e}_i -
(y-x)). \label{2.12}
\eeq
We can calculate the propagator at the discretized level by getting rid
of the $\del$-function by a Fourier transformation:
\beq
G_a (p;m_0) = \int \d x\; \e^{-i p (x-y)}G(x,y;m_0) =
\sum_{n=0}^\infty \e^{-m_0a\cdot n} \int \prod_{i=1}^n \d\hat{e}_i \;
\e^{-i a p\cdot \hat{e}_i}.
\label{2.13}
\eeq
We have
\beqn
\int \d\hat{e} \;\e^{-i p\cdot \hat{e}} =
2 \pi^{D/2}\left[
\frac{{\rm J}_{\frac{D-1}{2}}(pa)}{\left(pa/2)\right)^{\frac{D-1}{2}}} \right]
\equiv f(pa),
\eeqn
where
\beq
f(0) = {\rm Vol} (S^{D-1}),~~~~f(pa) \approx f(0)(1 - c^2 (pa)^2 +\cdots).
\label{2.14}
\eeq
The final expression for $G_a(p;m_0)$ is
\beq
G_a(p;m_0) = \sum_{n=0}^\infty \left(\e^{-m_0a} f(pa)\right)^n =
\frac{1}{1-\e^{-m_0a} f(pa)}.  \label{2.15}
\eeq
This is an exact expression for the regularized propagator.
We have to take $a \to 0$ to get the continuum limit and it is
seen that we get the free relativistic propagator if we at the same
time {\it renormalize the bare mass $m_0$} such that
\beq
\e^{-m_0 a}f(pa) \to 1-c^2\, m_{{\rm ph}}^2 a^2,~~~~{\rm i.e.}~~~
m_0 = \frac{\log f(0)}{a} + c^2\, m_{{\rm ph}}^2 a
\eeq
From eqs. \rf{2.14} and \rf{2.15}  it follows that:
\beq
G_a (p;m_0) \to \frac{1}{c^2\, a^2} \frac{1}{p^2+ m_{{\rm ph}}^2} =
\frac{1}{c^2 \, a^2} G_{c}(p;m_{{\rm ph}}), \label{2.16}
\eeq
where $G_{c}(p;m_{{\rm ph}})$ denotes the continuum propagator.
The divergent factor which relates $G_a$ and $G_{c}$
is a kind of wave function renormalization, but it has
a physical meaning since the power of $a$ which appears
reflects directly the short distance behavior of the propagator,
as we shall discuss in detail later.

It is worth  rephrasing the above results in terms of dimensionless
quantities, and in this way make the statistical mechanics aspect
more visible. Introduce $\m=m_0a$ and $q=pa$ and consider coordinates
in $R^D$ as dimensionless. The steps in the random walk will be of length
one and \rf{2.15} reads
\beq
G_\m (q) = \sum_{n=0}^\infty \e^{-\m n} f(q)^n =
\frac{1}{1-\e^{-\m}f(q)}. \label{2.17}
\eeq
$\m$ acts like a chemical potential
for inserting additional sections in the piecewise linear
walk. We have a {\it critical value $\m_c =\log f(0)$ of the chemical
potential $\m$}.
For $\m > \m_c$ the sum is convergent for all $q$ and the average number of
steps in the random walk is finite. For $\m < \m_c$ the sum
is divergent for some range of $q$ and the average length of the
paths not defined. For $\m \to \m_c$ very long paths will
dominate the sum and this is where we can take a continuum limit.
We can write
\beq
f(q) \sim \e^{\m_c} (1-c^2 q^2),~~~~G_\m (q)
\approx \frac{1}{\m-\m_c +c^2 q^2},
\label{2.18}
\eeq
and at this point we can introduce the physical length scale $a$,
and the physical momentum $p_{{\rm ph}}$ and the physical mass $m_{{\rm ph}}$,
both intended to be kept fixed for $a \to 0$, by
\beq
\m-\m_c = m_{{\rm ph}}^2 a^2,~~~~q =c p_{{\rm ph}} a. \label{2.19}
\eeq
This defines $a$ as a function of $\m$ by
\beq
a(\m) = \frac{1}{m_{{\rm ph}}} \sqrt{\m-\m_c}.
\eeq
Later we will later discuss relations like \rf{2.19} in great detail.

It should be emphasized that the critical behavior we have found this
way is universal. Any ``reasonable'' class of random paths
should result in the same scaling limit.
The piecewise linear paths are convenient because the
results, even at the discretized level, are Euclidean invariant.
If we choose to regularize the summation over all paths by
considering the sub-class of paths which can be formed
by links on a hyper-cubic lattice in $R^D$, the regularized
propagator will be:
\beqn
G_a=\sum_n \e^{-m_0a n}\sum_{P_n} \del (a \sum \hat{e}_i -(y-x)).
\eeqn
Again the regulatized propagator can be computed  by (lattice)
Fourier transformation and in the scaling limit one gets the same
results as for the piecewise linear random walks.

\subsection{One-dimensional gravity}

Let us now turn to a somewhat different quantization of the free propagator.
The action (2.1) has a beautiful
geometrical interpretation and from the discretized
point of view there was no problems associated with the
quantization, as explained above. However, the square root,
which appears in eq. \rf{2.2a} after a choice of parametrization,
makes it very difficult to use the action in formal continuum
manipulations. For this reason it might be preferable to use
a different action which at the classical level is equivalent with 
the one given in eq. \rf{2.1}:
\beq
S[x,g]_{(x,y)} =\frac{1}{\a'} \int_0^1 \d\xi \sqrt{g(\xi)} \left[
g^{ab}(\xi)  \frac{\prt x^\m}{\prt \xi^a}
\frac{\prt x^\m}{\prt \xi^b} + \m \right].
\label{2.3x}
\eeq
In eq. \rf{2.3x} $x^\m(\xi)$ denotes a path in $R^D$ such that
\beqn
x^\m(0) = x^\m,~~~~~x^\m (1) = y^\m,
\eeqn
and $g_{ab}$ is an {\it internal} metric on the one-dimensional
manifold given by the parametrization $\xi$. The indices $a,b$ can only
take the value $1$, but we have written the action in a general
covariant way. Like \rf{2.1} the action \rf{2.3x}
is invariant under the  reparametrization $\xi \to f(\xi)$ defined in
eq. \rf{2.3}  provided $g$ satisfies
\beq
g^{(f)} (f(\xi))=\frac{1}{\dot{f}^2} g(\xi).
\label{2.6x}
\eeq
This transformation rule ensures the invariance of internal distance:
\beq
\d s^2 = g_{ab} \d\xi^a \d\xi^b = g^{(f)}_{ab}\d f^a \d f^b;~~~~~a,b=1.
\label{2.7x}
\eeq
The classical equations
of motion, obtained by considering $g_{ab}$ and $x^\m$ as independent
variables, agree with the classical equations obtained 
from eq. \rf{2.1}:
\bea
\frac{\del S}{\del \sqrt{g}} &=& -\frac{1}{\sqrt{g}}
\left(\frac{ \d x^\m}{\d\xi}\right)^2 + \m =0   \nn\\
\frac{\del S}{\del x^\m} &=& -\frac{\d}{\d\xi}
\left(\frac{1}{\sqrt{g}}\frac{\d x^\m}{\d\xi}\right)=0. \nn
\eea
However, it is by no means obvious that the quantum systems defined
from eqs. \rf{2.1} and \rf{2.3x} agree. The quantum equivalent of
eq. \rf{2.10} is:
\beq
G(x,y) = \int \frac{\cD g_{ab}}{{\rm Vol (diff)}}
\int\cD_g x
~\e^{-S[x,g]_{(x,y)}}\, ,
\label{2.8x}
\eeq
where $x(0)=x$ and $x(1)=y$.
The integration variable is equivalence classes of metrics, i.e.
metrics which are related by reparametrization. This is indicated
by the symbolic division by the ``volume'' of the diffeomorphism group.

In order to define the functional integral in eq. \rf{2.8x} we first introduce
{\it a reparametrization invariant cut-off ``a''}, i.e. we
consider only paths $x(\xi)$ which have no structure below the
length scale $\d s =a$, $\d s$ given by eq. \rf{2.7x}. This can be achieved by
restricting ourselves to piecewise linear paths where the {\it internal
length} of the individual pieces is {\it a}. For a given smooth
metric $g_{ab}(\xi)$ and a given smooth path we can, if we want,
approximate the action by a corresponding action of a piecewise linear
path: first we discretize the manifold $[0,1]$ parametrized by $\xi$
according the above prescription. Here it is important to note
that the only reparametrization invariant quantity which characterizes
the manifold is the total length:
\beq
l = \int^1_0 \d\xi \,\sqrt{g(\xi)}
\label{2.9x}
\eeq
The length $l$ is clearly invariant, and for any metric $g_{ab}$ satisfying
\rf{2.9x} we can transform it to the constant metric
$g^{(c)}_{ab} = l^2 \del_{ab}$ by choosing the function $f$ in \rf{2.6x} as
\beq
f(\xi) = \frac{1}{l}\; \int_0^\xi \d\xi'\,\sqrt{g(\xi')}.
\label{2.10x}
\eeq
A given discretized path will always have the length $na$. For a
given $l$ we simply take $n = [l/a]$, and given $g_{ab}$ we can calculate
the points $\xi_i$ on the manifold parametrized by $\xi$
where the internal length from the boundary is $ia$:
\beq
i a = \int_0^{\xi_i} \d\xi \; \sqrt{g(\xi)}.
\label{2.11x}
\eeq
For a given metric we have discretized the manifold. The
continuum action  can now be approximated on the discretized manifold
in a natural way by using that $\d\xi(\xi_i) \approx a/\sqrt{g (\xi_i)}$ and
$\sqrt{g}g^{ab} = 1/\sqrt{g}$ in one dimension:
\beq
\int^1_0 \d\xi \left[\frac{1}{\sqrt{g}} \left(
\frac{\d x_\m}{\d\xi}\right)^2 +
\m \sqrt{g} \right]  \approx \sum_{i=1}^{n-1} a
\left[\frac{(x^\m(\xi_i) -
x^\m(\xi_{i-1}))^2}{a^2} + \m\right].
\label{2.12x}
\eeq
Note that the discretized action contains no explicit reference to
the metric $g_{ab}$. The $\xi_i$'s appear as one would have expected
it from a constant metric. But this is consistent with the observation
that any metric is equivalent to a constant metric and the fact
that the action reparametrization invariant.

We can now combine the information given above: the integration of
equivalence classes of metrics reduces in one dimension to an
integration over the length $l$ and in the discretized approach
this integration is replaced by a summation over $n = l/a$.
The weight of (discretized) configurations will be determined by
eq. \rf{2.12x}. Let us at this stage change
to dimensionless quantities, as advocated above.
By redefining $x$ and $\m$ we can dispose of  $\a'$ and
choose $a=1$, and the discretized version of eq. \rf{2.8x} will be
\beq
G_\m (x,y) = \sum_{n=1}^\infty \e^{-\m n}
\int \prod_{i=1}^{n}
\prod_{\m=1}^D \d x^\m_i\; \e^{-\sum_{i=1}^{n+1}
(x^\m_i-x^\m_{i-1})^2}
\label{2.14x}
\eeq
A few remarks about the formula: we have replaced the index $\xi_i$ with
$i$ since it is just a dummy label of an integration variable.
We have also defined $x_0 =x$ and $x_{n+1}=y$. Note
finally that the index $g$ in the measure $\cD_g \phi$ in formula \rf{2.8x}
is explicit present in the measure in eq. \rf{2.14x}: it is $n$.

Since the integrals in eq. \rf{2.14x} are Gaussian they
can be performed explicitly:
\beq
\int \prod_{i=1}^{n}\prod_{\m=1}^D
\d x^\m_i\; \e^{-\sum_{i=1}^{n+1}
 (x^\m_i-x^\m_{i-1})^2} =
\pi^{nD/2} \; \frac{\e^{-(x-y)^2/4n}}{n^{D/2}},
\label{2.15x}
\eeq
and we can write
\beq
G_\m(x,y) = \sum_{n=1}^\infty \frac{1}{n^{D/2}} \;
\e^{(\m_0-\m)n-(x-y)^2/4n^2},~~~~~\m_0 = D\log \sqrt{\pi}.
\label{2.16x}
\eeq
Again we see that there is a
critical point $\m_0$ above which $G_\m(x,y)$ is
convergent and below which it is divergent.

If we take the Fourier transformation of $G_\m(x,y)$ we
immediately get \rf{2.18} and we conclude that
the theories defined by eqs. \rf{2.1} and \rf{2.3x} equivalent 
both at the classical and the quantum level.

The reader is invited to compare the above calculation
of $G(x,y)$ defined by eq. \rf{2.8x} 
with a calculation performed entirely in  the continuum,
where one first has to introduce  parametrization
and ghosts and in the end a cut-off in order to calculate ill-defined
determinants. The advantage of being able to work
directly with equivalence classes of metrics should be obvious.

\subsection{Scaling relations}

The free relativistic quantum particle is described as the
scaling limit of the random walk. In the following we
will consider models which cannot be solved as completely
as was the case above. In general we will not be interested in
the complete solution at the discretized level, but only in the
scaling limit where we approach the critical point. Let us therefore
discuss scaling relations using the random walk as an example.
The relations to be derived will be valid in a much broader context
and we will use them many times in the rest of these lectures.

Consider a model for random walks from point $0$ to $x$ in $R^D$.
Step number $n$ will be characterized by an initial position $x_{n-1}$,
and an (unnormalized) probability distribution $P({v})$
for a step to $x_n = x_{n-1} + v$. Let us assume that $P$ is only a
function of $|v|$. These assumptions can (and will be) considerable
relaxed in the following, but they are convenient for a first
discussion.  If we use the notation $x_0=0$ and $x_{n+1}=x$ for a random
walk of $n$ steps from $0$ to $x$,
our generalized propagator can be written as:
\beq
G_\m (x) = \sum_n \e^{-\m n} \int \prod^{n}_{i=1} \d x_i
\prod_{i=1}^{n+1} P(|x_i-x_{i-1}|),~~~~\int \d v \;P(v) = \e^{\m_c}.
\label{2.20}
\eeq
Note that   $\int \d x\, G_\m (x)$ can be performed since
the integration over probability distributions reduces to a product
of single integrals:
\beq
\chi(\m) \equiv \int \d x\; G_\m (x) = \frac{1}{1-\e^{-(\m-\m_c)}}.
\label{2.20a}
\eeq
$\chi(\m)$ is called the {\it susceptibility}\footnote[1]{This notation
is borrowed from spin systems, where the spin susceptibility is the
second derivative of the free energy with respect to the magnetic
field, but also has the interpretation as the integral of the
spin-spin correlation function over space.}.
It is now clear that the critical point is $\m_c$. For $\m >\m_c$
\rf{2.20} is convergent for all $x$. For $\m < \m_c$ \rf{2.20}
is divergent for all $x$. One can solve the model given by eq. \rf{2.20}
in the scaling limit since it follows from the central limit theorem
that the convolution of $P$ many times reduces to the normal distribution,
i.e. we get precisely the Gaussian model considered in the last
subsection. Let us, however, discuss some general properties of
eq. \rf{2.20} which will be of use for the more general models to be
considered in the following.

\vspace{12pt}
\noindent {\bf Theorem:} $G_\m(x)$ falls off exponentially for $\m > \m_c$.

\vspace{12pt}
\noindent
$p(x) =G_\m(x)/\chi(\m)$ is the probability density for a random 
walk from 0 to $x$, i.e. $\Dl p(x) = p(x)\Dl x$ is the probability 
for a random walk from 0 to $x$ and we have the
the inequality $\Dl p(x+y) \geq \Dl p( x) \Dl p(y)$,
simply because the random walks from $0$ to $x+y$ which pass through
$x$ are a subset of all random walks from $0$ to $x+y$. This means that
$-\log \Dl p(x)$  is a {\it sub-additive} function of $|x|$:
\beq
-\log \Dl p(x) \leq -\log \Dl p(\a x)-\log \Dl p ((1-\a)x),~~~~0<\a<1.
\label{2.21}
\eeq
From the sub-additivity alone it follows that the following limit exists
\beq
\lim_{|x| \to \infty} \frac{-\log G_\m (x)}{|x|} = m(\m),~~~~m(\m)\geq 0.
\label{2.22}
\eeq
Since $G_\m(x)$ by the definition \rf{2.20} is a decreasing function of
$\m$ it follows that $m(\m)$ must decrease as $\m$ decreases towards
$\m_c$. Since $\int \d x \,G_\m (x)$ exists for $\m > \m_c$ we conclude
that $m(\m) \geq 0$.
It is now clear that $m^{-1}(\m)$ will serve as the correlation
length for the random walk and {\it we can only obtain a correlation
length which is large compared to the individual step length provided
$m(\m) \to 0$ for $\m \to \m_c$}. We will assume this is the case and
introduce the following critical exponents:
\beq
m(\m) \sim (\m-\m_c)^\n~~~~~~~~~~~~~~~~~{\rm for}~~\m\to\m_c. \label{2.23}
\eeq
\beq
G_\m(x) \sim \frac{1}{|x|^{D-2+\eta}}~~~~~~~~~~~~~~~~
{\rm for}~~1 \ll |x| < \frac{1}{m(\m)}.
\label{2.24}
\eeq
\beq
\int \d x \;G_\m(x) \sim \frac{1}{(\m-\m_c)^\g}~~~~~~~{\rm for}~~~\m \to \m_c.
\label{2.25}
\eeq
The mass exponent $\n$, the anomalous scaling dimension
$\eta$ and the susceptibility exponent $\g$ are not independent.
They satisfy {\it Fischer's scaling relation}:
\beq
\g = \n (2-\eta).  \label{2.26}
\eeq
The proof of this relation is simple. From the behavior
assumed for $G_\m(x)$ in eqs. \rf{2.23}-\rf{2.25} we can cut off
the integration over $x$ in the susceptibility at $1/m(\m)$,
i.e. eq. \rf{2.25} can be estimated as follows:
\beqn
\chi(\m) \sim \int_{|x| < 1/m(\m)} \d x \;\frac{1}{x^{D-2+\eta}} \sim
\frac{1}{m(\m)^{2-\eta}}.
\eeqn
Eq. \rf{2.26} follows from the definition of $\g$ and $\n$.

For our simple random walk case it follows from eq. \rf{2.20a}
that $\g=1$ and it follows from the results in the last subsection that
$\n =1/2$. This implies that $\eta =0$ and this is
why it is called the anomalous scaling dimension: If $\eta$ is different
from zero it is anomalous with respect to the free particle.

\begin{figure}
\input{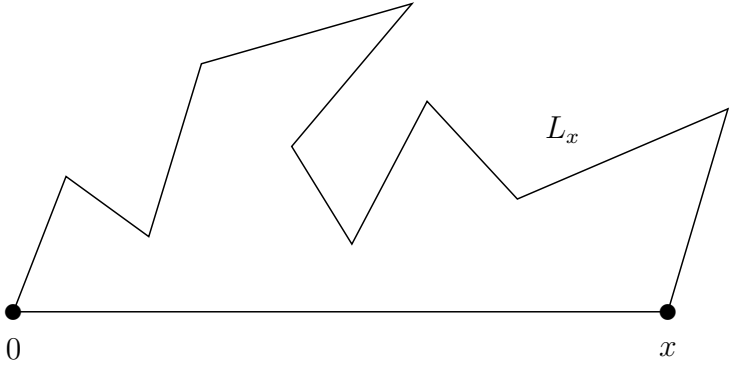}
\caption[2_1]{A typical piecewise linear path between $0$ and $x$.}
\label{2_1}
\end{figure}

The exponent $\n$ has a direct geometric meaning:
\beq
d_H = 1/\n,  \label{2.27}
\eeq
where $d_H$ denotes the {\it Hausdorff dimension} of the random walk.
It can be defined in the following way: Let $\bra L \ket_x$
denote the average length of a path from $0$ to $x$ in the ensemble
of paths defined by eq. \rf{2.20}. This is illustrated in fig.\,\ref{2_1}.
Let us define
\beq
\bra L \ket_x \sim  x^{d_H}~~~~{\rm for}~~~|x| \to \infty,~~~
m(\m)|x| ={\rm const.} \label{2.28}
\eeq
Note that for any power $k$ we can write for large $n$:
\beqn
\frac{\int \prod_{i=1}^n \d x_i \left(\sum^{n+1}_{i=1} |x_i-x_{i-1}|^k \right)
\prod^{n+1}_{i=1} P(x_i-x_{i-1})}{\int \prod^n_{i=1} \d x_i
\prod^{n+1}_{i=1} P(x_i-x_{i-1})}  \sim c_k n.
\eeqn
Let us use this for $k=1$:
\beq
\bra L \ket_x \equiv  \frac{\sum_n \e^{-\m n}
\int \prod^n_{i=1} \d x_i \left(\sum^{n+1}_{i=1}
|x_i-x_{i-1}|\right) \prod^{n+1}_{i=1} P(x_i-x_{i-1})}{
\sum_n \e^{-\m n} \int \prod^{n+1}_{i=1} \d x_i
\prod^{n+1}_{i=1}P(x_i-x_{i-1})},
\label{2.29}
\eeq
i.e. we can write:
\beq
\bra L \ket_x  \sim \bra n \ket_x =
- \frac{ \frac{\prt}{\prt \m} G_\m (x)}{G_\m (x)} \sim m'(\m) |x|.
\label{2.30}
\eeq
For $\m > \m_c$ fixed and $|x| \to \infty$ we just get $\bra L \ket \sim
|x|$ and the reason is clear:  $\bra n \ket_x$ is finite for fixed $|x|$
and just goes to infinity proportional to $|x|$. This implies that
only paths which are close to the straight line from 0 to $x$
contribute for $\m > \m_c$ . However, the coefficient
$m'(\m)$ diverges as $\m \to \m_c$  and we
are interested in the limit where $m(\m) |x|$ is constant. This is the limit
where we can introduce a physical ``lattice'' length $a(\m)$, a physical
length $x_{{\rm ph}}$ and a physical mass $m_{{\rm ph}}$ by
\beq
m(\m) = m_{{\rm ph}} a(\m),~~~x_{{\rm ph}} = x\, a(\m),~~~~{\rm i.e.}~~
a(\m) \sim (\m-\m_c)^\n,    \label{2.31}
\eeq
such that $x_{{\rm ph}}$ and $m_{{\rm ph}}$ are kept fixed for
$\m \to \m_c$, where $a(\m) \to 0$. In this limit we have
$(\m-\m_c)^\n \sim |x|^{-1}$ and
\beq
\bra L \ket_x \sim \frac{m'(\m)}{m(\m)} \sim \frac{1}{\m-\m_c} \sim x^{1/\n}.
\label{2.31a}
\eeq
Since the random walk described by \rf{2.20} has $\n=1/2$
we arrive at the well known result that $d_H=2$, but \rf{2.27}
will be valid if $\n \neq 1/2$. We will meet  such situations
for the smooth random walks considered in the next subsection.

It should also be emphasized that the ``physical'' length of
$\bra L \ket_x$ will diverge in the scaling limit if we define
$L_{{\rm ph}} = L a(\m)$. It has to be this way for dimensional
reasons:
\beq \bra L_{{\rm ph}} \ket \sim \frac{x_{{\rm ph}}^{1/\n}}{a(\m)^{(1-\n)/\n}}. \\
\eeq
The propagator itself will be singular in the scaling limit.
As already mentioned it can be viewed as a kind of wave function
renormalization. From the assumed short distance behavior \rf{2.24}
we get from \rf{2.31}
\beq
G_\m(x) \sim a^{D-2+\eta} G_c (x_{{\rm ph}},m_{{\rm ph}})~~~{\rm for}~~a \to 0,
\label{2.31b}
\eeq
or, for the Fourier transformed where $q = p_{{\rm ph}} a$:
\beq
G_\m (q) \sim a^{\eta -2} G_{c}(p_{{\rm ph}},m_{{\rm ph}}). \label{2.31c}
\eeq

\subsection{Smooth random walks}

In the last subsection we saw that ordinary random walks have
Hausdorff dimension $d_H =2$. This result is quite universal
as follows from the general expression
\rf{2.20} for a random walk. In the scaling limit the different random
walk representations all agreed with the direct discretization
of the action \rf{2.1}. Let us consider the first non-trivial
generalization of purely geometrical nature:
\beq
S[P(x,y)] = \m \int_{P(x,y)} \d l + \l \int_{P(x,y)} \d l \;|\k(l)|,
\label{2.40}
\eeq
where $k(l)$ denotes the curvature of the path $P$.

\vspace{12pt}

\begin{figure}
\input{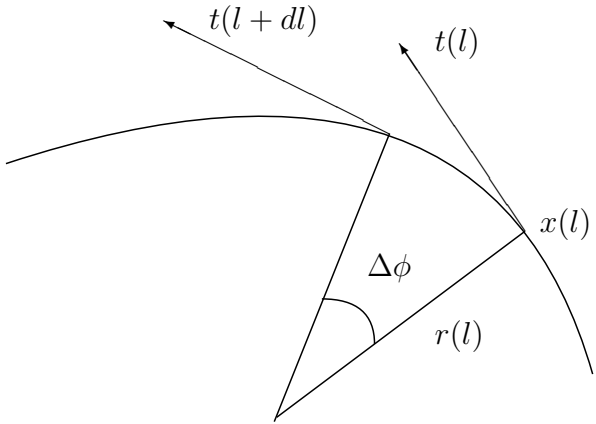}
\caption[2_2]{The geometry related to the calculation of
curvature. $r(l) = 1/\k(l)$.}
\label{2_2}
\end{figure}

Recall how the curvature of a curve in $R^D$ is defined. Let
$x^\m (\xi)$ be a parametrization of the curve. Use as $\xi$ the
length $l$ of the curve. Let $t^\m(l)$ denote the normalized
tangent of the curve:
\beq
t^\m = \frac{\d x^\m}{\d l} \equiv \dot{x}^\m,
~~~~~~t^\m t^\m =1,~~~~t^\m\dot{t}^\m =0.
\label{2.40a}
\eeq
It follows that $\dot{t}^\m$ is orthogonal to $t^\m$ and we write
\beq
\dot{t}^\m = \k n^\m, \label{2.41}
\eeq
where {\it $\k$ is the curvature and $n^\m$ the principal normal.}
It lies in the osculating plane of the curve. The geometry is
shown in fig.\,\ref{2_2}.
It is seen that
\beq
|\Delta t| = 2\sin \frac{\Delta \phi}{2} \approx \Delta \phi,
\label{2.42}
\eeq
$\Delta \phi$ is called the angle of contingency and we have
\beq
\k(l) = \frac{\d\phi}{\d l},~~~~~\k(l) = \frac{1}{r(l)},
\label{2.43}
\eeq
where $r(l)$ is the radius of curvature. It is the radius of the osculating
circle which is defined by the quadratic approximation to the curve
and has its center on the principal normal at the
distance $r(l)$ from $x^\m(l)$.

\vspace{12pt}

\begin{figure}
\input{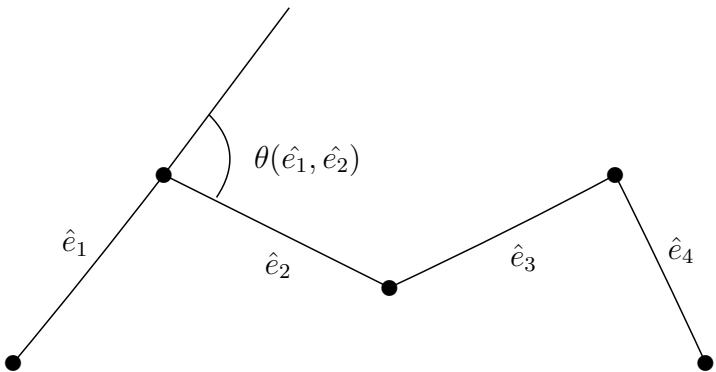}
\caption[2_3]{The angle $\th(\he_1,\he_2)$
between successive tangent vectors in
the piecewise linear random walk.}
\label{2_3}
\end{figure}

\begin{figure}
\input{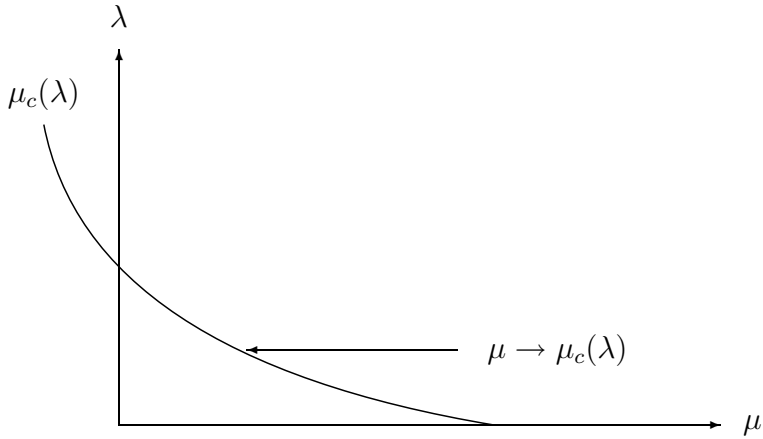}
\caption[2_4]{The phase diagram in the $(\m,\l)$-plane. The theory
is defined to the right of the critical line $\m_c(\l)$.}
\label{2_4}
\end{figure}

To quantize the theory defined by  the action \rf{2.40}
we have to perform the path integral\footnote{There is an extensive
literature on canonical quantization, saddle-point calculations and
large $d$ expansions of the theory. Ref. \cite{rw} is a very incomplete list.}.
 Again it is useful to
regularize the sum over all paths by restricting  the sum
to be over all piecewise linear paths, the length of the
individual paths being $a$ (which we choose equal 1) \cite{adjrw}.
If the path consists of $n$ linear pieces, the $i$'s piece
will be characterized by the a unit vector $\he_i$.
Although there is no universal definition of curvature
for such paths, eq. \rf{2.42} is a useful guide, where we simply take
$\Delta \phi_i$ to be difference $\th (\he_{i+1},\he_i)$ in angles between
$\he_i$ and $\he_{i+1}$. This is shown in fig.\,\ref{2_3}.
The discretized version of the action  \rf{2.40} reads:
\bea
S[P_n,\m,\l] &=& \m \sum_{i=1}^n |\he_i| +
\l \sum_{i=1}^{n-1}  2\sin \frac{\th(\he_{i+1},\he_i)}{2} \nn \\
&=& \m n + \l \sum_{i=1}^{n-1} f(\th(\he_{i+1},\he_i)), \label{2.44}
\eea
where we have already generalized the curvature term to a function
$f(\th)$ satisfying
\beq
f(0) =0,~~~~~f'(\th) >0~~~{\rm for}~~0 \leq \th \leq \pi.
\label{2.45}
\eeq
The path integral for the propagator from $0$ to $x$ can now be written as:
\bea
G(x;\m,\l) &=& \int \cD P(x) \; \e^{-S[P(x);\m,l]}  \label{2.46}  \\
           &\sim& \sum_n \e^{-\m n} \int\prod_{i=1}^n \d\he_i \;
           \prod_{i=1}^{n-1} \e^{-\l f(\th(\he_{i+1},\he_i))} \;
           \del( \sum\he_i-x). \nn
\eea
In order to solve this model let us introduce the probability
$K_\l (\he_2,\he_1)$ for a step $\he_2$ provided the
former step was $\he_1$:
\beq
K_\l (\he_2,\he_1) = \frac{1}{N_\l} \e^{-\l f(\th(\he_2,\he_1))},~~~
N_\l = \int \d\he_1\; \e^{-\l f(\th(\he_2,\he_1))}. \label{2.47}
\eeq
With this notation we can write:
\beq
G(x;\m,\l) \sim \sum_n \left(\e^{-\m}N_\l\right)^n \int \prod_{i=1}^n \d\he_i
\prod_{i=1}^{n-1} K_\l (\he_{i+1},\he_i) \; \del (\sum \he_i -x).
\label{2.48}
\eeq
We can calculate the susceptibility since the additional integration
over $x$ just  removes of the $\del$-function in eq. \rf{2.48} and
the integrations of $\he_i$ become trivial
($\int \d\he_1 K_\l (\he_2,\he_1) =1$):
\beq
\chi(\m,\l) = \int \d x\; G(x;\m,\l) \sim \frac{1}{1-\e^{-\m}N_\l}.
\label{2.49}
\eeq
For a fixed $\l$ it follows from eq. \rf{2.49} that the critical
point $\m_c(\l)$ is determined by:
\beq
\e^{\m_c(\l)} = N_\l,~~~~~{\rm i.e.}~~~\chi(\m,\l) \sim \frac{1}{\m-\m_c(\l)}.
\label{2.50}
\eeq
In the $(\m,\l)$-coupling constant plane we have a critical line $\m_c(\l)$,
as shown in fig.\,\ref{2_4}.
The model \rf{2.48} is defined to the
right of the critical line and  the scaling limit for a fixed $\l$ leads
to the susceptibility \rf{2.50}, {\it i.e. the susceptibility
exponent $\g=1$}. From the general arguments presented in the
last subsection it follows that $G(x;\m,\l)$ falls of exponentially
for large $|x|$ for any $(\m,\l)$ to the right of the critical line:
\beq
\lim_{|x|\to\infty} \frac{-\log G(x;\m ,\l)}{|x|} = m(\m,\l) .
\label{2.51}
\eeq
However, in this model we have in addition an independent
correlation between tangents:
\beq
\bra \he_n \cdot \he_1\ket \equiv \int \prod_{i=1}^n \d \he_i\;
[\he_n \cdot K(\he_n,\he_{n-1}) \cdots K(\he_2,\he_1) \he_1].
\label{2.52}
\eeq
For a fixed $\l$ the tangent-tangent correlation function falls of
exponentially with the number of steps:
\beq
\bra \he_n \cdot \he_1\ket \sim \e^{-n m_t(\l)},~~~~m(\l) > 0.
\label{2.52a}
\eeq
This implies  that we for a fixed $\l$  just have
an ordinary random walk in the scaling limit. After $n_0 \sim 1/m_t(\l)$ steps
the initial orientation of the a tangent will be lost. If we
group together $n_0$ steps we will have an ordinary
random walk where there is no correlation between successive groups of steps.
Can we ever get any non-trivial behavior? The proof of \rf{2.52a}
will show the way to a non-trivial random walk behavior.

\vspace{12pt}
In order to prove eq. \rf{2.52a} let us view $K_\l(\he_2,\he_1)$ as
a kernel for an operator $\hKl$ acting on $L^2 (S^{D-1})$:
\beq
(\hKl \phi) (\he_2) = \int \d\he_1\; K_\l (\he_2,\he_1) \phi(\he_1),~~~~~
\phi \in L^2 (S^{D-1}).
\label{2.53}
\eeq
The kernel is symmetric  and $K_\l(\he_2,\he_1) >0$ and $\hKl$ is
compact. The Perron-Fr\"{o}benius theorem tells us
that the largest eigenvalue is non-degenerate and that the corresponding
eigenfunction is the only one which can be chosen positive.
Since the constant function clearly  is an eigenfunction of $\hKl$ with
eigenvalue 1 we conclude that 1 is the largest eigenvalue. It is easy
to show that $-1$ is not an eigenvalue. Finally
$\phi_a (\he) = \hat{a}\cdot \he$ is an eigenfunction since
\beq
(\hKl \phi_a)(\he) \equiv \int \d\he_1\; K_\l(\he,\he_1)\; \hat{a}\cdot \he_1
=\a(\l)\; \hat{a}\cdot\hat{e} \, ,
\label{2.54}
\eeq
The  equality follows since the integral is linear
in $\hat{a}$ and invariant under
simultaneous rotation of $\hat{a}$ and  $\he$, i.e. proportional to
$\hat{a}\cdot \he$.  From Perron-Fr\"{o}benius it follows that $\a(\l) <1$,
and direct calculation shows:    
\beq\label{2.54a}
0\leq \a(\l) \to 1~~~~~{\rm for}~~~ \l \to \infty.
\eeq
We can directly apply \rf{2.54} in eq. \rf{2.52} to get
\beq
\bra \he_n \cdot \he_1\ket = \a^n(\l) = \e^{-n m_t(\l)},~~~~
m_t(\l)= \log \frac{1}{\a(\l)}.
\eeq

\vspace{12pt}

The only possibility to get a non-trivial scaling limit is to take
$\l \to \infty$ simultaneously with $\m \to \m_c(\l)$ since
the tangent-tangent correlator can only  approach
macroscopic distances for $\l \to \infty$.
At the same time it is necessary that $\m
\to \m_c(\l)$ since this is the only possibility for the
two point function to be non-trivial according to \rf{2.51}.
{\it Such a scaling limit exists}. In order to construct the limit
it is useful to introduce the propagator $G(\he_f,\he_i;x)$ which
depends explicitly on the first and last step in the random walk:
it is defined by \rf{2.48} except that there is no integration
over $\he_1 =\he_i$ and $\he_n = \he_f$. $G(\he_f,\he_i;x)$ can
be viewed as a kernel for an operator $\hG$ on $L^2(S^{D-1})$
precisely as $K_\l(\he_2,\he_1)$ is the kernel of $\hKl$. Let
$| 1 \ra $ denote the constant function:
$1= 1(\he).$ We have by definition:
\beq
G(x;\m,\l) = \la 1|\, \hG\, | 1 \ra \equiv
\int \d\he_f\d\he_i \; G(\he_f,\he_i;x).
\label{2.55}
\eeq
By Fourier transformation we can get rid of the $\del$-function in
\rf{2.48}. If we introduce the notation $\Dm = \m-\m_c(\l)$ and denote
the Fourier transformation of $G(\he_f,\he_i;x)$, $\hG (x)$ and $G(x;\m,\l)$
by $G(\he_f,\he_i;q)$, $\hG (q)$ and $G(q;\m,\l)$, we can write:
\beq
\hG(q) = \sum_n \e^{-\Dm n} \left[\e^{-iq\cdot \he} \hKl\right]^n =
\frac{1}{1-\e^{-\Dm-iq\cdot \he}\hKl}
\label{2.56}
\eeq
In this formula $q \cdot \he$ is viewed as a multiplication operator, i.e.:
\beq
(f(q\cdot \he) \phi) (\he) = f(q\cdot \he)\phi(\he).
\eeq
We have in analogy with \rf{2.55}:
\beq
G(q;\m,\l) = \la 1 | \hG (q) | 1\ra \equiv \int \d\he_f \d\he_i\,
G(\he_f,\he_i;q).
\label{2.57}
\eeq

While these expressions look somewhat formal, they allow a rather
transparent discussion of the scaling limit. Let us consider a
fixed $\l$ and take $\Dm$ and $q$ to zero. Consider the matrix
element:
\beq
\la 1 | 1-\e^{-\Dm -iq\cdot \he} \hKl | 1 \ra \approx
\la 1 | \Dm +iq\cdot \he + \oh (q \cdot \he)^2 | 1 \ra  = \Dm + c q^2.
\label{2.58a}
\eeq
Let us now introduce the ordinary scaling for the random walk:
\beq
\Dm \sim m_{{\rm ph}}^2 a^2,~~~~q \sim p_{{\rm ph}} a,~~~~
a(\m,\l) \sim  \sqrt{\m -\m_c(\l)}.
\label{2.58}
\eeq
For fixed $\l$ there is a finite mass gap from 1 to the next lowest
eigenvalue. This implies that the only matrix element which
contributes to \rf{2.57} to leading order in $a$ is \rf{2.58a}:
\bea
G(q;\m,\l) &=&\la 1 | \frac{1}{1-\e^{-\Dm-iq\cdot \he}\hKl} | 1\ra
\nn \\
& \sim &
\frac{1}{a^2}\left[ \frac{1}{p_{{\rm ph}}^2+m^2_{{\rm ph}}}
 + O (\sqrt{a})\right],
\label{2.59}
\eea
i.e. just the ordinary propagator. However, if $\l \to \infty$ the mass
gap goes to zero and more matrix elements will contribute to
$\la 1 | \hG (q) | 1\ra$ and the expansion in \rf{2.59} is no
longer valid. By Taylor expanding $\phi(\he)$ around a fixed vector
$\he_0$ and using that $e^{-\l f(\th(\he_0,\he))}$ will be peaked around
$\he_0$ one can show:
\beq
\hKl \to \e^{-c(\l) L^2},~~~ c(\l) \sim \l^{-2} ~~{\rm for} ~~\l \to \infty.
\label{2.60}
\eeq
where $L^2$ denotes the Beltrami-Laplace operator on $S^{D-1}$. Let
us now introduce a lattice length scale by
\beq
c(\l) = \l_{{\rm ph}} a(\l),~~~~m(\m,\l)\equiv m_{{\rm ph}} a(\l)= \Dm.
\label{2.61}
\eeq
The last equation fixes $\m$ as a function of $\l$ and defines
the approach to $\m_c(\l)$ for $\l \to \infty$.
From \rf{2.59} we get
\bea
G(q;\m,\l) &=&\la 1 | \frac{1}{1-\e^{-\Dm-iq\cdot \he}\hKl}
| 1\ra \nn \\
&\sim& \frac{1}{a(\l)} \;
\la 1 | \frac{1}{ \l_{{\rm ph}} L^2 + m_{{\rm ph}} + i p_{{\rm ph}}\cdot \he} | 1 \ra.
\label{2.62}
\eea
This is our final expression. The matrix element on the rhs is
expressed in terms of continuum variables, and the scaling factor in
front tells us that $\eta =1$ if we compare with eq. \rf{2.31c}.
Likewise eq. \rf{2.31} and $m(\m,\l) = \Dm$ shows
that $\n =1$, i.e. $d_H =1$. {\it We have a new class of smooth
random walks  ($d_H=1$).} We have already shown that $\g =1$
and the exponents $\g,\n$ and $\eta$ is our first example
of a set of non-trivial exponents. They satisfy Fischer's scaling relation.

The ordinary random walk has a stochastic interpretation as
a Brownian motion of a particle, i.e. each step is
performed according to some
probability distribution $P$, but is independent of the former
steps. With the extrinsic curvature term added the step also
depends on the direction of the former step. For a finite
coupling constant this does not change the universality class of the
random walk, but as we take $\l \to \infty$ we enter a new
class of random processes characterized by different critical
exponents. The interpretation of these is that
the {\it velocity, rather than the position of the particle itself,
is changed stochastically according to some probability distribution $P$}.
The path in an
ordinary random walk will be continuous, but, with probability
one, nowhere differentiable. If the velocity is stochastic the
typical path will be differentiable and the first derivative
continuous but nowhere differentiable. There is a number of
stochastic processes which have this feature. The well known
Ornstein-Uhlenbeck
process \cite{ornstein} is one of them. It can be shown that the propagation
of particles in such processes are described by propagators of
the type \rf{2.62}, but we have to refer to the original articles
for details \cite{adjrw}.

The progagator \rf{2.62} is also related to the propagation of
a spinning particle with infinitely many components. This
will be discussed at  the end of the next subsection.

\subsection{Fermionic random walks}

For the ordinary bosonic propagator we found the following representation:
\beq
\la 1 | \hG(q)  | 1\ra \sim  \frac{1}{\Dm + q^2} =
\frac{1}{a^2}\frac{1}{m^2_{{\rm ph}} + p_{{\rm ph}}^2}. \label{2.70}
\eeq
The fermionic propagator in $D$ dimensions is obtained by the change
\beq
\frac{1}{p^2+m^2} \to \frac{1}{i p^\m \g^\m +m},  \label{2.71}
\eeq
where the $D$ $\g$ matrices satisfy:
\beq
\{ \g^\m,\g^\n\} = 2 \del^{\m\n},~~~~~\left(\g^{\m}\right)^\dg = \g^\m
\label{2.72}
\eeq
The lowest dimensional representation of this so-called Clifford algebra
is by matrices of dimension $n = \left[D/2\right]$. In two dimensions
one can take the Pauli matrices $\sg_1$ and $\sg_2$. In three dimensions
one  can use all three Pauli matrices. The $\g$-matrices can be used 
to construct a representation of ${\rm SO}(D)$: The matrices
\beq
s^{\m\n} = \oh \sg^{\m\n} = \frac{i}{4}[\g^\m,\g^\n] \label{2.73}
\eeq
satisfy the commutation relations for the generators of ${\rm SO}(D)$
and rotations will be generated by
\beq
K(\om) = \e^{-i s^{\m\n}\om^{\m\n}/2}.   \label{2.74}
\eeq
These rotations act on {\it spinors}, i.e. vectors in $C^n$,
$n= \left[D/2\right]$. If
\beq
\om^{\m\n} = \th n^{\m\n}(\he_1,\he_2),~~~~~
n^{\m\n} = \frac{\he_1^\m\he_2^\n-\he_1^\n\he_2^\m}{2 \sin \th},
\label{2.75}
\eeq
where $\th$ is the angle between $\he_1$ and $\he_2$,
the rotation \rf{2.74} is a rotation with angle $\th$ in the plane spanned
by $\he_1$ and $\he_2$. An important aspect of the spinor representation
of the rotation group is that a $2\pi$-rotation gives -1. This
well known fact will be of outmost importance for the
fermionic random walk.

Let $x^\m(\xi)$ be a curve in $R^D$ and let $t^\m(\xi)$ be the
normalized tangent vector, defined by \rf{2.40a}-\rf{2.41}.
$t^\m$ and $\dot{t}^\m$ span the osculating plane of the curve
and we can write:
\beq
\om^{\m\n} \equiv \oh (t^\m \dot{t}^\n-t^\n\dot{t}^\m) = \k n^{\m\n}.
\eeq
Recall that the curvature $\k$ is related to the  angle of contingency
by \rf{2.43}. This implies that we can write
\beq
\om^{\m\n} \d l =  \d\th n^{\m\n} \label{2.76}
\eeq
where $n^{\m\n}$ is the antisymmetric tensor which defines the
osculating plane and $d\th$ is the angle between tangent vectors $t(l)$
and $t(l+\d l)$.  If we consider a discretized random walk  and as
usual denote the two successive unit vectors as $\he_i$ and $\he_{i+1}$
and the angle between them as $\th(\he_{i+1},\he_i)$ the
discrete analogy of \rf{2.76} is
\beq
\om^{\m\n}(\he_{i+1},\he_i)  =  \th (\he_{i+1},\he_i)
n^{\m\n} (\he_{i+1},\he_i) \label{2.77}
\eeq
Let $\psi\in C^n$ be a  spinor. Let us imagine it
``propagates'' along the given path in such a way that it is
always rotated according to the orientation of the curve, i.e. for
each discrete step we perform a rotation given by the rotation
matrix
\beq
K(\he_{i+1},\he_i) = \e^{-is^{\m\n}\om^{\m\n}(\he_{i+1},\he_i)}.
\label{2.78}
\eeq
The total rotation during a travel along the path $P_n$ will be given by
\beq
K(P_n)  = K(\he_n,\he_{n-1}) \cdots K(\he_2,\he_1). \label{2.79}
\eeq
A formal continuum version of $K(P_n)$ for a smooth path is
\beq
K(P) = \cP \e^{-\frac{i}{2}\int_P \d l\; s^{\m\n} \om^{\m\n} (l)} \label{2.80}
\eeq
where $\cP$ denotes the path ordered integral. This factor appears
in the famous {\it Strominger-Polyakov representation of the
fermionic propagator}\cite{polbook}:
\beq
G(x,y) = \int\cD P(x,y)\; \ e^{-m\int_{P(x,y)} \d l} K(P(x,y)).
\label{2.81}
\eeq
The definition is identical to the one for the bosonic particle
except for the  matrix $K$, which rotates a spinor ``along the curve''.
The factor $K(P)$ is rather ill defined and it has been
difficult to use this expression.
However, from the above definitions it is clear how to write down
a well defined discretized version of \rf{2.81} \cite{adjfrw}:
\beq
G_\m (\he_f,\he_i,x) = \sum_n \e^{-\m n} \int \prod^{n}_{i=1} \d\he_i \;
\prod_{i=0}^{n} K(\he_{i+1},\he_i) \;\del(\sum \he_i -x),
\label{2.82}
\eeq
where $\he_0=\he_i$ and $\he_f=\he_{n+1}$.
This expression is very similar to \rf{2.48} for the bosonic particle
and we can directly use the machinery developed in the last subsection.
The matrix $K(\he_2,\he_1)$ can be viewed as the kernel for an
operator $\hK$ which acts on wave functions belonging to
$\cH=L(S^{D-1})\times C^n$, i.e. spinors on $S^{D-1}$.
If we normalize $K$ we have:
\beq
(\hK \psi) (\he) \equiv \int \d\he_1 \;K(\he,\he_1)\;\psi(\he_1),~~~~~~
\int \d\he_1 \; K(\he,\he_1) = \hat{1}.
\label{2.83}
\eeq
In the same way we can view $G_\m(\he_f,\he_i,x)$ as the kernel
of an operator $\hG (x)$ on $\cH$. We get rid of the $\del$-function
in eq. \rf{2.82} by Fourier transformation and express the Fourier
transformed $\hG (q)$ as
\beq
\hG_\m (q)  = \frac{1}{1-\e^{-\m -iq\cdot\he} \hK}   \label{2.84}
\eeq
in the same way as for the scalar particle (see eq. \rf{2.56}).

Let us now define the following scaling limit:
\beq
\m = m_{{\rm ph}}a,~~q = p_{{\rm ph}} a, \label{2.85a}
\eeq
i.e. the propagator becomes:
\beq
\hG_\m(q) = \frac{1}{a}\; \left[\frac{1}{\frac{1-\hK}{a} +
m_{{\rm ph}} + i p_{{\rm ph}}\cdot \he + O(a)}\right]. \label{2.85}
\eeq
As usual $p\cdot\he$ should be viewed as an multiplication operator.
As for the scalar particle it can be shown that $1$ is the largest
eigenvalue and that there is a gap to the next lowest eigenvalue
(there is no extrinsic curvature which allows us to tune the gap
to zero). It follows that only eigenvectors corresponding
to the eigenvalue $1$ will propagate for long paths and
contribute to matrix elements of ${\hG}_\m (q)$ for $a\to 0$.
For the scalar particle we only got a non-trivial result from
the scaling ansatz \rf{2.85a} in the limit where the extrinsic
curvature term reduced the mass gap to zero. {\it Here it is
different because the eigenvalue 1 is degenerate}. It
is $2n$-times degenerate and the operator ${\hG}_\m(q)$ acting
on the Hilbert space $\cH$ reduces in the scaling limit to
$2n\times 2n$ matrix  acting on the eigenspace $V$ corresponding
to the eigenvalue 1.

\vspace{12pt}
\noindent
{\bf Theorem:} The constant vectors and the column vectors of
the matrix $\g^\m \he^\m$ span the $2n$ dimensional eigenspace $V$
of $\hK$ corresponding to eigenvalue 1.

\vspace{12pt}
\noindent
That constant vectors on $C^n$ are eigenfunctions of $\hK$ of eigenvalue
1 is clear. By definition of the rotation we have:
\beq
K(\he_2,\he_1) \g\cdot \he_1 K^{-1} (\he_2,he_1) = \g\cdot \he_2.
\label{2.86}
\eeq
This shows that the columns of the matrix $\g\cdot \he$ are
eigenfunctions of eigenvalue 1 since:
\beqn
\hK \g\cdot \he = \int \d\he_1 \; K(\he,\he_1)\;\g\cdot \he_1 =
\g\cdot \he \int \d\he_1 K(\he,\he_1) = \g\cdot \he.
\eeqn
We leave it as an interesting exercise to show that there are no other
eigenvectors corresponding to the eigenvalue 1 (see \cite{adjfrw}).

\vspace{12pt}

The projection operators $P^{\pm} = \frac{1}{\sqrt{2}} ( 1 \pm \g\cdot \he)$
commutes with $\hK$ and split $\cH$ into ``$\pm$-chirality spaces'':
\beqn
\cH = \cH_+ \oplus \cH_-,~~~~~~V=V_+\oplus V_-.
\eeqn

\vspace{12pt}
\noindent
{\bf Theorem:} On $V_\pm$ the multiplication operators $\he^\m$
can be replaced by $\pm\g^\m$-matrices.

\vspace{12pt}
\noindent
The proof is simply by calculation: Let $\psi(\he)$ and $\phi(\he)$
be two vectors in $\cH$. The scalar product on $\cH$ (and $V$) is
\beqn
\la \phi | \psi\ra = \int \d\he\;  \phi^* (\he)_\a  \psi(\he)_\a
\eeqn
where $\a$ is a spinor index. Let us now apply this to $n$ vectors
$v^{(\b)}_\a(\he) = P^+_{\a\b}(\he)$. These are orthonormal vectors on $V_+$.
It follows that
\beqn
\la v^{(\b_1)} | \he^\m | v^{\b_2}\ra =
\int \d\he  \; {v}^{*(\b_1)}_\a(\he)  \he^\m v^{(\b_2)}_\a (\he) =
\frac{1}{D} \g^\m_{\b_1\b_2}.
\eeqn
We can finally state that the scaling limit \rf{2.85} is
given by
\beq
\hG_\m (q) \to \frac{1}{a}
\pmatrix{\frac{1}{m_{{\rm ph}} -ip_{{\rm ph}}\cdot \g} & 0 \cr
         0 & \frac{1}{m_{{\rm ph}} +ip_{{\rm ph}}\cdot \g}}.
\label{2.87}
\eeq
We recognize two copies of the Dirac operator corresponding to $V_\pm$.
This doubling is needed by the Nielsen-Ninomiya theorem.

\begin{figure}
\input{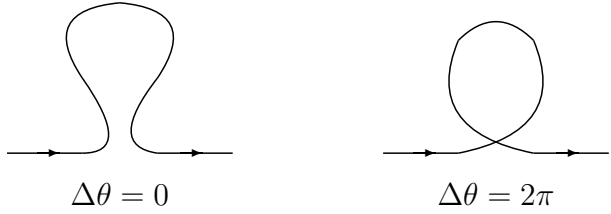}
\caption[2_5]{The cancellation between two paths. For the first we get
a phase factor $e^{i\Delta \th/2} = 1$, while the second gives
a phase factor $e^{i\Delta \th /2} = -1$.}
\label{2_5}
\end{figure}

On the way we have determined the critical exponents for
the fermionic random walk. From  eq. \rf{2.85} and 
$\hG_\m(q=0) = \int \d x \,\hG(x)$ we get $\g=1$. From $m(\m) = \m$ we get
$\n =1$ and from $G_\m (q) \sim 1/a$ we get $\eta =1$. {\it The fermionic
random walk has the same critical exponents as the smooth random walk}.
This implies that $d_H =1$ and effectively we have smooth paths. However,
we have no extrinsic curvature term to produce the smoothness. Rather,
it comes about because of cancellations between a large number
of bosonic paths. To show this let us consider two dimensions
(but the mechanism is the same in higher dimensions).
From \rf{2.76} it follows that we can write:
\beq
\om^{\m\n}s^{\m\n} = \frac{\d\th}{\d l} \sg_3   \label{2.88}
\eeq
and the eigenvalues of $\sg_3$ is precisely the split of $\cH$ in
$\cH_\pm$. On one of these spaces $K(P)$ \rf{2.80} becomes
a phase factor and \rf{2.81} allows a scalar interpretation:
\beq
G(x) = \int \cD P(x) \; \e^{-m \int \d l +\frac{i}{2} \int \d\th},
\label{2.89}
\eeq
where $\th(l)$ is the angle of the tangent relative to a
fixed direction in the plane. {\it It is now essential that we
have spinors since the $-1$ resulting from a $2\pi$ rotation
leads to a cancellation between intersecting and non-intersecting paths}
as shown in fig.\,\ref{2_5}. Only smooth paths survive this cancellation.

Even if the mechanism for producing the smooth paths seems very
different for the bosonic and the fermionic particles
there is a connection. As is rather clear from the derivation
of the bosonic propagator \rf{2.62} the tangent vectors are
effectively performing a random walk on $S^{D-1}$. This is
just what one would expect from a classical spin and this
is made clear from the expression for the operator $\hKl$
in the scaling limit:
\beq\label{2.90a}
\hKl \sim \e^{-\l_{{\rm ph}} L^2 a}.
\eeq
Since $\hKl$ is the {\it transfer matrix} for proper
time evolution along the path and since the Hausdorff dimension
is one, $L^2$ can be viewed as the continuum Hamiltonian, and
we know that it is precisely the Hamiltonian for a
classical spinning particle.  This interpretation is
substantiated by canonical quantization of the particle
with extrinsic curvature. The Lagrangian contains
higher derivatives and this implies that $\dot{x}/|\dot{x}|$ will
be a new additional canonical coordinate which will serve as a
the spin. We have to refer to \cite{ply} for further discussion.
Here it is worth emphasizing that this result is quite natural
in the discretized approach. In addition it is seen that
the wave function of such a classical spinning particle
is an infinite component object as is clear from eq. \rf{2.62}.
It is closely related to the infinite component spinor
introduced by Majorana in an attempt to avoid the negative
eigenstates of the Dirac operator.
From this point of view
it is possible to consider the Dirac propagator in a more
a more general context.
If the matrix $K(\he_1,\he_2)$ had been generalized to contain
an extrinsic curvature term, one would have obtained
a scaling limit similar to \rf{2.62}, the
only difference being that
\beq\label{2.91}
\l_{{\rm ph}} L^2 \to \l_{{\rm ph}} (J^2-s^2),
\eeq
where $s^2$ is the spin operator while $J^2$ refers to the
total angular momentum operator. For $\l_{{\rm ph}} \to \infty$
it will project to the finite dimensional subspace which we obtained above
for the ordinary Dirac operator but in a limiting process where the
Hausdorff dimension always is one.

It is natural to ask the following question: Although  we have found
a nice discretized version of \rf{2.81} and have shown that
it leads to the Dirac propagator in the scaling limit, \rf{2.81}
itself is not a path integral of an action. Is there a path integral
which leads to \rf{2.81}? The answer is yes! The supersymmetric
generalization of the bosonic action \rf{2.3x}. The supersymmetric partner
of the field $x^\m$ is a Grassmann variable $\psi^\m$ while the
supersymmetric partner of $e(\xi) \equiv \sqrt{g(\xi)}$ is a
``gravitino'' field $\chi(\xi)$ and we have
\bea
S_B &=& \int \d\xi \;\left[\frac{1}{e}\dot{x}^2 + e\right] \to  \nn \\
S_F &=& \int \d\xi \left[\frac{1}{e} \dot{x}^2 - \psi^\m \dot{\psi}^\m +
\frac{1}{e} \chi \psi^\m \dot{x}^\m +e -
\oh \chi \left(\frac{\d}{\d\xi}\right)^{-1} \chi \right].  \label{2.90}
\eea
It is possible to show that the path integral over the
Grassmann variable $\chi$ and $\psi$ results in the factor $K(P)$
in \rf{2.80}. Details can be found in \cite{polbook}.
\newsection{Random surfaces and strings} \label{surface}

\subsection{Definition of the model}

The theory of random paths described the relativistic particle.
We expect the theory of random surfaces to describe the
relativistic string. The strings sweep out
a surface while they  propagate. 
The path integral is a sum over all such surfaces with a
weight given by the classical action. It is our goal to define
this sum and analyze it in detail.

As for the relativistic particle we have two actions which are
equivalent at the classical level. The first action
is geometrical, only determined by the area spanned
between the starting position and the end position of the closed
string\footnote{For simplicity we consider here only the theory
of closed strings.}. Let $M(l_i)$ denote the manifold
with boundaries $l_i$, $i=1,...,n$, $F(l_i)$ a corresponding surface
in $R^D$  and $x^\m(\xi)$ the coordinates of $F$ in $R^D$. The action is
\bea
S[F(l_i)] &=& \m \int_{F(l_i)} \d A(F) \label{3.1} \\
&=& \m \int_{M(l_i)} \d^2 \xi \;
\sqrt{\left(\frac{\prt x^\m}{\prt \xi^1}\right)^2
\left(\frac{\prt x^\m}{\prt \xi^2}\right)^2-
\left(\frac{\prt x^\m}{\prt \xi^1}\frac{\prt x^\m}{\prt \xi^2}\right)^2},
\nn
\eea
An alternative description is obtained by introducing an {\it internal
metric} $g_{ab}$, $a,b=1,2$ on $M(l_i)$ and use the following action:
\beq
S[g,x] = \frac{1}{\a'} \int_{M(l_i)} \d^2\xi\,\sqrt{g}
\left[g^{ab}\frac{\prt x^\m}{\prt \xi^a}\frac{\prt x^\m}{\prt \xi^b}
+ \m \right].
\label{3.2}
\eeq
The classical equations for the actions \rf{3.1} and \rf{3.2} agree,
but it is
not at all obvious that the quantum theories are identical.
In the case of strings we have many  ``natural'' objects, in contrast to
the situation for the free particle where one only has the two-point function.
A surface can join an arbitrary number of strings and one is led to
consider the $n$-loop amplitude $G(l_1,...,l_n)$ between
the $n$ loops or strings. The formal path integral expression for
the $n$-loop is written as:
\beq
G(l_1,\ldots,l_n) = \int \cD F(l_i) \; \e^{-S[F(l_i)]},
\label{3.3}
\eeq
where the integration is over {\it physical distinct surfaces $F$
in $R^D$}, or
\beq
G(l_1,\ldots,l_n) = \int_{M(l_i)}
\frac{\cD g_{ab}}{{\rm Vol(diff)}}\int \cD_g x \; \e^{-S[g,x]}.
\label{3.4}
\eeq
where the integration is over all equivalence classes of metrics on
$M(l_i)$ and all embeddings $x^\m (\xi)$.

In order to define \rf{3.3} and \rf{3.4} we have to introduce
a {\it reparametization invariant cut-off}. We follow the procedure
outlined for the relativistic particle. In the case \rf{3.3}
it amounts to use as a building block a smallest triangle in $R^D$
and glue together these triangles in all possible ways
to surfaces with the given boundary conditions.
Alternatively one could consider the hyper-cubic
lattice surfaces where the surfaces are made of plaquettes 
\cite{weingarten,dfj}.
Many of the results we will obtain in the following are valid
(and easier to prove) for these models than for the model defined
by eq. \rf{3.4}. However, eq. \rf{3.4} relates closer to
quantum gravity, since it is just two-dimensional gravity
coupled to $D$ free scalar fields $x^\m$, and for this reason
it is   convenient to
use here eq. \rf{3.4} rather than eq. \rf{3.3} as  we are
going to consider quantum gravity in some detail
\cite{adf,david1,kkm}
(again there is a large number of articles describing
this approach \cite{jkp,bd,afkp,djkp}, just to mention some of 
the articles which concentrated on computer simulations of 
the model).

In the case of the random walk the reparametrization invariant cut-off
introduced was related to the shortest distance. In the two dimensional
case it is natural to combine shortest distance and smallest area in
a single cut-off. The fundamental building block will in this way
be an equilateral triangle with edge length $a$. At this point
we encounter a new problem compared with the one-dimensional
situation: The gluing is in no way unique. In the process of gluing
together triangles to form a two-dimensional manifold the order
of a given vertex (i.e. the number of triangles to which the
vertex belongs) is almost arbitrary.
In the case where we use eq. \rf{3.3} and glue together triangles
directly in $R^D$ the answer is clear: We should include all {\it distinct
 different}\footnote{By different ways of gluing we have in mind that the
resulting (abstract) triangulations are different in the way defined below.}
ways of gluing compatible with the boundary conditions
since we will get  different surfaces in $R^D$. But also
for the model \rf{3.4} where the triangles are defined
with respect to the internal metric the freedom of gluing will
go hand in hand with the fact that a closed surface,
apart from the total area (the equivalence of the total
length of the path), also has a new {\it local} invariant: $R(\xi)$ :
the Gaussian curvature. $R(\xi)$ cannot be changed by a reparametrization
of the surface.
In the following it will be argued that the sum over triangulations
in a precise way captures this new degree of freedom.

\vspace{12pt}

\begin{figure}
\input{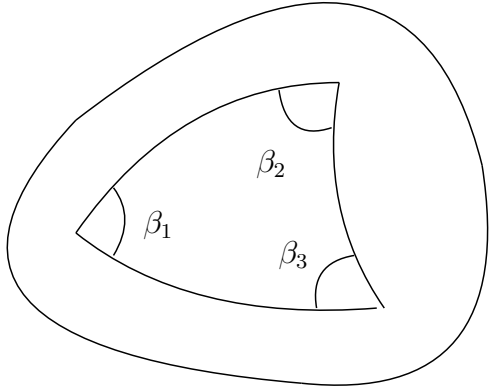}
\caption[3_1]{A geodesic triangle on a smooth surface.}
\label{3_1}
\end{figure}

Let us make a short digression and discuss curvature. The curvature
(or Riemann) tensor $R^d_{abc}$ can be defined in terms of the metric. It
describes the deviation from flat space. This is manifest in the
formula for  parallel transport of a vector $S^a$
around an infinitesimal closed curve:
\beq
\Delta S_a = \oh R^d_{~abc} S_d \oint x^c \d x^b. \label{3.5}
\eeq
The once contracted tensor $R_{ab}$ is called the Ricci tensor and
the scalar obtained by contracting the Ricci tensor is called
the scalar curvature $R$:
\beqn
R_{ab} = R^c_{~acb},~~~~~~R = R^a_{~a}.
\eeqn
In two dimensions there is only one independent component:
\beq
R_{abcd} = \oh (g_{ac}g_{bd} - g_{ad}g_{bc}) R,~~~~R_{ab} =\oh g_{ab}R,~~~~
R=2 K.
\label{3.6}
\eeq
where $K$ denotes the Gaussian curvature on the surface. $K$ has
the simple geometrical interpretation as being the product of
the principal curvatures associated the normal planes intersecting
the surface which is assumed to be embedded in $R^D$ in order
for this interpretation to make sense. But $R$ itself
is of course independent of this embedding. Eq. \rf{3.5} simplifies:
\beq
\Delta S_a = \oh R S^b \d A_{ab},~~~~~~d\th = \oh R \d A = K \d A, \label{3.7}
\eeq
where $\d A_{ab}= \oh \oint(x_a\d x_b-x_b\d x_a)$ is the area tensor
and $d\th$ the infinitesimal angle by which $S^a$ has been rotated
during the parallel transport. There is a nice integrated version
of this relation, known as (one of the versions of) the Gauss-Bonnet
theorem: Let $T$ be a geodesic triangle on the surface, i.e. a triangle
where the sides are geodesic curves, with angles $\b_1,\b_2$ and $\b_3$.
The sum of the angles is no longer $\pi$ but the deviation from
the Euclidean value is given by the integral of the Gaussian curvature
over the interior of the triangle:
\beq
\int_T K\, \d A  = \b_1+\b_2+\b_3 -\pi \equiv \ep_T,  \label{3.8}
\eeq
where $\ep_T$ is called the excess angle of the triangle. The
situation is illustrated in fig.\,\ref{3_1}. Parallel transport
of a vector along the boundary of the triangle will rotate the vector
by the angle $\ep_T$.

\begin{figure}
\input{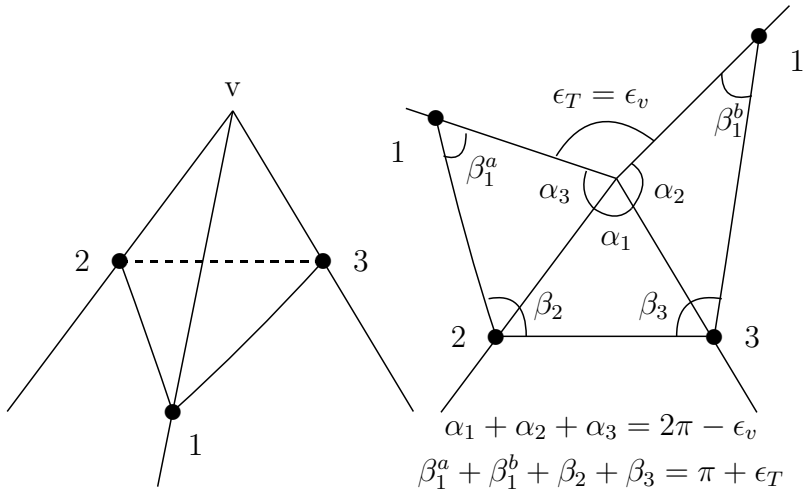}
\caption[3_2]{A geodesic triangle on a piecewise linear
surface. The vertex $v$ is an interior point of the triangle.
(fig.\,\ref{3_2}a).
Fig.\,\ref{3_2}b shows the excess angle after the piecewise linear
neighborhood has been cut open along a link and unfolded in the
plane.}
\label{3_2}
\end{figure}

Due to Regge \cite{regge} we can define curvature and parallel transport
in a natural way on piecewise linear surfaces.
The curvature cannot be located in the interior
of the triangles since we view the interior as flat.
Since the curvature is defined as an intrinsic
geometric quantity it is clearly
{\it bending invariant}\footnote{Historically the Gaussian curvature
was defined in terms of the principal curvatures $\k_1$ and $\k_2$,
which explicitly depended on the embedding. It came as a surprise that
$K = \k_1\k_2$ is independent of the embedding (bending invariant).
This is a ``Theorema egregium'', a ``most excellent theorem'',
wrote Gauss.}. Since we can bend the surface around an edge
without changing anything we cannot use the edges either.
In this way we are lead to locate the curvature
of the piecewise linear surfaces at the vertices. To each vertex $v$
we associate a deficit angle $\ep_v$ by:
\beq
 \ep_v = 2\pi -\sum_{t \ni v} \a_v(t), \label{3.9}
\eeq
where the summation is over the $v$-angles of the triangles to which
$v$ belongs. This is illustrated in fig.\,\ref{3_2} where
a geodesic triangle with the vertex $v$ of the piecewise
linear manifold is shown.
It follows from fig.\,\ref{3_2}
that $\ep_T=\ep_v$ on the piecewise linear surface, i.e. we can write:
\beq
\int_T K\, \d A  \sim \ep_v,~~~~~\int K\, \d A \sim \sum_v \ep_v.  
\label{3.10}
\eeq
The lhs of the equations are intended to be valid for
smooth surfaces, the rhs for piecewise linear surfaces.
In the last equation we have generalized the formula to any region
where the boundary is a piecewise geodesic curve and the
summation is intended to be over interior vertices in the
triangulation. In particular the formula will be valid for
closed surfaces.

Let us split $\ep_v$ in a curvature term and an area term by
assigning the following area to the vertex $v$:
\beq
A_v = \frac{1}{3} \sum_{t \ni v} A_t,~~~~~K_v = \frac{\ep_v}{A_v},
\label{3.11}
\eeq
i.e. the area of each triangle is distributed equally among its three
vertices. With these definitions we have finally:
\beq
\int \d A \sim \sum_v A_v,~~~~~\int K\,\d A  \sim \sum_v  K_v A_v.
\label{3.12}
\eeq
Regge originally intended to use these formulas by constructing
a sensible sequence of piecewise linear approximations
to a {\it given smooth surface} such that
\beq
\sum_v A_v f_v \to \int \d A(\xi) f(\xi),  \label{3.14}
\eeq
\beq
\sum_v  f_vK_v \,A \to \int  f(\xi) \,K(\xi)\, \d A(\xi) .  \label{3.15}
\eeq

On the same grid one can discretize the covariant action
of the free scalar fields in a natural way. For a given triangulation
we denote the vertices by indices $i,j,k,...$. The triangulation
is characterized by its coincidence matrix, which specifies the
neighbor vertices to a given vertex (see below for definition), 
and by the length $l_{ij}$
of the links between vertices $i$ and $j$. A natural coordinate
system is introduced on the piecewise linear surface by assigning
coordinates $y_i$ to the vertices $i$ such that $l_{ij} = |y_i-y_j|$.
The $y_i$'s  live in some ambient space $R^n$. The interior of the
the triangle $(ijk)$ is parametrized by barycentric coordinates
and for a field $\phi$ defined at the vertices by $\phi_i$
we use in the same way the linear extension to the interior of the triangle:
\beq
y = \xi^1 y_i+ \xi^2y_j+(1-\xi^1-\xi^2)y_k  \label{3.16}
\eeq
\beq
\phi(y) = \xi^1 \phi_i + \xi^2\phi_j+ (1-\xi^1-\xi^2)\phi_k \label{3.17}
\eeq
The metric will be defined by
\beq
g_{ab} = \frac{\prt y^\a}{\prt \xi^a}\frac{\prt y^\a}{\prt \xi^b},
~~~a,b =1,2 \label{3.18}
\eeq
and it is straightforward to calculate $g_{ab}$ and
$g^{ab} \prt_a \phi \prt_b \phi$
\beq
g_{ab} = \pmatrix{l_{ik}^2 &\oh (l_{ik}^2+l_{jk}^2-l_{ij}^2) \cr
                  \oh (l_{ik}^2+l_{jk}^2-l_{ij}^2) & l^2_{jk} },
\label{3.19}
\eeq
\bea
g^{ab}\prt_a\phi\prt_b \phi &=& \frac{1}{g}\left[
l_{ij}^2 (\phi_i-\phi_k)(\phi_j-\phi_k)\right. + \nn \\
& &\left. l_{jk}^2 (\phi_j-\phi_i)(\phi_k-\phi_i) +
l_{ik}^2 (\phi_i-\phi_j)(\phi_k-\phi_j)\right], \label{3.20}
\eea
where
\beq
g =( 2 \Delta_{ijk})^2 = \oh \left( l_{ik}^2 l_{jk}^2 + l_{ij}^2l_{kj}^2 +
l_{ik}^2l_{ij}^2 -\oh (l_{ij}^4+l_{ik}^4+l_{jk}^4) \right).
\eeq

\vspace{12pt}

After this digression let us return to the problem of regularizing
the integration of Riemannian structures of two dimensional manifolds.
The reparametrization invariant regularization suggested above
consists of constructing all piecewise linear manifolds obtainable
by gluing together equilateral triangles of side-length $a$ and assigning
to these the metric structure offered to us by Regge calculus. In the case
of equilateral triangles the formulas above simplify a lot: Let $n_v$ denote
the order of vertex $v$ in a given triangulation, i.e. the number of
triangles which contain vertex $v$. From eq. \rf{3.11} we get:
\beq
A_v = \frac{1}{3} n_v \;\frac{\sqrt{3}}{4} a^2,~~~~~R_vA_v
= \frac{2\pi}{3} (6-n_v).
\label{3.21}
\eeq
Let us for simplicity of notation absorb $\sqrt{3}/4$ in $a^2$ and put
the resulting $a=1$ in the following. The formulas for curvature and
for the Gaussian action then read:
\beq
A_v = \frac{1}{3}  n_v, ~~~~~R_v = 2\pi \frac{6-n_v}{n_v},
\label{3.22}
\eeq
\beq
\sum_v A_v = N_T,~~~\sum_v A_v R_v = 4\pi \chi,~~~~\chi= N_T-N_L+N_V.
\label{3.23}
\eeq
Here $N_T,N_L$ and $N_V$ denote the number  of triangles, links and vertices
in the triangulation, and $\chi$ denotes the
Euler characteristic of the manifold. \rf{2.23}
is the discretized version of the Gauss-Bonnet theorem mentioned
above\footnote{It is easy to prove eq. \rf{2.23} 
from Euler's formula which states
that $\chi = N_P-N_L+N_V$ for any polygon net covering a surface of topology
characterized by $\chi$. If one use Euler's formula on a triangulation
($2N_T=3N_L$) and combine it with \rf{3.22} one arrive at the formula
in \rf{3.23}.}. Finally the action of the free Gaussian field 
given by eq. \rf{3.20}
in the case of equilateral triangles becomes
\beq
\int \d^2\xi \,\sqrt{g} g^{ab} \prt_a \phi \prt_b \phi \to
\sum_{(ij)} (\phi_i -\phi_j)^2         \label{3.24}
\eeq
From \rf{3.22} one observes that different triangulations,
i.e. triangulations which  cannot be mapped onto each other
by a simple relabeling of the vertices, lead to different local
curvature assignments and consequently  inequivalent metric structures.
In two dimensions a closed manifold is characterized entirely
by its Euler number. Given a manifold we want to integrate over
equivalence classes of metrics. Since all the different triangulations
we can construct by gluing the equilateral triangles together
correspond to inequivalent  metrics it is clear that one should sum
over all such triangulations. By this prescription one approximates
a continuous integration over metrics by the summation over
a grid of points in the space of inequivalent metrics.
{\it The conjecture is that this grid becomes uniformly dense when
the number of triangles $N_T$ of the triangulations goes to
infinity}. We shall later on verify this conjecture.

Although we have used Regge's prescription for assigning curvature,
the philosophy outlined is very different from
the one which motivated Regge. In the classical Regge calculus
the objective was to approximate a given smooth surface
by a piecewise linear manifold. A fixed triangulation was chosen
and the link length treated as the dynamical variable which should
be adjusted to get the best approximation to the given
manifold\footnote{The given smooth
manifold in the Regge approach was the one given by solution to
Einsteins equations with suitable boundary conditions and it would
be an extremum of the Einstein-Hilbert action. An important
feature in this context is the convergence \rf{3.15} (in higher
dimensions where the Einstein-Hilbert action is not a topological
invariance), which ensured a good approximation to the action if the
triangulations were chosen well in accordance with the geometry
of the problem.}. In particular, different link assignment
will not necessarily result in a different metric assignment, as is
clear by considering triangulations of the plane. Clearly there is a
lot of room for moving the vertices around (and thereby changing the
link length)  without changing the metric at all. Integration over
link length is not an integration over equivalence classes
of metrics, but involves a highly non-trivial Jacobian.
Here we are not interested in approximating specific manifolds,
but in using different triangulations to label different
equivalence classes of metrics.

The regularized definition of the multi-loop Green functions
\rf{3.4} reads:
\beq
G_\m (l_1,\ldots,l_n) = \sum_{T \in \cT(l_1,\ldots,l_n)} \e^{-\m N_T }
\int \prod_{i\in T/{\{l_1,\ldots,l_n\}}} \d x_i\;
\e^{-\sum_{(ij)} (x_i - x_j)^2}. \label{3.25}
\eeq
In this formula $T$ denotes an abstract triangulation, defined
by its vertices $i$ and a table which tell us the neighbor vertices
to a given $i$. This information is encoded in the {\it coincidence
matrix} $\tilde{C}_T$ which is an $N_V\times N_V$ matrix where the $(ij)$
entry is $-1$ if $i$ and $j$ are neighbors, $0$ if they are not neighbors
and $n_v$, the order of the vertex, on the diagonal. The links
are defined to be pairs $(ij)$ of neighbor vertices and the
triangles are defined as triples of neighbors $(ijk)$ such that
each link $(ij)$, which is not a boundary link, belongs to precisely
two triangles  $(ijk_1)$ and $(ijk_2)$.
The summation is over {\it different} triangulations. Two triangulations
are considered as identical\footnote{Often the word {\it equivalent} is
used for such two triangulations. However, we are going to reserve
this notation to triangulations which have a common subdivision.}
if there  is a map between  the vertices compatible
with the assignment of links and triangles.
$l_i$ has a two-fold meaning as a fixed polygon loop
in target space $R^D$ and an abstract boundary in the triangulation
$T$ and $\cT(l_1,...,l_n)$ denotes a suitable class of triangulations
with the given boundaries. Usually we have in mind all triangulations
of a given topology $\chi$. However, occasionally it is convenient
to enlarge the class of simplexes considered, such that they strictly
speaking do not form a combinatorial manifold. Local ``irregularities''
of this kind should not be important, since they are related
to short distance effects which should not play any role
in the continuum.  More serious is the restriction on topology.
Formula \rf{3.25} is very tantalizing in the sense that it
has no reference to topology. Is it possible that eq. \rf{3.25} provides
a non-perturbative definition of the summation over topologies?
It has always been an annoying aspect of the continuum formula
\rf{3.4} that we only know how to interpret it for a given
manifold, i.e. a given topology in the two-dimensional case.
A summation over different topologies has to be performed by
hand. It turns out that {\it eq. \rf{3.25}  can be used to
study the summation over topologies, but not directly as it stands}.
A special limit, {\it the double scaling limit} has to be taken.
This will be discussed later. At the moment we will always restrict
the class of triangulations $\cT$ to mean triangulations with a
fixed topology, usually the simplest, the spherical topology.
In the following we will also use the notation spherical topology
for surfaces with boundaries where we recover the sphere after
closing the boundary.

It is often useful to consider a number of special cases of \rf{3.25}.
If there is no loops at all we talk about {\it the partition
function for closed surfaces}:
\beq
Z(\m) =\sum_{T \in \cT} \frac{1}{S_T} \e^{-\m N_T }
\int \prod_{i\in T/\{i_0\}} \d x_i\;
\e^{-\sum_{(ij)} (x_i - x_j)^2}. \label{3.26}
\eeq
In this formula is included an additional symmetry factor $S_T$ for
the triangulation. It is similar to the additional factor which appears
in vacuum Feynman diagrams, and it reflects the additional
symmetry which can be present for surfaces without a marked
boundary: A permutation of the vertices might leave unchanged
the links and triangles and in this way not change the surface.
$S_T$ is equal to the order of the
automorphism group of the graph $T$. A vertex $i_0$ is excluded
from the integration in order to kill the mode associated with
translational invariance. The Gaussian integration is
independent of this choice and alternatively one could have chosen
to fix the center of mass.

\begin{figure}
\input{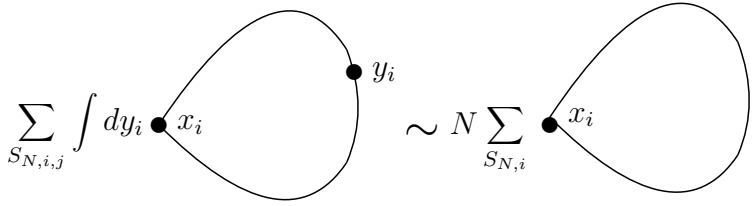}
\caption[3_4]{To each surface $S_{N,i}$ with one marked point
$x_i$ corresponds $N-1$ surfaces $S_{N,i,j}$ with two marked points
since we can put $y_i$ at the $N_1$ other vertices. This is true up to
symmetry factors.}
\label{3_4}
\end{figure}

Another limiting case arises if we contract the loops to points, i.e.
marked vertices. Strictly speaking this cannot be done in a
continuous way on the triangulations. We denote the
$n$-point function $G(x_1,\ldots,x_n)$. It will be given by
\beq
G_\m (x_1,\ldots,x_n) = \sum_{T \in \cT(i_1,\ldots,i_n)}
\frac{\e^{-\m N_T } }{S_T}\,
\int \prod_{i\in T/{\{i_1,\ldots,i_n\}}} \d x_i\;
\e^{-\sum_{(ij)} (x_i - x_j)^2} \label{3.27}
\eeq
where the symmetry factor can be different from zero for the
1- and 2-point function. Note that the 1-point function (which
by translational invariance is independent of the target space
point $x_1$) is (essentially) equal to (minus) the derivative
of the partition
function with respect to $\m$, while the integral of the 2-point
function is (essentially)
equal to the double derivative of the partition function:
\beq
G_\m (x_1) \sim - Z'(\m),~~~~~~~\int \d x \;G_\m(x,y) \sim Z''(\m).
\label{3.28}
\eeq
The equations follow from the observation that differentiation
of $Z(\m)$ multiplies each triangulation with a factor $N_T$
coming from $\e^{-\m N_T}$.
The 1-point function is the summation over marked triangulations,
but there are $N_V$ of these for each triangulation without a
marked vertex, up to symmetry factors which play no role
for the generic large triangulation. Relations like \rf{3.28}
will be valid in the limit where triangulations with large $N_T$ dominate.
This is illustrated in fig.\,\ref{3_4}.
As we shall see shortly the model will have
{\it a critical point} $\m_c$. For $\m$ above the critical point
all $G_\m$ will be analytic functions of $\m$, but at the
critical point they will contain non-analytic parts. These
are the universal parts which have our interest. They are determined
by the large $N_T$ part of the triangulations since finite $N_T$'s
only produce analytic contributions. The word ``essentially'' above and
the symbol ``$\sim$'' in \rf{3.28} refers to this non-analytic
part determined by the large $N_T$'s. The line of argument can be extended
to the $n$-point function.  Let the generalized susceptibility
$\chi^{(n)}(\m)$ be defined as the integral over $n-1$ of the arguments.
$\chi^{(n)}(\m)$ is ``essentially'' equal to the derivative
of $\chi^{(n-1)}(\m)$ since the derivative brings down a factor $N_T$ in the
definition \rf{3.27} while $\chi^{(n)}(\m)$ contains an additional
marked point which  produces a factor $N_V$ ($\approx N_T/2$)
in the counting of surfaces. It is summarized in the formulas:
\beq
\chi^{(n)}(\m) \equiv \int \d x_1\cdots \d x_{n-1} \; G_\m(x_1,\ldots,x_n)
\sim (-1)^n \frac{\d^n}{\d \m^n} Z(\m). \label{3.28a}
\eeq

\subsection{Physical observable}

Eqs. \rf{3.25} -\rf{3.28} defined the regularized loop-functions.
The following theorem ensures the existence of a critical
point $\m_c$ like for the random walk:

\vspace{12pt}
\noindent {\bf Theorem:} For surfaces of spherical topology
exist a critical point $\m_c$ such that $G_\m(\{l_i\})$
is convergent for $\m > \m_c$ and divergent for $\m < \m_c$.
$\m_c$ is independent of the boundary loops $\{l_i\}$.

\vspace{12pt}
\noindent {\bf Conjecture:} The theorem is true for surfaces
of any fixed topology $\chi$ and $\m_c$ is the same for all $\chi$.

\vspace{12pt}
The theorem will not be proven here, but it is not difficult
to show that $Z(\m)$ is well defined for $\m$ sufficiently large.
Two steps are needed. First we have to bound the Gaussian
integral for a given triangulation. In fact it is not difficult
to show that there exists constants $c_l$ and $c_u$ such that
\beq
\e^{c_l N_T} \leq   \int \prod_{i\in T/\{i_0\}} \d x_i\;
\e^{-\sum_{(ij)} (x_i - x_j)^2} \leq \e^{c_u N_T}. \label{3.29}
\eeq
Secondly we have to use that the number of triangulations of a
fixed topology is exponentially bounded. Let us denote the
number of triangulations with topology $\chi$,
which can be constructed from  $N$ triangles as $\cN(N,\chi)$:
Constants $d_l$ and $d_u$  exist such that
\beq
\e^{d_l N_T} \leq \cN(N_T,\chi) \leq \e^{d_u N_T}. \label{3.30}
\eeq
Combining eqs. \rf{3.30} and \rf{3.29} we conclude that for any
$\chi$:
\beq
c_l+d_l \leq\m_c  \leq c_u+d_u. \label{3.31}
\eeq
For the rest of the theorem we refer to the original article \cite{adf}.
The conjecture that $\m_c$ is independent of $\chi$ is almost certainly
true. It has been proven for sums over triangulations coupled
to matter with central charge $c \leq 1$.

It is worth to notice that the exponential bound on the
number of triangulations plays an important role for the existence of the
critical point. If we try to define the summation
over all topologies directly from eq. \rf{3.25} we will fail due to
the entropy of triangulations. The number of triangulations
of $N$ triangles $\cN(N)$, with no restriction on
topology grows faster than factorially and from eq. \rf{3.29}
it follows that the sum \rf{3.25} is ill defined for any choice of $\m$!
Later we will discuss some attempts to make sense of the sum \rf{3.25}
after all.

After the existence of a critical point is established
it is of interest to study the critical behavior of the
$n$-loop and $n$-point functions when we approach the
critical point.

The most important quantity in this context is the mass
gap since it determines the possible scaling.
The mass gap can be defined by the exponential decay of 
the two-loop function as the distance $d$ between the loops goes
to infinity:
\beq
m(\m) =- \lim_{d\to \infty} \frac{G_\m(l,l_d)}{d}, \label{3.32}
\eeq
where $l_d$ denotes the loop $l$ displaced a distance $d$.

\vspace{12pt}
\noindent{\bf Theorem:} The two-loop function falls off exponentially
with the distance between the two loops.

\begin{figure}
\input{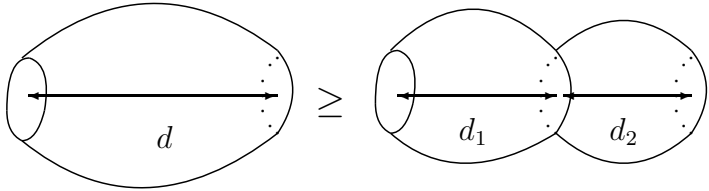}
\caption[3_3]{The sum over surfaces from $l$ to $l_d$ is larger than
the restricted sum there is a ``bottle neck'' at distance $d_1$.}
\label{3_3}
\end{figure}

\vspace{12pt}
The argument is the same as for the random walk: since
$G_\m (l,l_{d_1+d_2})$ is the unnormalized probability density for
a propagation of a string from $l$ to $l_{d_1+d_2}$ it will,
multiplied by an appropriate normalization factor
which converts it into a probability, be larger
than the corresponding product of $G_\m (l,l_{d_1})$ and
$G_\m (l_{d_1},l_{d_+d_2})$, since this product (correctly normalized)
imposes the constrain that the surfaces from $l$ to $l_{d_1+d_2}$
should pass through $l_{d_1}$. This is illustrated in fig.\,\ref{3_3}.
I.e. {\it the (correctly normalized) two-loop function is
sub-additive} and since the normalization factor is independent
of $d$ the limit \rf{3.32} exists.
It follows from the definition that $m(\m)$ is an increasing  function
of $\m$. Unfortunately there exists no proof that $m(\m)$ goes
to zero for $\m \to \m_c$. {\it We will assume this is the case}.
We can then define the same critical exponents and fractal
dimensions as for the random walk, but let us first introduce
a new critical exponent $\rho$ corresponding to the string
tension. We define the string tension $\sg(\m)$ by  the
exponential decay  of the 1-loop function $G_\m(l_A)$ for a large
planar loop $l_A$ which encloses an area $A \sim l^2$:
\beq
G_\m (l_A) \sim A^\b \e^{-\sg(\m) A}. \label{3.33a}
\eeq
The proof that the $G_\m(l_A)$ falls off exponentially is
again based on sub-additivity: Let $A$ denote both the planar region enclosed
by $l_A$ and its area. The number of surfaces
which has $l_A$ as boundary is larger than the sub-ensemble
where we divide $A$ in two pieces with area $A_1$ and $A_2$ by introducing a
new boundary in the middle of $A$ which we force the surfaces to respect.
We conclude that $G_\m (l_A) \geq G_\m(l_{A_1}) G_\m (l_{A_2})$
and this implies \rf{3.33a}

Why is $\sg(\m)$ called the string tension? We can view $G_\m(l_A)$ as
the partition function $Z(A)$ for an ensemble of surfaces which are
allowed to fluctuate, but where the boundary, i.e. the frame, is fixed.
The Gibbs free energy of the system is $F(A) = -\log Z(A)$.
The string tension is defined by change in free energy per unit
area if we change the area $A$
\beq
\Delta F = \sg \Delta A.
\eeq
From \rf{3.33a} it will precisely be our $\sg(\m)$ for large $A$.

Let us now introduce the scaling parameters:

\begin{itemize}
\item[{(1):}] The critical mass exponent $\n$ is defined by the assumed
scaling of $m(\m)$ to zero. This allows us to introduce the
physical mass and a length scale which goes to zero at the critical
point.
\beq
m(\m) \sim (\m -\m_c)^\n,~~~~~m(\m) = m_{{\rm ph}} a(\m). \label{3.34}
\eeq
\item[{(2):}] The short distance behavior of the 2-point function is
characterized by the anomalous scaling exponent $\eta$:
\beq
G_\m (x,y) \sim  |x-y|^{2-d-\eta},~~~~~1\ll |x-y| \ll 1/m(\m),
\label{3.35}
\eeq
\item[(3)]
The susceptibility $\chi(\m)$ and the susceptibility exponent $\g_s$
is defined by:
\beq
\chi(\m) = \int \d x\; G_\m(x,y) \sim \frac{1}{(\m-\m_c)^{\g_s}}. \label{3.36}
\eeq
\item[(4)] The {\it string tension} $\sg(\m)$ is defined by the
exponential decay of the 1-loop function and we assume
\beq
~\sg(\m) \sim (\m-\m_c)^\rho.
\label{3.37}
\eeq
From dimensional arguments we expect  $\rho = 2\n$. However, we will
prove that $\sg(\m)$ does not scale to zero for the simplest model.
\item[(5)] The extrinsic Hausdorff dimension $d_H$ of the ensemble of
surfaces is defined  by
\beq
\bra  Area  \ket_r \sim \bra N_T \ket_r \sim r^{d_H}, \label{3.38}
\eeq
where the average area is over an ensemble of surfaces with
two marked points a distance $r$ apart in target space $R^D$.
\end{itemize}

Let the topology of the surface be spherical except for
possible boundaries. If the mass scales to zero it follows, by
arguments identical to the ones presented for the random walk, that
\beq
\g_s = \n (2-\eta),~~~~~~~d_H =1/\n,  \label{3.39}
\eeq
i.e. Fischer's scaling relation and the relation between the mass
exponent and the Hausdorff dimension. For most systems one has that
$0\leq \eta \leq 2$. In ordinary statistical systems $\eta =2$ at the
infinite temperature limit, while $\eta= 0$ is the Gaussian approximation.
For such systems it is clear that $\n >0$ implies $\g >0$.
In the following we will often encounter systems where $\g_s < 0$.
It is important to realize that such systems exist. As
a simple model one can take the closed random
walk with two marked points $x_i$ and $y_j$ kept fixed in
$R^D$. It has {\it some} analogy with the 2-point function for
random surfaces\footnote{It does  {\it not} describe the propagation
of the desired geometrical object, the particle, between $x$ and $y$,
while the two-point function for the closed string indeed describes
the propagation of the string between $x$ and $y$.}. Since it is clearly the
product of two ordinary random walks between $x$ and $y$
the mass, which is determined by the exponential decay at large
distances, will be twice that of the ordinary random walk, i.e.
it has the same critical exponent $\n$. The short distance
behavior will be
\beq
\tilde{G}_\m(x) \sim \frac{1}{|x|^{2(d-2)}},~~~~{\rm i.e.}~~~\eta = d-2.
\label{3.40}
\eeq
For $d>4$ it follows that $\eta >2$. 
It is easy to show directly that $\g_s = 2-d/2$, i.e.
Fischer's scaling relation is valid even if $\g < 0$.

There exists a theorem which indicates that it might be difficult
to obtain a positive $\g_s$ for random surfaces.

\vspace{12pt}
\noindent {\bf Theorem:} $\g_s \leq 1/2$ for the random surfaces model
\rf{3.25} with spherical topology.

\vspace{12pt}
\noindent {\bf Conjecture:} $\g_s >0$ implies $\g_s =1/2$ for the random
surface model \rf{3.25} with spherical topology.

\vspace{12pt}

The theorem {\it and the conjecture} are known to be true
for the hyper-cubic random surface model \cite{dfj}. Rather than giving the
rigorous arguments let me present the underlying geometrical reason for
the theorem and the conjecture. But first a remark about the
technical point which should be dealt with if the  arguments
should be made exact. In order to apply the cutting and sewing arguments below
one would have to introduce
somewhat more complicated objects than the $n$-point functions
we have been considering until now:
the correct objects to consider are $n$-loop functions
where each boundary loop consists of, say, three links. Fix the center
of mass of the boundary vertices of loop $i$ to be $x_i$, the
corresponding point in the $n$-point function, but
integrate over the positions of the boundary vertices compatible
with these constraints. The corresponding $n$-loop function
will be a function of $x_i$, precisely as the $n$-point function
and we expect that they coincide in the scaling limit where
the ``bare'' distances $x_i$ all scale to infinity while
the ``physical'' distances $x_{{\rm ph}} = x\; a(\m)$ stay fixed as $\m \to\m_c$,
i.e. $a(\m) \to 0$. The reason is that any contribution where
the distances between the boundary vertices are larger than $1$
will be exponentially suppressed by the Gaussian action and
in the scaling limit distances of order 1 mean physical distances
of order $a(\m)$, i.e. of the order of the ``lattice spacing''.
Had we considered surfaces made of plaquettes living on the
hyper-cubic lattice  or surfaces in $R^D$ build directly from the gluing
of equilateral triangles these problems would be absent and the
arguments to be presented would be exact in the sense that one would not
have to integrate over positions of the boundary vertices. In the
following we will ignore these complications for the sake of argument.

Consider the generalized susceptibility
$\chi^{(n)}(\m)$ as defined by eq. \rf{3.28a}.   For $n >2$ we have the
obvious geometrical inequality (see fig.\,\ref{3_5})
\beq
\chi^{(n)} (\m) \geq \left(\chi^{(2)}(\m)\right)^n  \label{3.41}.
\eeq
\begin{figure}
\input{3_5.tex}
\caption[3_5]{The sum $\sum_{S_n} \int\cdots$
over all surfaces with $n$ punctures is larger
than  the restricted sum $\sum_{S'_n} \int\cdots$
over over surfaces $S'_n$ with $n$ punctures
where the surfaces $S'_n$ are characterized by a ``joint'' where
the $n$ ``bubbles'' connected to the $x_i$'s get together. Since we integrate
over the position of the joint this effectively implies a factorization
in integrated 2-point functions from $x_i$ to the joint.}
\label{3_5}
\end{figure}
Assume $\g_s >0$. Eq. \rf{3.36} implies that the 2-point function diverges
as $\m \to \m_c$, i.e. the non-analytic part dominates and for
$\m \to \m_c$ it is legal to use eqs. \rf{3.28a} and \rf{3.36} in \rf{3.41}:
\beq
\frac{c_n}{(\m-\m_c)^{\g_s+n-2} }\geq \frac{c_2}{(\m-\m_c)^{n\g_s}}
~~~~{\rm for}~~~\m \to \m_c. \label{3.42}
\eeq
We conclude that
\beq
\g_s \leq 1-\frac{1}{n-1},~~~~~~n > 2. \label{3.43}
\eeq
It is tempting to apply the formula for $n=2$, in which case
we get $\g_s \leq 0$. However, \rf{3.41} is not valid for $n=2$
since there is not a unique decomposition in ``joints'', as illustrated
in fig.\,\ref{3_6} which gives the correct decomposition for
$n=2$.

Let us consider the following two random surface models:
In one model we allow the gluing of triangles such that
a minimal loop-length on the surface can be two. This can only happen
if the surface is pinched in a bottle neck consisting of these two
loops. Such bottle necks are precisely what
we have in mind in fig.\,\ref{3_6}. We call this class of triangulations
$\cT_2$. The other class of triangulations  differs from $\cT_2$ only
by not allowing such two-loops\footnote{In terms of the dual $\phi^3$
graphs the difference is that self-energy diagrams are excluded. In both
cases $\phi^3$ tadpole diagrams are excluded. They correspond to
one-loop diagrams.}. In this class, which is denoted $\cT_3$
we can still have bottle necks, the only difference is that the length
of the bottle neck loop will be three. In the scaling limit we clearly
expect no difference between the random surface models constructed
from $\cT_2$ and $\cT_3$ since the bottle neck loops anyway will be
of the order of the cut-off. Denote the $n$-loop and $n$-point functions
of the two models by $G_\m$ and $\bar{G}_{\bm}$.
Up to the technical complications mentioned
above (but rigorous for the other classes of random surface models mentioned)
we have the following identity for the 1-point function, (or more
precisely, for the one-loop function where the boundary
consists of two links in the way described above).
\bea
G_\m &=& \sum_{\bT \in \cT_3} \e^{-\m N_{\bT}}
         \left(1+G_\m)\right)^{N_{\bar{L}}}\nn
     = \sum_{\bT \in \cT_3} \e^{-\bm N_{\bT}} = \bG_\m. \label{3.44}  \\
\bm &=& \m -\frac{3}{2} \log (1+G_\m) \label{3.45}
\eea
The interpretation of the following: In the class of triangulations
$\cT_2$ each link serves as a potential source of a bottle neck
from which a new {\it baby universe} can grow. For a given
triangulation $T \in \cT_2$ we can cut away the maximal size
baby universes and close the corresponding two-link boundary. This
will leave us with a triangulation which belongs to $\cT_2$. In this way
we get all triangulations of $\cT_2$ with one boundary by
summing over $\cT_3$ and for each link either do nothing
or add a whole one-loop universe, i.e. $G_\m$ itself.
Finally we note that $N_{\bar{L}} = 3N_{\bT}/2$.
\begin{figure}
\input{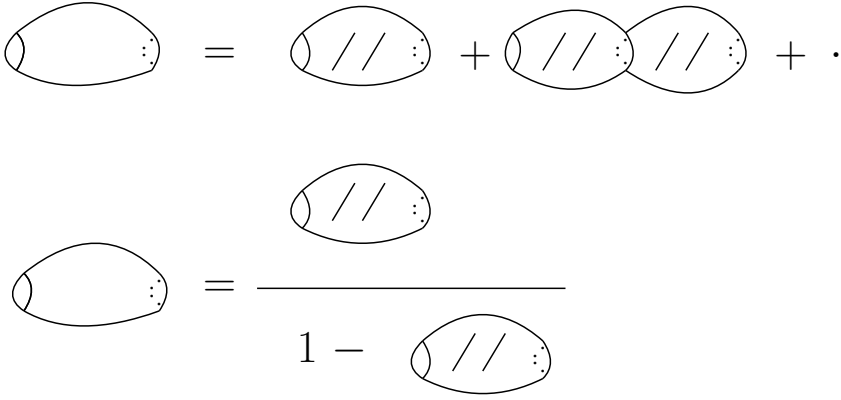}
\caption[3_6]{The decomposition of the 2-loop function for the
class of triangulations in parts which cannot be cut in two
part along a 2-loop bottle neck such that the two boundary loops
are separated. It is possible to perform the sum over the ``irreducible''
components as shown in the lower part of the figure.}
\label{3_6}
\end{figure}

Eqs. \rf{3.44} and \rf{3.45} define the relation between the two models.
Let us use that $\g_s < 1$, i.e. that
$G_\m= {\rm const.} + (\m-\m_c)^{1-\g_s}$
is finite at the critical point. Differentiating eqs. \rf{3.44} and \rf{3.45}
after $\bm$ and $\m$ we get
\bea
\frac{\d\m}{\d\bm} &=& 1-\frac{3}{2}
\frac{\bchi(\bm)}{1+\bG_{\bm}} \label{3.46a} \\
\chi(\m) &=& \frac{\bchi(\bm)}{1 - \frac{3}{2}
\frac{\bchi(\bm)}{1+\bG_{\bm}}},
\label{3.46}
\eea
i.e. the algebraic version of the fig.\,\ref{3_6}. The factor
$3/2(1+G_\m)$ multiplying each bottle neck is a combinatorial
factor associated with the outgrow of the baby universes {\it at}
the bottle neck.

By universality the two models based on
$\cT_2$ and $\cT_3$ have the same critical exponents.
Let us assume that $\g_s >0$. It implies that $\chi(\m) \to \infty$
for $\m \to \m_c$.  The same is true for $\bchi(\bm)$ for
$\bm \to \bm_c$, the critical point of the $\cT_3$ model.
We conclude from \rf{3.46} that $\bm(\m_c) > \bm_c$
since $\bchi(\bm(\m_c)) = 3(1+G_{\m_c})/2 < \infty$ and we can Taylor
expand the rhs of \rf{3.46} around $\bm(\m_c)$:
\beq
\chi (\m) \sim \frac{1}{\bm-\bm(\m_c)} \sim \frac{1}{\sqrt{\m -\m_c}}.
\label{3.47}
\eeq
The last equation follows from \rf{3.46a} {\it which shows that
the transformation from $\bm$ to $\m$ is non-analytic in $\m_c$}:
At the critical point we have $d\m/d\bm =0$, and since
$\bchi(\bm)/(1+\bG_{\bm})$ is monotonic decreasing we have $d^2\m/d\bm^2 >0$:
\beq
\m -\m_c = {\rm const.} (\bm-\bm(\m_c)^2,~~~~~\bm-\bm(\m_c) =
{\rm const.} \sqrt{\m - \m_c}. \label{3.48}
\eeq

The above line of arguments is rigorous for the hyper-cubic random surface
model \cite{dfj}. For the Gaussian model considered it is plausible but not
completely proven due to the technical assumptions mentioned. In a later
section we shall see that in more elaborate theories with more than
one coupling constant it is possible to find a loop-hole
in the argument, and indeed a different critical behavior.

\vspace{12pt}

The arguments presented above can easily be generalized.
The important relation \rf{3.46} is valid not only for $\chi(\m)$
but for the  Fourier transformed $G_\m(p)$ of the two-point function
$G_\m(x,y)$. Recall that
$$\chi(\m) = \int \d x \;G_\m(x,y) = G_\m (p=0).$$
The extension of \rf{3.46} to $p\neq 0$ has the simple graphical
interpretation that a momentum $p$ is flowing through the
bubbles in fig.\,\ref{3_6}. We have
\beq\label{3x.100}
G_\m(p) =  \frac{\bG_{\bm} (p)}{1 - \frac{3}{2}
\frac{\bG_{\bm}(p)}{1+\bG_{\bm}}},
\eeq
Since $\bm$ did not approach $\bm_c$ for $\m \to \m_c$ we can in
this region expand $\bG_{\bm} (p)$ as
\beq\label{3x.102}
\bG_{\bm}(p^2) = \bchi(\bm) - c\,p^2+\cdots
\eeq
where $c$ is constant. The rhs of \rf{3x.100} can be expanded
around $p=0$ and after the use  of \rf{3.46} and little algebra this leads to
\beq\label{3x.101}
G_\m(p) \sim \frac{1}{\chi(\m)^{-1} + c p^2 +\cdots}.
\eeq
This relation shows that  {\it if $\g_s >0$}, i.e. $\chi(\m)$
diverges for $\m \to \m_c$, we have:
\beq\label{3x.102a}
m(\m) \sim \chi(\m)^{-\oh},~~~~~{\rm i.e.}~~\n= \oh \g_s~~~
{\rm and}~~\eta =0,
\eeq
where the last relation follows from Fischer's scaling relation
$\g_s = \n(2-\eta)$. In fact \rf{3x.102a} shows that {\it for
strings embedded in $R^D$ the
 mass $m(\m)$ scales to zero  if and only if $\chi(\m)$ is divergent
at the critical point}.

\vspace{12pt}
A as final application of technique leading to \rf{3.46}
let us consider the string tension. We consider a large loop $l_A$
where the boundary length $|l_ A| \sim \sqrt{A}$ and the sum of all
random surfaces with this loop as boundary. Again we can cut away
two-loops and we get directly the analogue of \rf{3.44}\footnote{It is
assumed that one cannot have two-loops directly at the boundary.
If this assumption is dropped there will  be a
perimeter term $(1+G_\m)^{|l_A|}$ on the rhs of \rf{3x.103}.
This term plays no role for the string tension argument.}
\beq\label{3x.103}
G_\m (l_A) = \bG_{\bm} (l_A),~~~~~{\rm i.e.}~~\sg(\m) = \bar{\sg}(\bm).
\eeq
{\it This relation tells us that the string tension does not scale
to zero if $\g_s > 0$} since $\bm$ does not go to $\bm_c$ for $\m \to \m_c$
in that case.   Let us instead write:
\beq\label{3x.104}
\sg(\m) = \sg_0 + c (\m-\m_c)^\rho.
\eeq
This still defines $\rho$ as a critical exponent and differentiating
\rf{3x.103} after $\m$ and using \rf{3.46a} :
\beq
\frac{d \sg}{d \m} \sim \chi(\m),~~~~~{\rm i.e.} ~~\rho=1-\g_s.
\eeq
If we combine this with $\g_s=1/2$ we get $\rho = 2\n$ as one would
expect from dimensional analysis if the string tension was scaling.
It seems still to be satisfied with the definition \rf{3x.104}.
We will return to this  definition when we consider strings
with extrinsic curvature.

\subsection{Non-scaling of the string tension}

We have defined the string tension $\sg(\m)$  as the exponential decay of the
one-loop Green function for large loops (see \rf{3.33a}). Above we presented
arguments in favor of a non-scaling string tension. The arguments
did to constitute a proof since we had to make certain technical
assumptions. It is therefore important that there exists a simple
rigorous proof of the non-scaling of the string tension \cite{ad}.

\vspace{12pt}
\noindent{\bf Theorem:} $\sg(\m) > 0$ for all $\m \geq \m_c$.

\begin{figure}
\input{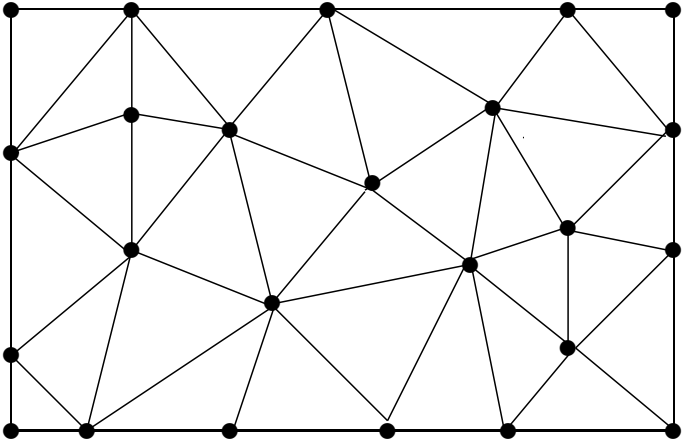}
\caption[3_7]{A typical triangulation of a square loop. When mapped into
$R^D$ we integrate over the interior vertices, while the boundary is
kept fixed.}
\label{3_7}
\end{figure}

\begin{figure}
\input{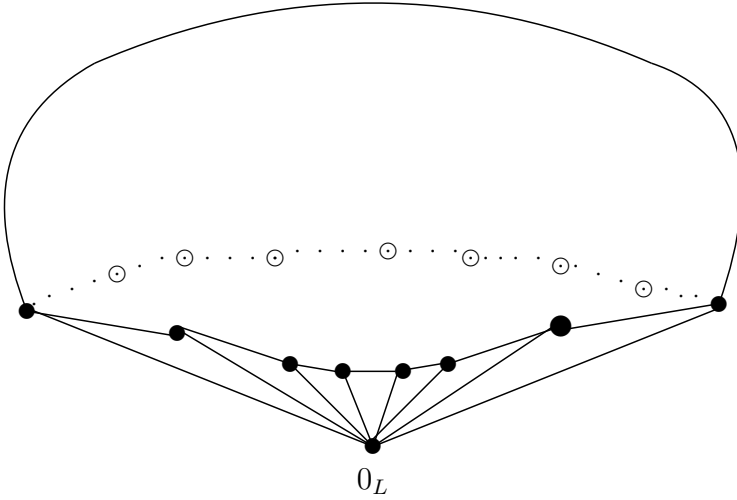}
\caption[3_8]{The surface obtained by contraction the boundary loop
of a graph like the one in fig.\,\ref{3_7} to one point.}
\label{3_8}
\end{figure}

\vspace{12pt}
\noindent
Let us for simplicity assume that we have
a large square loop $l_{L^2}$ of area $L^2$~:
\beq
G_\m (l_{L^2}) \sim \e^{-\sg(\m) L^2} ~~~ {\rm for} ~~~ L \to \infty.
\label{3x.a13}
\eeq
It is  possible to bound
the Green function $ G_\m (l_{L^2})$ in the following way:
The points at the boundary are kept fixed and are not integrated over,
as it is also assumed in the general notation for the loop Green functions.
It is natural to imagine that the density of boundary points is proportional
to the length of the perimeter but it is not essential for the following.
Let $T$ be one of the triangulations (see fig.\,\ref{3_7}) in the sum
\rf{3.25}.
The action
\beq
S[x,T] = \sum_{(i,j)} (x_i-x_j)^2 \label{3x.a23}
\eeq
can be bounded because of the following decomposition~:
\beq
S [x,T]=S_{min}(T,l_{L^2})+S[x',T'] \label{3x.a24}
\eeq
For the given (abstract) triangulation $T$ we let
$S_{min}(T,l_{L^2})$ denote the
minimum of $S[x,T]$ as a function of the coordinates $x^\m(i)$ of
the  vertices $i \in T/\prt T$.  $T'$ denotes the triangulation where
all boundary points are identified. For the surface embedded in $R^D$
we can view it as a contraction of the boundary loop
$l_{L^2}$  to a single point $0_L$ of order $|\partial T| \propto L$.
This is illustrated in fig.\,\ref{3_8}.
The decomposition \rf{3x.a24} follows from the quadratic nature
\rf{3x.a23}  of $S[x,T]$.

The loop Green function $G_\m (l_{L^2})$ for
the large square loop $l_{L^2}$ can now be written as
\beq
G_\m (l_{L^2})= \sum_{T \in \cT (l_{L^2})}
\e^{- S_{min}(T,l_{L^2})-\m N_T} \,\,
\int \prod_{i \in T/\partial T} \d x (i)\; \e^{- S[x,T']}.
\label{3x.a28}
\eeq

Next we note that the sum of squares of the length of any two sides of
a triangle is $\geq$ 2 times its area. It follows that
\beq
S_{min} (T,l_{L^2}) \geq 2 L^2
\label{3x.a29}
\eeq
and from \rf{3x.a28} we can write
\beq
G_\m (l_{L^2}) \leq \e^{-2 L^2}\, G_\m (0_L)
\label{3x.a30}
\eeq
In eq. \rf{3x.a30}  $G_\m (0_L)$ denotes the loop Green function where
   the loop $l_{L^2}$ is contracted to one point of order
   $|l_{L^2}| \propto L$.
   Since $\m_c$ is independent of boundaries this Green function
   has the same critical point as ordinary Green functions like
   $G_\m (l_{L^2})$ and it can be bounded by
\beq
G_\m (0_L) \leq \e^{c(\m) L}
\label{3x.a31}
\eeq
where $c(\m)$ is finite for $\m > \m_c$.  This is a consequence
of $\g_{s} \leq 1/2$, which implies that the one-point function
is finite at the critical point.

From the definition of the string tension it finally follows that
\beq
\sigma (\m) \geq 2.
\label{3x.a32}
\eeq

People performing strong coupling expansions will recognize estimates like
\rf{3x.a30} as typical strong coupling estimates. What usually happens
is that the function $G_\m (0_L)$, which is based on a strong coupling
approximation, becomes dominant before one reaches $\m_c$.
However, in this case we can control it all the way down to $\m_c$
since $\g_s \leq 1/2$.

What are the consequences of this non-scaling of the string tension?
We have to assume that $m(\m)$ scales in order to take the continuum
limit.  As discussed above this uniquely fixes how the
lattice spacing scales to zero as a function of $\m$:
$a(\m) \sim m(\m)$. It follows that the physical string tension
scales to infinity since we have (for dimensional reasons)~:
\beq
\sigma_{{\rm ph}} a^2(\m) =\sigma (\m) \geq 2 .
\label{3x.a34}
\eeq
  Since the physical string tension scales to infinity fluctuations
  including any surfaces having an area different from the minimal
  area for the given boundary $l_{L^2}$ will be strongly suppressed.
  When we approach the critical point $\m_c$ we will be left with a class
  of surfaces consisting of a minimal surface, depending on the Green loop
  function in question, and singular, spiky, branched polymers growing out
  everywhere on this surface. Such polymers are essentially one-dimensional
  objects with no or very little area.

For large dimensions $D$ there is little doubt that this picture is
correct. It might still be that it is  too coarse an approximation
to consider the surfaces strictly as polymers for lower dimensions.
According to the
theory of such polymers the generic values of $\g_s$ and $\n$ for
polymers are $\g_s=1/2$ and $\n=1/4$ as we will now explain.

\subsection{Branched polymers}

\begin{figure}
\input{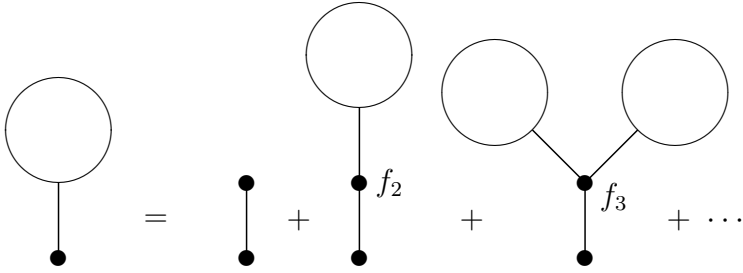}
\caption[3_10]{The equation for rooted branched polymers.}
\label{3_10}
\end{figure}

\begin{figure}
\input{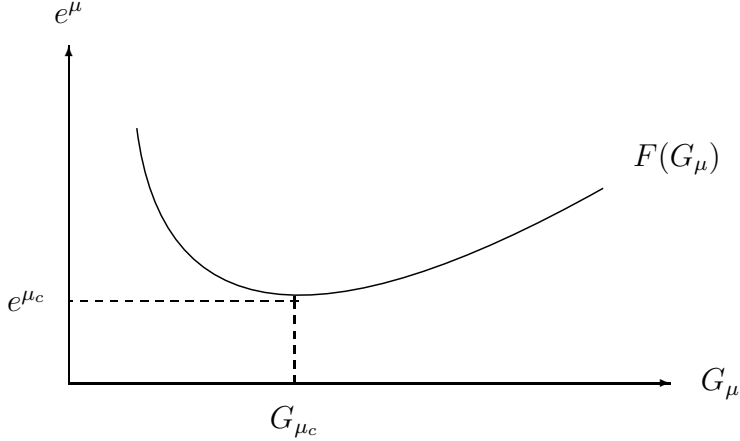}
\caption[3_11]{The graphic solution of the branched polymer model.}
\label{3_11}
\end{figure}

Let us return to eq. \rf{3.46} as shown in fig.\,\ref{3_6}.
If $\g_s =1/2$ we concluded that the modified theory based on the
class $\cT_3$ of triangulations  is not critical for $\m \to \m_c$
in the model based on $\cT_2$. This implies that the
individual bubbles in fig.\,\ref{3_6} are not critical, i.e. they are
of lattice size. The only way the number of triangles can
grow to infinity is by the successive gluing of bubbles. All dynamics
lie in this gluing and it seems that we get a
perfect model of this dynamics by consider a model of {\it branched
polymers}: Each individual bubble is represented as a link
with an associated  chemical potential which we denote $\m$ as in the
original model and with a weight factor $f_n$ associated
with the joining of $n$ bubbles at a vertex $v_n$.
In the original surface theory we have
the possibility of gluing $n$ bubbles to $n$ links which share
two vertices. This is the motivation for introducing
the factor $f_n$ associated with a branched polymer vertex $v_n$
of order $n$.  If we consider only spherical surfaces (as we will
do in the following) {\it the branched polymer graphs will be
tree graphs}. There can be no loops.
The partition function for the branched polymers  can be written as
\beq\label{3x.200}
Z(\m) = \sum_{BP}\frac{1}{S_{BP}} \e^{-\m N_{BP}} \prod_v f_{n(v)}
\eeq
where the summation is over all branched polymers, i.e all tree graphs,
and the product is over all vertices in the tree graphs. It is convenient
to consider {\it rooted trees}, i.e. with one marked link, since the
symmetry factor $S_{BP}$ in this case drops out.
It corresponds to the one-loop function $G_\m$
considered for the full surface theory, and we will also denote the
corresponding $BP$ function $G_\m$. It satisfies the self-consistent
equation \cite{adf}:
\beq\label{3x.201}
G_\m = \e^{-\m}(1 + f_2G_\m+ f_3 G_\m^2 \cdots),
\eeq
which is shown in fig.\,\ref{3_10}.  We can solve for $\m$
as a function of $G_\m$:
\beq\label{3x.202}
\e^\m = \frac{1 + f_2G_\m+ f_3 G_\m^2 \cdots}{G_\m}\equiv F(G_\m).
\eeq
This relation is shown in  fig.\,\ref{3_11} and for the weights $f_n$
positive\footnote{There is a loop-hole in this argument if we
allow infinite branching and the weights $f_n$ in addition satisfy
certain convergence relations which move the critical point $\m_c$
out to the radius of convergence of the rhs of eq. \rf{3x.202}.
I refer to the original articles for discussion \cite{adf}.}
we conclude that the
lowest value $\m_c$ of $\m$ for which eq. \rf{3x.202} has a solution
is the minimum of $F(G_\m)$:
\beq\label{3x.203}
\m-\m_c  \sim c (G_{\m_c}-G_\m)^2,~~~~~{\rm i.e.}~~
G_\m = G_{\m_c} - \tilde{c} \sqrt{\m-\m_c}.
\eeq
Since $G_\m$ is the one-point function it is expected to have the
critical behavior $G_\m \sim (\m-\m_c)^{1-\g_s}$ and from \rf{3x.203}
we conclude that {\it the generic value of $\g_s$ for branched
polymers is 1/2}.

One can consider branched polymer models directly in $R^D$, for instance
with Gaussian interactions between the vertices. As long as the
branched polymers are tree graphs it is possible to perform
the Gaussian integration explicitly, as for the random walk, and
for the calculation of $\g_s$ one immediately gets the above
considered model. Finally $\n=1/4$ and $\eta=0$. This
has the following interpretation: Consider a branched
polymer propagating from $x$ to $y$. For a given branched
polymer there is a unique path of minimal length
along the links going from $x$ to $y$. This path can be viewed
as a random walk path and the summation over branched polymers
is a summation over all random walks from $x$ to $y$ where
each vertex can be the source of an outgrowth of a rooted branched
polymer. This observation  allows us to solve the
problem as a random walk problem, only is the chemical
potential $e^{-\m}$ renormalized by the factor $(1+G_\m)$,
i.e. we get a random walk with $\bm$ given by:
\beq\label{3x.204}
\e^{-\bm}=\e^{-\m}(1+G_\m),
\eeq
i.e. expanding around $\m_c$:
\beq\label{3x.204a}
\bm- \bm_c \sim \left(1 -\frac{1}{1+G_\m}\, \frac{\d G_\m}{\d\m}\right)
 (\m-\m_c) \sim \sqrt{\m-\m_c}.
\eeq
The exponential decay of the random walk is given by
\beq\label{3x.205}
m(\bm) \sim (\bm-\bm_c)^{\oh} \sim (\m-\m_c)^{\oq}.
\eeq
The exponent $\n=1/4$ is due to the non-analyticity of the
coupling constant transformation \rf{3x.204}-\rf{3x.204a}, a phenomena
we have encountered a number of times by now.
Since we still have $\n = 1/d_H$ we conclude that the
Hausdorff dimension of the ensemble of branched polymers is 4, the ``product''
of two random walks.

\subsection{Extrinsic curvature terms (I)}

It is natural to ask if there are more elaborate random surface
models where both the mass and the string tension scales.
One can view the situation much as in the random walk case.
The generic random walk had $\n=1/2$, but by tuning an extrinsic
curvature term to infinity it was possible to enter a new universality
class of smooth random walks. We saw that this universality
class was related to spinning particles in the sense that
one got the same critical exponents as for the Dirac particle
and that the random walk with extrinsic curvature term could be
viewed as a classical spinning particle, the tangent vector of the
path playing the role of the classical spin. For the Dirac particle
one could further relate it to an underlying world line supersymmetry.

Could a similar scenario be present for random surfaces? The theory
of random surfaces is not so well understood yet and we cannot
present analytic arguments in the same detail as was the case for the
random walk. However, in many ways the situation is identical to
that of the random walk \cite{adfj1,adfj2}.
First, it is possible to add extrinsic curvature terms to the
action in a natural way. In the context of statistical mechanics
of membranes these terms are  well known  (see the lectures of
Peliti and Nelson)\footnote{ 
Also from the point of view of the continuum string theory there 
is a vast literature and many different 
motivations for including extrinsic curvature terms. 
In \cite{extrinsic}  a very incomplete list of references is provided.}.
The terms suppress the
branched polymer outgrows and their presence could result in a phase
transition where the new phase is characterized by smoother surfaces.
In the case of the
random walks we had to take the bare coupling constant of the extrinsic
curvature term to
infinity in order to reach a new phase.
For the random surfaces there are strong indications
that the transition takes place for a finite value of the
coupling constant. In addition numerical simulations indicate
that {\it both the mass and the string tension scales
to zero at the transition point}. Therefore this point is of
interest if we want to discuss continuum limits of the
random surfaces theory.
One could further ask if there is any hint of extrinsic
curvature terms  coming from a fermionic surface theory.
The answer is yes. An old result of Wiegmann \cite{wiegmann} shows that
the integration over fermionic variables will produce such terms.
To be more precise the results are the following: If we consider
a fermionic string theory, i.e. a string theory where
we have local worldsheet supersymmetry, i.e. our
bosonic variables $x^\m (\xi)$ have supersymmtric partners $\psi^\m (\xi)$,
one can explicitly integrate out the fermions and arrive at the
effective bosonic string action:
\bea
S_{eff}&=& S_{bos}+ \tau \int \d^2\xi \sqrt{h} \left[
(D_a n^\m_i)^2 + (e_a^\m\prt_b e_c^\m)^2\right]+\nn \\
&& \frac{ik}{8} \Tr \left( \oh \int \d^2\xi \;
\d A\wedge A + \frac{1}{3\pi} \int_D d^3 \xi A\wedge A \wedge A \right).
\label{3.61}
\eea
The terms in eq. \rf{3.61} refers explicitly to the {\it extrinsic}
geometry of the surface defined by the bosonic variables $x^\m(\xi)$.
$\tau$ depends on the fermionic representation, $k$ is an integer,
$e^\m_a$, $a=1,2$ are unit tangent vectors of the surface,
$n^\m_i$ are $D-2$ normal vectors and
$D_a$ the covariant derivative with respect to the generic {\it normal
bundle} associated with the surface:
\beq
A^a_{ij} = n^\m_i\prt^a n^\m_j,~~~~~A^a = A^a_{ij}M_{ij},~~~M_{ij}\in
{\rm so}(D-2).
\label{3.62}
\eeq
The Lie algebra elements $A^a$ will be generators of parallel transport
in the ($D$-$2$)-dimensional vector space orthogonal to the tangent space.
The last term in eq. \rf{3.61} is  a WZW-like term. $\int \d^3 \xi$ is the
integration over a three-dimensional manifold which has the two-dimensional
manifold as its boundary. This term, which in Euclidean space
is a pure phase factor, is analogous to the term $\cP \exp i
\int \om^{\m\n}s^{\m\n}/2$ for the Dirac particle. It is a very interesting
question whether it serves to cancel the contributions
between various rough surfaces in the path integral, leaving only
smoother ones, as was the case for the fermionic particle.
{\it The term $(D_a n^\m_i)^2$ definitely acts in favor of smoother
surfaces}.

\vspace{12pt}
To see this recall the following facts from the classical
theory of surfaces embedded in $D$ dimensions.
Let $h_{ab}$ denote the induced metric and $\Gamma_{ab}^{~~c}$
the corresponding connection:
\beqn
h_{ab} \equiv \prt_ax^\m\prt_bx^\m,~~~~
\Gamma_{ab;c}=\oh\,
(\prt_ah_{bc}+\prt_bh_{ac}-\prt_ch_{ab})=
\prt_a\prt_bx^\m \prt_cx^\m.
\eeqn
where the indices $a,b,c...$ are lowered and raised
with $h_{ab}$ and the
inverse $h^{ab}$. The second fundamental form is given by
\beqn
K_{i;ab} = -\prt_a n^\m_i \prt_bx^\m= n^\m_i \prt_a\prt_b x^\m.
\eeqn
The basic equations which are satisfied {\it if} $x^\m(\xi)$ represents
a surface embedded in $R^D$ are the Gauss-Weingarten equations:
\bea
\prt_a\prt_b x^\m &=& \Gamma_{ab}^{~~c}\prt_cx^\m+ K_{i;ab}n^\m_i \nn \\
\prt_a n_i^\m &=& - n^\n_i\prt_a n^\n_j-K_{i;ab}h^{bc}\prt_cx^\m\nn
\eea
Using the definition \rf{3.62} the last equation reads:
\beqn
D_{a;ij} n^\m_j=-K_{i;ab}h^{bc}\prt_cx^\m,~~~~~
D_{a;ij}\equiv \prt_a\del_{ij}+A_{a;ij}.
\eeqn
Let $\tilde{D}_a$ denote the ordinary covariant derivative
with respect to the connection $\Gamma^{~~c}_{ab}$ and $\Box \equiv
\tilde{D}_a\tilde{D}^a$. Using the covariant constance of $h_{ab}$:
$\tilde{D}^ah_{ab} = 0$ one can check that:
\beqn
\Box x^\m = -\frac{1}{\sqrt{h}}
\prt_a \sqrt{h} h^{ab} \prt_b x^\m=
h^{ab}(\prt_a\prt_b- \Gamma^{~~c}_{ab} \prt_c)x^\m =
K_{i;a}^a n^\m_i.
\eeqn
The final result is that
\beq
(\Box x^\m)^2 = (h^{ab}K_{i;ab})^2 =
(h^{ab}D_an^\m_i D_bn^\m_i)^2.
\label{3.63}
\eeq
The principal curvatures $\k_1$ and $\k_2$
of the embedded surface is determined by  the second fundamental form:
If we define the {\it mean curvature} $H = (\k_1+\k_2)/2$ and the Gauss
curvature (as already mentioned) $K =R/2 = \k_1\k_2$, we have
\beq
( h^{ab}K_{i;ab})^2 = 4H^2,~~~~~( h^{ab}K_{i;ab})^2
-h^{ad}h^{cb}K_{i;ab}K_{i;cd}= R/2 \label{3.64}
\eeq

\vspace{12pt}

To summarize the situation we seemingly get a number of terms
which can act to produce smoother bosonic surfaces if we integrate
out the fermionic degrees of freedom for a string with world sheet
supersymmetry. Here we will consider only the simplest of these terms:
the extrinsic curvature term.

We have two different versions of the action available: one which
refers exclusively to the geometry of the surface $F$ embedded in
$R^D$ and one which is a hybrid between terms referring to
extrinsic and intrinsic geometry:
\beq
S[F;\m,\l] = \int_F \d A (\m + \l H^2), \label{3.65}
\eeq
\beq
S[g,x;\m,\l] = \int \d^2\xi \sqrt{g}\left[\m+
 g^{ab}\prt_a x^\m \prt_b x^\m + \l g^{ab} D_an^\m_i D_bn_i^\m \right].
\label{3.66}
\eeq
In the last equation the induced metric enters via the equation
of Weingarten. Let us for simplicity discuss the situation in
$D=3$ (the results are easily generalized to any dimensions) and
let us choose the action given by eq. \rf{3.66} 
since it is easier to use in numerical
simulations. In $D=3$ there is only one normal $n^\m$ to the
surface and $D_a n_i$ reduces to $\prt_a n$. If we want to
discretize the system there is no unique way to include such
higher derivative terms but as in the case of the random walk
it can be done in a natural way. Let $T$ be an abstract triangulation
and let $i$ be a vertex and $(ijk)$ a triangle.
$x_i$ is the coordinate in $R^3$ and $(x_ix_jx_k)$ will define
a triangle $\triangle$ in $R^3$ and a normal  $n_\triangle$.
For a given triangulation $T$, i.e. a given choice of
equivalence class of internal metrics $g$, the discretized version of
eq. \rf{3.66} is
\beq
S[T,x;\m,\l] = \m N_T+ \sum_{(ij)}(x_i-x_j)^2 + 
\l \sum_{(\triangle_i,\triangle_j)}
(n_{\triangle_i}-n_{\triangle_j})^2, \label{3.67}
\eeq
and the for the path integral we can write:
\bea
Z[\m,\l]& =& \int \frac{\cD g_{ab}}{{\rm Vol(diff)}}\int\cD x
\; \e^{-S[g,x;\m,\l]}\nn\\
& \to &
\sum_{T\in \cT}\frac{1}{S_T} \, \int \prod_{i \in T/\{i_0\}} \d x
\;\e^{-S[T,x;\m,\l]}\, . \label{3.68}
\eea

\begin{figure}
\input{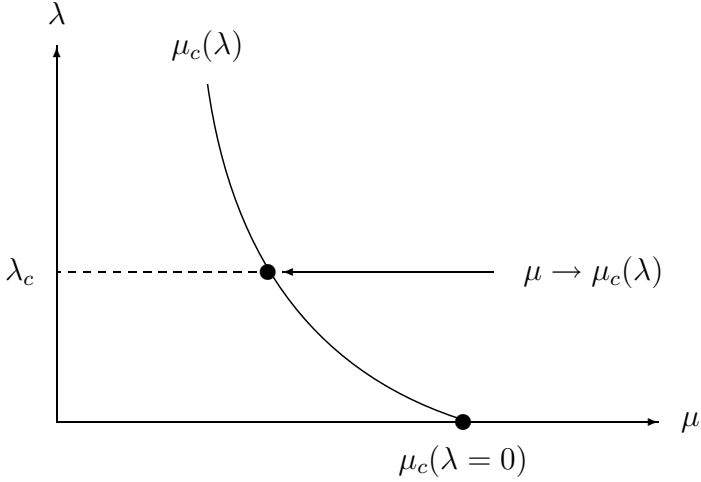}
\caption[3_9]{The phase diagram for the model \rf{3.68}}
\label{3_9}
\end{figure}

This is a theory with two coupling constants, as for the random walk
with extrinsic curvature. Qualitatively the phase diagram looks identical
to the diagram for the random walk except that in the two-dimensional
system we have {\it the possibility for a phase transition for a finite
value of $\l$} as shown in fig.\,\ref{3_9}.
Will there be a transition
for a finite value  $\l_c$ of $\l$? It is not known rigorously, but
extensive numerical simulations seem to support the idea that
we have  such a transition and that surfaces for $\l >\l_c$
are flat \cite{numex}. Could this be the transition we asked for where the
string tension scales to zero together with the mass.
Again numerical simulations seem to support this idea \cite{tension,aijp}!

Let us briefly discuss the possible scaling at the critical point.
For $\l=0$ we have seen that the string tension $\sg(\m,\l=0)$
does not scale to zero. Let $\m_c(\l)$ denote the critical line
in the $(\m,\l)$-plane (see fig.\,\ref{3_9}) and let $\Dm=\m-\m_c(\l)$
parametrize the approach to $\m_c(\l)$. Let us assume the string
tension has the form:
\beq
\sg(\m,\l) = \sg_0 (\l) + c(\l)\Dm^{2\n(\l)}~~~~~{\rm for} ~~~\Dm \to 0.
\label{3.69}
\eeq
Since $\sg_0(0) > 0$ we need to have $\sg_0(\l) \to 0$ for $\l \to \l_c$
in order to have a scaling of the string tension.
Let us finally assume that
\beq
\sg_0(\l) \sim (\l_c -\l)^\a ~~~~{\rm for} ~~~\l \to \l_c.
\label{3.70}
\eeq
As discussed above a number of times
the continuum limit is dictated by the exponential decay of
our correlators. Here we consider two: the one-loop function
$G_{\m,\l} (l_A)$ which falls off like $e^{-\sg(\m,\l)A}$ and the
two-point function $G_{\m,\l}(x,y)$ which falls off like
$e^{-m(\m,\l) |x-y|}$. There is no reason not to
expect the mass $m(\m,\l)$ to scale to zero for $\Dm \to 0$ for
all $\l \leq \l_c$ and we can write at $\l_c$:
\bea
\sg(\m,\l_c) A& =& c(\l_c) \Dm^{2\n(\l_c)} A = \sg_{{\rm ph}} A_{{\rm ph}}     \nn \\
m(\m,\l_c) x &  = &\tilde{c}(\l_c) \Dm^{\tilde{\n}(\l_c)} x =
m_{{\rm ph}} x_{{\rm ph}}.  \label{3.71}
\eea
The scaling limit is one where $(\cdot)_{{\rm ph}}$ are kept fixed
while $A$ and $x$ goes to infinity as $\Dm \to 0$.
We can introduce two different exponents $\n(\l)$ and $\tilde{\n}(\l)$.
But as we have already seen for the string without extrinsic curvature
they have to agree if the string tension scales:
$\n(\l_c) = \tilde{\n}(\l_c)$ and they will be related to the Hausdorff
dimension:  $\n = 1/d_H$.

From \rf{3.70} and \rf{3.71} it is thus possible to define a
consistent scaling limit for $\l \to \l_c$ and $\Dm \to 0$
such that $m^2(\m,\l) \to 0$ and $\sg(\m,\l) \to 0$ while the
ratio stays constant. The only requirement is that
\beq
\Delta \l^\a \sim \Dm^{2\n}.
\eeq

\vspace{12pt}

It should be emphasized that it is indeed possible to measure the
exponents $\n(\l)$, $\tilde{\n}(\l)$ and $\a$ by Monte Carlo simulations
of the statistical system \rf{3.68} \cite{aijp}. 
In such simulations it is convenient
to transform from the ``grand canonical'' ensemble \rf{3.68}
to a ``canonical'' ensemble where the number $N$ of triangles is fixed.
This is a Legendre transformation in $\m$ and $N$. There is no space
to discuss the details, but it should be mentioned that a very nice
feature of this transformation is that the string tension
$\sg(\m,\l)$ just becomes the expectation value of the simplest local
observable, the Gaussian action itself:
\beqn
\sg(\m,\l) = \lim_{N \to \infty} \frac{\la S_G \ra_N -\frac{D}{2}(N-1)}{A},
\eeqn
where the average is taken in the canonical ensemble and the formula
is true apart from subleading corrections in $1/A$.

The numerical simulations lead to a value of $\n$ close to $0.25$
i.e. $d_H \approx 4$ (more precisely $3.5 < d_H < 4.5$). It
is also possible to measure the string susceptibility at the
critical point\cite{abjpx}:  $ \g_s(\l_c) \approx  1/4$. The same values were
obtained  some years ago in a related hyper-cubic model \cite{bbm} and
suggests universality between the two models.
We will return to the interpretation of  $ \g_s(\l_c) \approx  1/4$
in the context of conformal field theories with central charge $c>1$
coupled to two-dimensional  quantum gravity.

\subsection{Supersymmetric random surfaces}

Part of the motivation to consider the surfaces with extrinsic curvature
came from the study of surfaces with worldsheet supersymmetry.
Is it possible
to implement the local supersymmetry at a discretized level?
The answer is no. Local supersymmetry cannot be put on a lattice since
the generators relate to translations and rotations which are
explicitly broken by the presence of the  lattice. However, if we
consider the Green-Schwarz formulation of the superstring the situation
is somewhat different. In this formulation there is no local
worldsheet supersymmetry, but space-time supersymmetry. This is
no problem in the discretized approach which is Euclidean
invariant. In fact one can directly write down an action which is
supersymmetric \cite{ms,av}.

Recall the continuum formulation of the Green-Schwarz superstring:
The simplest supersymmetric action is obtained by the replacement:
\beq\label{3.500}
\prt_a x^\m \to \Pi^\m = \prt_ax^\m -i\bar{\th} \g^\m \prt_a \th ,
\eeq
in the bosonic string action:
\beq\label{3.501}
\int \d^2\xi \sqrt{g} g^{ab}\prt_a x^\m \prt_a x^\m \to
\int \d^2\xi \sqrt{g} g^{ab}\prt_a \Pi^\m \prt_a \Pi^\m.
\eeq
This action possesses an obvious supersymmetry if $\th(\xi)$ like
$x^\m$ is a worldsheet scalar, but an anticommuting space-time spinor. The
global supersymmetry is generated by the infinitesimal transformations:
\beq\label{3.502}
x^\m \to x^\m -\frac{i}{2}(\bar{\th}\g^\m \ep-\ep^* \g^\m \th),~~~~~~
\th \to \th + \ep.
\eeq
At the discretized level we have assigned a bosonic variable $x_i$ to
each vertex in a given  triangulation $T$. We now assign an
additional fermionic variable $\th_i$ to the vertex and the
discretized supersymmetric action is:
\bea
S_T[x,\th]= (x_i^\m-x_j^\m - \Omega^\m_{ij})^2 \label{3.503} \\
\Omega^\m_{ij} \equiv \frac{i}{2}(\bar{\th}_i \g^\m \th_j -
\bar{\th}_j \g^\m \th_i) \label{3.504}
\eea
The path integral would now be written as a summation over all triangulations
and for a given triangulation the integration is over all $x^\m_i$'s and
$\th_i$'s:
\beq\label{3.505}
Z(\m) = \sum_{T\in \cT} \frac{1}{S_T} \int \prod_{i\in T/\{i_0\}}
\d x_i \d\th_i \d\bar{\th}_i \e^{-S_T[x,\th]}.
\eeq

The problem with the above action at the continuum level
is that it is not understood if it leads to  the theory we want. Even at the
classical level the continuum system seems quite impenetrable.
The Dirac bracket prescription leads to complicated expressions which
seem impossible to disentangle.
A naive counting of degrees of freedom
shows that in spite of the global supersymmetry there is not
a perfect match between the fermionic and bosonic degrees of
freedom.  In dimensions $D=3,4,6$ and 10 it is possible
to add an additional term to the action which simplifies the
situation drastically:
\beq\label{3.506}
\frac{i}{2}\int \d^2\xi \;\ep^{ab} \prt_a x^\m (\bar{\th}\g^\m \prt_b \th-
\prt_b \bar{\th} \g^\m \th).
\eeq
This term has the correct symmetry properties in the
above mentioned dimensions if the fermions are chosen as Majorana
spinors in $D=3$ and 6, Majorana or Weyl spinors in $D=4$ and
Majorana-Weyl spinors in $D=10$. In all of these cases the
resulting fermionic degrees of freedom will be $2(D-2)$. If
we naively state that the bosonic degrees of freedom will be
$D-2$, since two degrees will be absorbed in reparametrizations
of the surface, we still have an incorrect number of degrees of
freedom. It can be shown that eqs. \rf{3.501} and \rf{3.506}
together lead to an additional {\it local} fermionic symmetry
on the worldsheet, the so-called $\k$-symmetry,
 which effectively allows a decoupling of $D-2$ fermionic degrees of freedom.
Unfortunately the local nature of the $\k$ symmetry makes
it difficult to enforce at the discretized level. A suggestion
for a discretized version of \rf{3.506} for a given metric,
i.e. a given triangulation, has been \cite{ms}:
\beq\label{3.507}
i \sum_{\triangle (ijk) \in T} \ep_{ijk}x^\m_i\Omega^\m_{jk}.
\eeq
However, we have no exact $\k$-symmetry even with this term,
so from this point of view it is not clear that we need to add
the term.

To summarize, it is nice that one can write down a regularized
version of a superstring partition function,
which even at the discretized level has space-time supersymmetry.
However, it is not based on the action used in the continuum.
Nevertheless one should be aware that the continuum action
\rf{3.501} was discarded  not because it was wrong, but
because a  simpler
alternative arose after adding the term
given by eq. \rf{3.506} to the action \rf{3.501}. The action \rf{3.501}
might still serve well in a non-perturbative framework, like the       
one presented here.
\newsection{Matrix models and two-dimensional quantum gravity}\label{smallc}

\subsection{Matrix models}
For the bosonic string without extrinsic curvature term 
the results of the discretization was somewhat
disappointing in the sense that we did not find a string theory.
However, it is worth to  recall that it is possible to
calculate $\g_s$ using continuum methods \cite{kpz}:
\beq
\g_{s}= \frac{c-1-\sqrt{(c-1)(c-25)}}{12}.
\label{5.0}
\eeq
The formula is not valid for $c >1$. The
bosonic string has $c=D$ and  we are probing a most difficult
region.

For $c <1$ some results are known from Liouville theory \cite{ddk}, 
e.g. \rf{5.0}. The concept of
discretization works very well for general covariant theories
in the same region.
A number of aspects of the theories can be solved both by continuum
methods and directly at the discretized level. Historically
many of the results were obtain first using the discretized
approach. The exact solution
allows one to study in detail the scaling limit. In this section
I will outline how to solve some of the theories with $c \leq 0$ coupled to
quantum gravity, using very elementary tools.

The continuum action of two-dimensional quantum gravity coupled to
matter is given by:
\beq
S_{{\rm eh}}[g;\m,G]+S_m[\phi,g;\l] = \int \d^2 \xi \sqrt{g}
\left[ \m -\frac{1}{4\pi G}R(\xi) + \cL_m(\phi,g;\l) \right].
\label{5x.5}
\eeq
Here $S_{{\rm eh}}$ denotes the Einstein-Hilbert action,
$\cL_m$  an invariant Lagrangian density,
$\phi(\xi)$ a matter field and $\l$ a coupling constant.
$\m$ and $G$ are the  cosmological and the
gravitational coupling constants, respectivly.
Since the integration
over the curvature term is a topological invariant for closed two-dimensional
surfaces, only the cosmological term will play a role, except if we sum
over topologies. If we consider a specific manifold,
i.e. if the topology is fixed and characterized by its
Euler characteristic $\chi$, and if $A$ denotes the area of the manifold
for an equivalent class of metrics, the Einstein-Hilbert action is 
\beq
S_{{\rm eh}}[g;\m,G] =\m A - \frac{\chi}{G}. 
\label{5x.5a}
\eeq
The quantum partition function can be written as
\beq
Z[\m,G,\l]=
\int \frac{\cD g_{ab} (\xi)}{{\rm Vol(diff)}} \; \e^{-S_{{\rm eh}}[g;\m,G]}
\int \cD_g \phi(\xi) \; \e^{-S_m[\phi,g;\l]}.
\label{5x.6}
\eeq
We have already considered the discretization of \rf{5x.6}
in the case of $D$ Gaussian fields $\phi^\m$.
Let us consider pure two-dimensional quantum gravity, i.e.
eq. \rf{5x.6} without any matter fields $\phi$.  Recall the
discretized translation of the gravity part:
\beq
\int \frac{\cD g_{ab} (\xi)}{{\rm Vol(diff)}} \to \sum_{T\in \cT}
\label{5x.7}
\eeq
\beq
S_{{\rm eh}}[g;\m,G] \to  \m N_T -\frac{\chi(T)}{G},
\label{5x.8}
\eeq
and the partition function \rf{5x.6} can be written as:
\beq
Z[\m,G] = \sum_{T \in \cT} \e^{-\m N_T+ \chi(T)/G}
\label{5x.9}
\eeq
As usual the summation is over a suitable class of (abstract) triangulations.
As noted in the last section eq. \rf{5x.9} is not well defined since we
can split it as follows
\beq
Z[\m,G] =  \sum_K \e^{-\m K}
\sum_{T \in \cT_K} \e^{\chi(T)/G},
\label{5x.10}
\eeq
where $\cT_K$ denotes the abstract triangulations constructed
from $K$ triangles. The number of inequivalent
triangulations which can be build from 
$K$ triangle grows faster than $K!$. For a given
$K$ the Euler characteristic $\chi(T) \geq -K/2$ and it is clear that
the sum \rf{5x.10} is divergent for all $\m$. In the continuum
it is not known how to define the path integral \rf{5x.6} except for a fixed
topology. Eq. \rf{5x.10} shows that even in the discretized approach
where we have introduced a cut-off we can still not define such a sum
in a straightforward manner. We can split \rf{5x.10} in
a sum over triangulations of fixed topology:
\beq
Z[\m,G] =  \sum_\chi \e^{\chi/G} \sum_K^\infty \e^{-\m K}
\sum_{T \in \cT_K(\chi)}.
\label{5x.11}
\eeq
The number of triangulations of fixed topology has an exponential bound:
\beq
\sum_{T \in \cT_K(\chi)}\equiv \cN(K,\chi) =  
\e^{\m_c K}K^{\g_\chi -3} (1 + O(1/K)),
\label{5x.12}
\eeq
contrary to the factorial bound valid  when all topologies are
included. As a consequence we can make sense of eq. \rf{5x.11} if we
restrict the topology:
\beq
Z_\chi(\m) = \sum_K \e^{-\m K} \cN(K,\chi).
\label{5x.13}
\eeq
For a given $\chi$ the critical cosmological term is the minimal
value of $\m$ for which the sum \rf{5x.13} is convergent. This value
is precisely the value $\m_c$ in \rf{5x.12} and it can be shown
to be independent of $\chi$. For a fixed topology we can
try to define a continuum limit by approaching $\m_c$ from above.
By combining eqs. \rf{5x.12} and \rf{5x.13}
we see that sufficiently high derivatives of $Z_\chi (\m)$ will
diverge for $\m \to \m_c$. If we define
\beq
\bra K^n\ket_\chi  =
\frac{\sum_K K^n\;\e^{-\m K} \sum_{T \in \cT_K(\chi)} }{Z_\chi (\m)}
= (-1)^n \frac{1}{Z_\chi (\m)} \;
\frac{\d^n Z_\chi}{\d \m^n }
\label{5x.14}
\eeq
we find for $\m \to \m_c$ and $n > 2-\g_\chi$:
\beq
\bra K^n \ket_\chi \sim \frac{1}{(\m-\m_c)^{\g_\chi -2+n}}.
\label{5x.15}
\eeq
This is an indication that large $K$ will dominate for $\m \to \m_c$
and that it makes sense to introduce a scaling parameter $a$ such
that $A = K a^2$ is viewed as the {\it physical area} of our two-dimensional
world.

\begin{figure}
\input{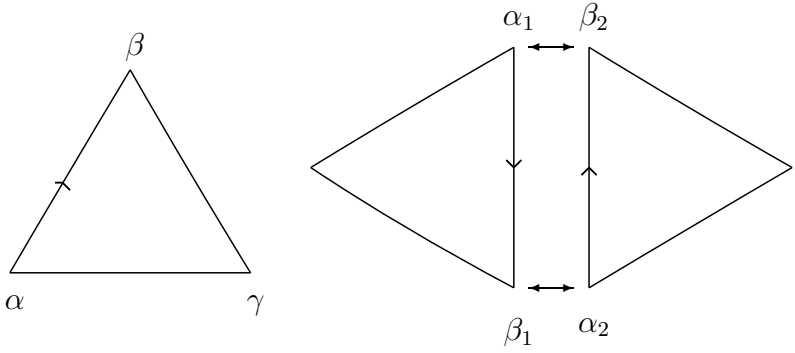}
\caption[5_1]{The matrix representation of triangles which converts the
gluing along links to a Wick contraction.}
\label{5_1}
\end{figure}

While these results have been obtained by the mathematicians \cite{tutte}
by explicit counting the number of ways to glue together
triangles to form closed combinatorial manifolds,
it is convenient from the
point of view of physics to make this counting ``automatic''.
This is done by representing the triangles by means of 
Hermitian matrices \cite{biz,david2}:
Label the vertices of the $i^{th}$
triangle by abstract indices $\a_i,\b_i,\g_i$
and attach an Hermitian matrix $\phi_{\a_i\b_i}$
to the oriented link from $\a_i$ to $\b_i$.
In this way we can attach the scalar quantity
\beq
\phi_{\a_i\b_i}\phi_{\b_i\g_i}\phi_{\g_i\a_i}=\Tr \phi^3
\label{5.8}
\eeq
to each of the $K$ triangles. The Gaussian integral
\beq
\int \d\phi \;\e^{-\oh \Tr \phi^2 } \frac{1}{K!} \left(\frac{1}{3}
\Tr \phi^3 \right)^K
\label{5.9}
\eeq
where
\beq
\d\phi \equiv \prod_{\a \leq \b} \d {\Re} \phi_{\a\b}
\prod_{\a < \b} \d {\Im} \phi_{\a\b}
\label{5.10}
\eeq
can be performed by doing all possible Wick contractions of $\phi$-fields.
This corresponds to performing all possible gluings of surfaces of $K$
triangles. The reason being that each Wick contraction in the
Gaussian integral glues together two links:
\beq\label{5.11}
\la \phi_{\a_i\b_i} \phi_{\a_j\b_j} \ra \equiv
\int \d \phi \;\e^{-\oh |\phi_{\a\b}|^2} \phi_{\a_i\b_i} \phi_{\a_j\b_j}
= \del_{\a_i\b_j} \del_{\b_i\a_j}.
\eeq
This is illustrated in fig.\,\ref{5_1}.
After all Wick contractions are performed on the lhs of eq. \rf{5.9} 
the $K$ triangles have been glued together in all possible ways.
The surfaces created in this
way will consist of disconnected parts, but we get the
connected  graphs\footnote{Among the connected graphs created in such an
unrestricted gluing will be graphs with one-loops and two-loops. These
are strictly speaking not combinatorial manifolds but they have a
clear identification as ``surfaces'' with a specific $\chi$. As already
mentioned we do not expect such short distance phenomena to play any role
in the scaling limit and this is substantiated by the known
fact that although the value of $\m_c$ in eq. \rf{5x.12} will
depend on the particular class of graphs we consider (one-loops excluded,
two-loops excluded etc), the exponent $\g_s$ will not.
This is in accordance with the general behavior in the theory of
critical phenomena: the positions of the critical points are not universal,
only the critical exponents.} by taking the logarithm of all graphs.
Furthermore  we can
calculate the contribution from a particular graph constituting a closed
surface: in the process of successive gluing we pick up a factor $N$,
$N$ being the number of indices, whenever a vertex becomes an internal
vertex in the  process of gluing together 
links by Wick contractions\footnote{It is important to stress
that $N$ is a formal expansion parameter which
should always be taken to $\infty$ at the end of a calculation since
the analogy with surfaces is based on the fact that the indices of the
different vertices are independent.}.
This means that we  get a total factor $N^V$, where $V$ is the
number of vertices. If we make the substitution
\beq
\Tr \phi^3 \to \frac{g}{\sqrt{N}} \Tr \phi^3
\label{5.12}
\eeq
it is seen that the factors of $N$ for $K$ triangles combine
to $N^{V-K/2}= N^\chi$, since the Euler characteristic
for a triangulation of $K$ triangles, $L$ links and $V$ vertices is
\beq
\chi= V-L+K= V -K/2.
\label{5.13}
\eeq
By multiplying each triangle $\Tr \phi^3$ by $1/\sqrt{N}$
the weight of a given triangulation only depends on its topology.
In addition the sum over all triangulations exponentiates. Collecting
this information we can write:
\beq
Z(\m,G)=\log \frac{Z(g,N)}{Z(0,N)}
\label{5.14}
\eeq
where $Z(\m,G)$ is defined by \rf{5x.11} and
\beq
Z(g,N)=\int \d\phi \;\exp\left(-\oh \Tr \phi^2 +
\frac{g}{3\sqrt{N}}\Tr \phi^3\right)
\label{5.16}
\eeq
provided we make the identification:
\beq
\frac{1}{G} = \log N,~~~~~ \m = -\log g.
\label{5.15}
\eeq
Below it is shown that eq. \rf{5.16} allows a $1/N^2$ expansion.
This expansion is therefore identical to the topological expansion
\rf{5x.11} of the random surfaces and eq. \rf{5.16} is an attempt to
perform a summation of this expansion. A glance at eq. \rf{5x.15} gives
an idea of the physics involved in this summation: 
Since $\g_s \leq 1/2$ the partition function $Z(\m)$ is finite 
at the critical point $\m_c$ and for $Z(\m)$ itself we get:
\beq
Z(\m,G) = \sum_\chi \frac{\e^{\chi/G}}{(\m-\m_c)^{\g_s(\chi) -2}}
+ {\rm less~singular~terms}.
\label{5.15a}
\eeq
The only way we can imagine at all a
summation over $\chi$ for $\m \to \m_c$ is to conjecture that
$\g_s(\chi)-2 = \tilde{c} \chi$ and that the {\it bare} gravitational
coupling constant $G$ is renormalized:
\beq
\frac{1}{G_R} = \frac{1}{G} - \log \frac{1}{(\m-\m_c)^{\tilde{c}}}.
\label{5.15b}
\eeq
{\it Although the Einstein-Hilbert action is topological in
two dimensions it will play a non-trivial role if we
attempt to perform the summation over topologies.}
The conjecture $\g_s(\chi) -2 = \tilde{c} \chi$ turns out to be true,
and $\tilde{c} >0$. This implies that the continuum limit $\m \to \m_c$
forces $G \to 0$ if we want to perform the summmation \rf{5.15a}.
This limit is called {\it the double scaling limit.} \cite{double}.
If we introduce the renormalized cosmological constant $\L$
by
\beq
\m-\m_c = \L a^2 \label{5.15c}
\eeq
this $a$ is precisely the lattice $a$ which goes into the
discretizations of triangles by $A_{ph} = N_T a^2$, where $A$ is the
continuum area of the surface. Equation \rf{5.15c} tells us that
that cosmological term has an additive renormalization.
In terms of $N$ the double scaling limit reads:
\beq
N a^{2\tilde{c}} = \e^{\frac{1}{G_R}}~~~~{\rm for} ~~~a \to 0, ~~N\to \infty
\label{5.15d}
\eeq
We will later verify the conjecture
\beq
\g_s(\chi) - 2= \tilde{c} \chi,~~~~({\rm in~fact~}\tilde{c}=5/4.)
\label{5.15e}
\eeq

A few comments are necessary at this point. Clearly a formula like \rf{5.16}
makes no sense as it stands. The matrix integral is not convergent
and we have only defined  and identified it with a
sum over triangulations in a perturbative sense, i.e. by expanding
$\exp (g \Tr \phi^3/(3\sqrt{N})$ in a power series,
and performing the resulting
Gaussian integral, i.e. gluing the triangles together via the Wick
contractions. It is of course a very interesting question if it is possible
to make sense of the integral \rf{5.16} in a non-perturbative way.
The original
continuum functional integral which led to \rf{5x.11} is  vague when it comes
to  the question of summing over different topologies. In the case of
string theory, and in the case of random surfaces which are intended
to be a regularized
version of string theory, {\it we have to sum over all topologies} in the way
indicated in \rf{5x.11} by unitarity: A closed
string can split in two, which can later join again, in this way changing
the topology of the surface from that of a sphere (with two boundaries)
to that of a torus (with two boundaries). In the case of gravity
it is not clear that such a change is required if we disregard any
connection between gravity and string theory. It is nevertheless tempting
to assume that  the summation should be performed. In the general case
of random surfaces we have already mentioned that our regularized
(discretized) approach has little to say about the sum over topologies:
It is a non-perturbative regularization of the string path integral
for a fixed topology, but perturbative in topology. In the special case
of strings in $d=0$, i.e. pure two-dimensional gravity, {\it a closed formula
like \rf{5.16} seems to offer some  possibility for a non-perturbative
definition of the sum over all topologies}. An obvious first suggestion 
is to define the functional integral by analytic
continuation. By the rotation $\phi \to e^{i\pi/6} \phi$
we can define a matrix integral
which has the {\it same} perturbative expansion as \rf{5.16}
\beq
\tilde{Z}(g,N) = \e^{-i\pi N^2/6} \int \d\phi\;
\exp\left(-\e^{i\pi/3} \Tr \phi^2 + i\frac{g}{3\sqrt{N}} \Tr \phi\right).
\label{5.17}
\eeq
Contrary to \rf{5.16} this integral is  well defined. If we expand the
interaction term in powers of $g$ and perform the Gaussian integrals we
get identical results to the ``Wick-gluing'' underlying the formal expression
\rf{5.16}. The problem with an expression like \rf{5.17} is that
we cannot be sure it is real. In fact it is not: it can be shown that
it contains a complex part which is {\it non-perturbative} in the
coupling constant $g$: it has the form
\beq
{\Im}\tilde Z(g,N) \sim e^{-{\rm const.}/g}
\label{5.18}
\eeq
i.e. it will never show up in the perturbative expansion in $g$.
It seems from these considerations that we have not yet succeeded
in a satisfatory definition of a non-perturbative
summation over topology, but it shows the potential power of the
discretized approach that one is able to discuss these questions at all .

A second, and much more simple minded remark is that the problem
with the definition of the matrix integral \rf{5.16} is not due
to the unboundedness of the $\Tr \phi^3$ term. Although from a geometrical
point of view it is natural to use triangles as building blocks,
in the context of two-dimensional quantum gravity  one could 
use squares, pentagons
etc.. Had we chosen to glue together squares, we would have a term
$g \Tr \phi^4/N$ in the action instead of the cubic term, but it would
still  appear with the wrong sign, i.e. the action would
be unbounded from below, since we want all ``surfaces''
build from squares to appear with a positive weight in our
functional integral. In general the gluing of $n$-gons
will be generated by the matrix integral
\beq
Z(g_n,N) = \int \d \phi \exp \left(-\oh \Tr \phi^2+ \frac{g_n}{nN^{n/2}}
\Tr \phi^n\right)
\label{5.19}
\eeq

For the purpose of a general (perturbative) analysis of the matrix integral
\rf{5.16} it is convenient to consider the generalization of
\rf{5.19} to an arbitrary set of coupling constants $\{g_i\}$:
\beq
Z(g_1,g_2,\ldots)= \int \d \phi\; \e^{-N \Tr V(\phi)}
\label{5.19a}
\eeq
where
\beq
V(\{g_i\}) = \sum_{n=1}^{\infty} \frac{g_n}{n}  \phi^n.
\label{5.20}
\eeq
In eq. \rf{5.20} we have of  convenience scaled $\phi \to \sqrt{N}\phi$.
In this way the topological nature of the expansion is still preserved:
All two-dimensional complexes of Euler characteristic $\chi$ will
have a factor $N^{\chi}$ associated with them. The interpretation of
\rf{5.19a} is intended to be as before: we have in mind a Gaussian integral
around which we expand, i.e. $g_2 >0$ and $g_n \leq 0$ with
the sign convention used in \rf{5.20}.
The convenience of considering an arbitrary
potential is that the general coupling constants $g_n$ act as sources
for terms like $\Tr \phi^n$, and by differentiating $Z$ with respect to
$g_n$ we can calculate expectation values of ``observables''
like $\Tr \phi^n$. $\bra \Tr \phi^n/N \ket$ has the following
obvious interpretation: It represent the summation over all
``surfaces'' which have a $n$-sided polygon as boundary. This follows from the
gluing procedure realized by Wick contractions of the Gaussian integrals.
In a similar way
\beq
\frac{1}{N^2} \bra \Tr \phi^n \Tr \phi^m \ket- \frac{1}{N^2}
\bra\Tr \phi^n\ket \bra\Tr \phi^m \ket
\label{5.21}
\eeq
will represent the sum over all connected two-dimensional complexes
which connect
one boundary consisting of $n$ links with another boundary consisting
of $m$ links. Since two-dimensional quantum gravity describes the
amplitude between one-dimensional geometries such expectation values
are precisely what we are looking for (in the end  we will of course
have to take some kind of scaling limit in order to make
contact with continuum physics). Let us  define the generating
functional for connected loop correlators. The expectation
value of an arbitrary observable is defined by
\beq
\bra Q(\phi) \ket \equiv \frac{1}{Z} \int \d \phi\;  
\e^{-N \Tr V(\phi)}\; Q(\phi).
\label{5.22}
\eeq
The generating function for $s$-loop correlators, which we will
also, somewhat inaccurate, denote the $s$-loop correlator, is defined by
\beq
W(z_1,\ldots,z_s) \equiv N^{s-2} \sum_{k_1,\ldots ,k_s}^{\infty}
\frac{ \bra \Tr \phi^{k_1} \cdots \Tr \phi^{k_s}\ket_{conn}}{z^{k_1+1}
\cdots z^{k_s +1}}
\label{5.23}
\eeq
where $conn$ refers  to the connected part as defined by \rf{5.21},
or its generalization to more correlators. One can rewrite \rf{5.23} as
\beq
W(z_1,\ldots,z_s) = N^{s-2} \bra \Tr \frac{1}{z_1-\phi}\;\cdots\;
\Tr \frac{1}{z_n-\phi}\ket_{conn}.
\label{5.24}
\eeq
In particular, we can drop the index $conn$ for the 1-loop correlator:
\beq
W(z) = \frac{1}{N} \bra \Tr \frac{1}{z-\phi} \ket = \frac{1}{N}
\sum_{k=1}^{\infty}\frac{\bra \Tr \phi^k\ket }{z^{k+1}}=
\frac{\d F}{\d V(z)}
\label{5.25}
\eeq
where we have introduced {\it the free energy $F$} by
\beq
Z(\{g_i\}) = \e^{N^2 F(\{g_i\})}
\label{5.26}
\eeq
and the so-called {\it loop insertion operator} by
\beq
\frac{\d}{\d V(z)} \equiv -\sum_{k=1}^{\infty} \frac{k}{z^{k+1}}\;
\frac{\d}{\d g_k}.
\label{5.27}
\eeq
The name ``loop insertion operator'' is natural since it follows from
the definition \rf{5.23} that
\beq
W(z_1,\ldots,z_s) = \frac{\d}{\d V(z_s)}\frac{\d}{\d V(z_{s-1})}
\cdots\frac{\d}{\d V(z_2)}W(z_1)
\label{5.28}
\eeq
and this equation shows that if the 1-loop operator is known for an
arbitrary potential, all multi-loop correlators can be calculated.

The 1-loop correlator is related to the density $\rho(\l)$
of eigenvalues defined by the matrix integral as follows :
\beq
\rho(\l) = \bra \sum_{i=1}^N \del(\l-\l_i) \ket
\label{*rho1}
\eeq
where $\l_i$, $i=1,\ldots,N$ denote the $N$ eigenvalues of the matrix
$\phi$.
With this definition we have
\beq
\frac{1}{N} \bra \Tr \phi^n \ket =
\int^\infty_{-\infty} \d\l\; \rho(\l) \l^n, ~~~\forall n \geq 0.
\label{5.28a}
\eeq
and therefore
\beq
W(z) = \int_{-\infty}^\infty \d\l\; \frac{\rho(\l)}{z-\l}
\label{*cut}
\eeq
For $N\to \infty$ exist, as we shall see,
consistent solutions where the support of $\rho$ is confined to a finite
interval $[y,x]$ on the real axis. In this case $W(z)$ will be
an analytic function in the complex plane, except for a cut at the
support of $\rho$ and it follows from
 Schwartz's reflection principle that
\beq
2\pi i\rho(\l) = \lim_{\ep \to 0} W(\l+i\ep)-W(\l-i\ep)
\label{*cut1}
\eeq

\subsection{The loop equations}

\begin{figure}
\input{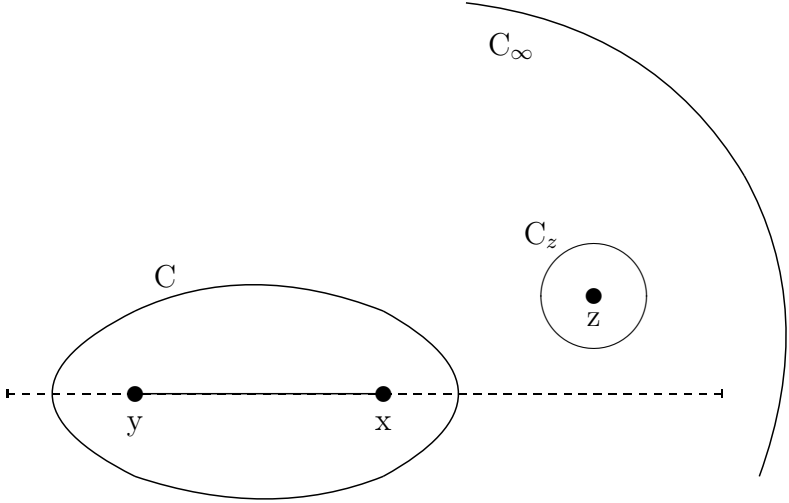}
\caption[5_2]{The integration contour $C$ and the cut from $y$ to $x$.
When deforming the contour to infinity we get two contributions:
one from the circle $C_\infty$ and one from the circle $C_z$ around
the pole $z$.}
\label{5_2}
\end{figure}

Amazingly few assumptions enter in the derivation of a set of equations
which allow us to solve, as an expansion in large $N$, the above defined
matrix model \cite{wadia,migdal,david3,jy}. 
Let us explore the invariance of the matrix integral
\rf{5.19a} under field redefinitions of the type:
\beq
\phi \to \phi+\ep \phi^n.
\label{*4.29}
\eeq
We consider $\ep$ as an infinitesimal parameter and one can prove
that to first order in $\ep$ the measure $\d \phi$ defined by eq. \rf{5.10}
will transform like
\beq
\d\phi \to \d\phi (1 + \ep \sum_{k=0}^n \Tr \phi^k
\; \Tr \phi^{n-k} ).
\label{*4.30}
\eeq
The action will change as
\beq
\Tr V (\phi) \to \Tr V(\phi) + \ep  \Tr \phi^{n} V'(\phi).
\label{*4.31}
\eeq
We can use these formulas to study the transformation of the measure
under more general field redefinitions of the form
\beq
\phi \to \phi +\ep \sum_{k=0}^\infty \frac{\phi^k}{p^{k+1}} =
\phi+ \ep \frac{1}{p-\phi}.
\label{*4.32}
\eeq
This kind of field redefinitions only make sense if $p$ is
chosen on the real axis outside the support of the
eigenvalues of $\phi$. As mentioned above we will verify that
this scenario is realized for $N \to \infty$.
Under the transformation \rf{*4.32}
the measure and the action change as
\beq
\d \phi \to \d\phi \left( 1+ \ep  \Tr \frac{1}{p-\phi}\;\Tr \frac{1}{p-\phi}
\right)
\label{*4.33}
\eeq
\beq
\Tr V(\phi) \to \Tr V(\phi) +
\ep \Tr \left ( \frac{1}{p-\phi} V'(\phi)\right).
\label{*4.34}
\eeq
The integral \rf{5.19a} will be invariant under a  redefinition of the
integration variables by eq. \rf{*4.32} and the change of measure
and action has to cancel to first order in $\ep$. By use of eqs. \rf{*4.33}
and \rf{*4.34} this leads to the following equation:
\beq
\int \d\phi \left\{ \left( \Tr \frac{1}{p-\phi} \right)^2 -
N\Tr \left( \frac{1}{p-\phi} V'(\phi) \right)\right\}\,\e^{-N\Tr V(\phi)}=0.
\label{*4.35}
\eeq
The first term in this equation is by definition
\beq
N^2 W (p) W(p) +  W(p,p).
\label{*4.36}
\eeq
The second term in eq. \rf{*4.35}
can be written as an integral over the the 1-loop correlator
by means of the density of eigenvalues:
\beq
\frac{1}{N} \bra \Tr \frac{V'(\phi)}{p-\phi} \ket =
\int \d\l \,\rho(\l)\, \frac{V'(\l)}{p-\l} =
\oint_C \dcmp \,\frac{V'(\om)}{p-\om} W(\om). \label{*4.37}
\eeq
The second equality is obtained by the rewriting
\beqn
\int \d\l\, \rho(\l) \left[ \oint_C \dcmp \,\frac{1}{\l-\om}\,
\frac{V'(\om)}{p-\om} \right] =
\oint_C \dcmp\, \frac{V'(\om)}{p-\om} \left[ \int \d\l\, \rho(\l)\,
\frac{1}{\l-\om} \right]
\eeqn
where the curve $C$ should enclose $\l$ but not $p$.
However, for the change of  the order of integration it is essential
that we can choose $C$ such that it encloses {\it all} eigenvalues
and not $p$. This is possible if the density $\rho$ has compact
support on the real axis.
With this assumption we finally can write \rf{*4.35} in the standard
form, known as the loop equation:
\beq
\oint_C \frac{d \om}{2 \pi i} \frac{V'(\om)}{z-\om} W(\om) =
W(z)^2 + \frac{1}{N^2} W(z,z)
\label{*loop}
\eeq
where $z$ is outside the interval $[y,x]$ on the real axis.
In addition, since we have seen that $W(z)$  is analytic outside the
support of $\rho$, eq. \rf{*loop} will be valid in the rest of the complex
plane by analytic continuation, again provided $C$ does not enclose $z$
(see fig.\,\ref{5_2}).

We can solve the loop equation by standard contour integration if we ignore
the last term on the rhs of eq. \rf{*loop}. This is in
accordance with a perturbative expansion in $1/N$, since the correlators
are normalized to order $O(1)$ in $N$. The approximation where we
ignore the last term on the rhs of \rf{*loop} is called the
large $N$ approximation. Translating back to the language of surfaces
the $1/N^2$ expansion is the  expansion in topology of the manifolds,
and the large $N$ approximation implies a restriction to
spherical topology.

Let us for simplicity assume that the potential is quadratic:
\beq
V(\phi)= \oh  \phi^2. \label{*4.38}
\eeq
From the definition \rf{5.25}  $W(z)$ falls off like
\beq
W_0(z) =\frac{1}{z}+ O(z^{-2}) \label{*4.39}
\eeq
and $C$ can be deformed to $\infty$ only picking up a simple pole ($ zW_0(z)$)
at $z$,  (see fig.\,\ref{5_2}). The contour at infinity
contributes with
\beq
\oint_{C_\infty} \dcmp \frac{1}{z-\om} = -1
\label{*4.40}
\eeq
and eq. \rf{*loop} reads
\beq
z W_0(z) - 1 =W_0(z)^2,~~~~
W_0(z) =\oh\left( z-\sqrt{(z-2)(z+2)}\right)
\label{*4.41}
\eeq
$W_0(z)$ has a square root cut at the
real axis ($[y,x]=[-2,2]$) and according to \rf{*cut1} the eigenvalue
density is related to $W_0(\l+i\ep)-W_0(\l-i\ep)$ and we have that
\beq
\rho(\l)= \frac{1}{2\pi} \sqrt{(2-\l)(2+\l)}
\label{*4.42}
\eeq
which is Wiener's famous semicircle law.

For a general potential we can find a solution which has essentially the same
structure as for the quadratic potential.
By deforming the contour $C$ out to infinity
we pick up a pole term $V'(z) W_0(z)$ at $z$, and at $C_\infty$ we
get the following integral
\beq
\oint_{C_\infty} \dcmpV\; W_0(\om), \label{*4.43a}
\eeq
which is a polynomium $Q(z)$ in $z$, as is seen by expanding
the integrand in powers of $1/\om$:
\beq
\frac{1}{z-\om}=-\sum_{j=1}^\infty \frac{z^{j-1}}{\om^j},~~~
W_0(\om) = \sum_{k=1}^\infty \frac{W_k}{\om^{k+1}},
\label{*4.43}
\eeq
Only the term with total power $-1$ will contribute,
i.e. if $V$ is a polynomium of power $n$ then $Q(z)$ will be of order $n-2$:
\beq
Q(z;g_i;W_k)= -\sum_{i=2}^n g_i \sum_{\stackrel{j+k =n-1,}{j>0,k>0}}z^jW_k.
\label{*4.43b}
\eeq
It follows that the loop equation
does not determine uniquely the $n-2$ ``moments''
$W_1,\ldots,W_{n-2}$, and the polynomium $Q(z)$ is consequently an
arbitrary polynomium of order $n-2$ which enters into the resulting
second order equation for $W_0(z)$
\bea
W_0(z)^2&=& V'(z)W_0(z) -Q(W_k,z,g_i), \nn \\
W_0(z)&=&\oh (V'(z)-\sqrt{(V'(z))^2-4Q(z)} \label{*4.44}
\eea
Not every choice of $Q(z)$ is allowed since $W_0(z)$ cannot have a
cut away from the real axis by construction. Let us here assume that
the eigenvalue density is as close in structure as possible to
\rf{*4.42}:
\beq
\sqrt{(V'(z))^2-4Q(z)}=M(z)\sqrt{(z-x)(z-y)}
\label{*4.45}
\eeq
where $M(z)$ is a polynomium of degree $n-2$. The eigenvalue density is
\beq
\rho(\l) \sim M(\l) \sqrt{(y-\l)(x-\l)}, \label{*4.46}
\eeq
and in this sense it is close to Wiener eigenvalue distribution \rf{*4.42}.
Nevertheless an interesting critical behavior
appears when zeros of $M(\l)$ accumulate near $x$ (or $y$) \cite{kazakov}.
We can  write:
\beq
W_0(z)= \oh \left( V'(z) -M(z) \sqrt{(z-x)(z-y)} \right)
\label{*sphere2}
\eeq
and the requirement  that $W_0(z)= O(1/z)$ uniquely determines $M(z)$ since
\rf{*sphere2} allows us to write
\beq
M(z) = \oint_{C_\infty} \dcmp \frac{M(\om)}{z-\om} =
\oint_{C_\infty} \dcmpV \frac{1}{\sqrt{(\om-x)(\om-y)}},\label{*M}
\eeq
as the part of the integral which involves $W_0(z)$ vanishes.
By expanding the last integrand in powers of $1/(\om-x)$ we can write:
\beqn
M(z) = \sum_{k=1}^\infty M_k (z-x)^{k-1},
\eeqn
\beq
M_k[x,y;g_i] = \oint_C \dcmp \frac{V'(\om)}{(\om-x)^{k+\oh}(\om-y)^{\oh}}.
\label{*M1}
\eeq
$W_0(z)$ can now be expressed in terms of the ``moments'' $M_k$:
\beq
W_0(z) = \oh \left(V'(z) -\sqrt{(z-x)(z-y)}
\sum_{k=1}^\infty M_k[x,y;g_i](z-x)^{k-1}\right). \label{*sphere1}
\eeq
The endpoints $x$ and $y$ of the cut is determined selfconsistently
by the following boundary conditions
\beq
M_{-1}[x,y,g_i] =2,~~~~M_{0}[x,y;g_i]=0, \label{*B1}
\eeq
These two equations follows  by contracting the
contour in the last integral in \rf{*M} to $C$. In this way we pick up
a pole term at $z$ which precisely cancels $V'(z)$ in \rf{*sphere2}
and we get yet another expression for $W_0(z)$:
\beq
W_0(z) = \sqrt{(z-x)(z-y)} \oint_C \dcmpV \frac{1}{\sqrt{(\om-x)(\om-y)}}.
\label{*sphere3}
\eeq
By expanding in $1/z$ and using that $W_0(z)$
contains no constant term and $W_1=1$ we get \rf{*B1}.

\subsection{Complete solution to leading order in $1/N^2$}

In principle the complete solution at spherical level is given by
\rf{*sphere1}-\rf{*B1}. These equations define $W_0(z)$ and
we can apply the loop inserting operator to obtain any multi-loop
correlator. Quite surprisingly one can obtain an explicit formula
if we change the matrix model slightly. Instead of Hermitian matrices
we use general complex matrices and consider a potential:
\beq
V(\pdp) = \sum_{n=1}^\infty \frac{g_n}{n} \Tr (\pdp)^n.
\label{5.30}
\eeq
The one-loop correlator is given by:
\beq
W_0(z) \equiv \frac{1}{N} \sum_n \frac{\bra \Tr (\pdp)^n \ket}{z^{2k+1}}.
\label{5.30a}
\eeq
This model has a surface representation \cite{morris}. $\Tr (\pdp)^n$ represents
a $2n$-gon where the boundary links have alternating black
and white colors, corresponding the $\phi$ and $\phi^\dagger$.
The effect of Gaussian integration
with respect to the complex matrices is
to glue together such ``checker-board'' polygons just as
Hermitean matrices glued together ordinary polygons. The only additional
rule is that only ``white'' and ``black'' links can be glued together
since $\bra \pdp \ket \ne 0$ while $\bra \phi^2 \ket=0$ and
$\bra \phi^\dagger\phi^\dagger \ket =0$.
Such short distance differences in gluing should be unimportant
in the continuum limit.

We can write down the loop equations for this model. Since the potential is
symmetric with respect to $\phi\to -\phi$ we have that $y=-x$ and we can
write:
\beq
\oint_C\dcmpxV W(\om) = W^2(z)+\frac{W(z,z)}{N^2},~~~~~
\oint_C\dcmpx \frac{\om V'(\om)}{\sqrt{\om^2-x^2}}=2,
\label{5.31}
\eeq
where the last equation expresses that $W_0(z) \sim 1/z$ for
$|z| \to \infty$.

The solution can be written as
\beqn
W_0 (z) = \oh\left(V'(z) - M(z) \sqrt{z^2-x^2}\right),
\eeqn
\beq
M(z) = \oint_{C_\infty} \dcmpx
\frac{\om V'(\om)}{(\om^2-z^2)\sqrt{\om^2-x^2}}
\label{5.33}
\eeq
If we expand the denominator of $M(z)$ in powers of $z^2/\om^2$ we can write:
\beqn
M(z) = \sum_{k=1}^\infty M_k[x,g_i] (z^2-x^2)^{k-1},
\eeqn
\beq
M_k[x,g_i] =
\oint_C \dcmpx \frac{\om V'(\om)}{(\om^2-x^2)^{k+1/2}},
\label{5.35}
\eeq
and the position $x$ of the cut is determined by the last equation
in \rf{5.31}:
\beq
M_0[x,g_i] = 2.
\label{5.35a}
\eeq

{\it While these formulas look rather complicated it is a  pleasant surprise
that things simplify a lot for the higher loop correlators.}
In order to apply the loop insertion operator we note that
\beq
\frac{\d}{\d V(z)} \equiv -\sum_k \frac{k}{z^{2k+1}} \;\frac{\d}{\d g_k} =
\frac{\prt }{\prt V(z)}
- \frac{x^2}{M_1 (z^2-x^2)^{3/2}}\frac{\d}{\d x^2}.
\label{5.36}
\eeq
where
\beq
\frac{\prt }{\prt V(z)}\equiv -\sum_k \frac{k}{z^{2k+1}}
\;\frac{\prt}{\prt g_k} ,~~~~~~~
\frac{\prt V'(\om)}{\prt V(z)}  = \frac{2\om z}{z^2-\om^2}.
\label{5.37}
\eeq
It is now straightforward, though tedious, to apply the loop insertion
operator. We find \cite{jy,ajm}:
\beq
W_0(z_1,z_2) = \frac{1}{4(z_1^2-z_2^2)^2}
\left[ z_2^2 \sqrt{\frac{z_1^2-x^2}{z_2^2-x^2}}+
       z_1^2 \sqrt{\frac{z_2^2-x^2}{z_1^2-x^2}}-2z_1z_2 \right].
\label{5.39}
\eeq
\beq
W_0(z_1,z_2,z_3) = \frac{x^4}{16M_1} \;
\frac{1}{\sqrt{(z_1^2-x^2)(z_2^2-x^2)(z_3^2-x^2)}}.
\label{5.40}
\eeq
\beq
W_0(z_1,\ldots,z_s) = \left(\frac{1}{M_1}\frac{d}{d x^2} \right)^{s-3}
\frac{1}{2x^2 M_1} \prod_{k=1}^s \frac{x^2}{2 (z_k^2-x^2)^{3/2}}.
\label{5.41}
\eeq
Notice the following \cite{ajm,ackm} (see also \cite{acm,akm,ak}):
\begin{itemize}
\item[(1):] The above formulas are valid for {\it any} potential $V$.
All dependence on the coupling constants are hidden in $M_1$ and $x$.
\item[(2):] From \rf{5.35} it follows that $dM_k/dx^2 = (k+1/2)M_{k+1}$.
For an arbitrary potential $V$ and $s>1$ the s-loop correlator
$W_0(z_1,\ldots,z_s)$ is a simple
algebraic function of $z_k$ and only depends on $x$ and $M_1,\ldots,M_{s-2}$.
\item[(3):] The same statements are true for the Hermitean matrix model,
except that we havetwo independent end points $x$ and $y$ and therefore two
independent set of ``moment'' $M_k$ and $J_k$ defined by:
\beq
M_k[x,y,g_i] = \oint_C \dcmp \frac{V'(\om)}{(\om-x)^{k+1/2} \sqrt{\om-y}},
~~~~
J_k[x,y,g_i] = M_k[y,x,g_i].
\label{5.42}
\eeq
\item[(4):] If we iterate the loop equation after $1/N^2$ this simplicity
continue to hold. For genus $g$ surfaces the Hermitean s-loop
correlator $W_g (z_1,...,z_s)$ is a simple rational function of
$(z_k-y)$ and $(z_k-x)$ and depends apart from $x$ and $y$ on
at most $2(3g-2+s)$ moment $M_k,J_k$, $k\leq 3g-2+s$. This is
even true for $s=0$ if $g>0$. As an example it can be shown that
the free energy $F_g$ for genus $g=1$ is
\beq
F_{g=1} = -\frac{1}{24} \log M_1J_1 (x-y)^4.
\label{5.43}
\eeq
This formula is valid for all potentials $V(\phi)$!
\end{itemize}

The proofs of the above statements are all trivial (but sometimes
tedious) and involve nothing beyond elementary linear algebra.

\subsection{The scaling limit}

In the case of the simplest potential ($\Tr \phi^3$ for the Hermitean
matrix model, $\Tr (\pdp)^2$ for the complex matrix model), we have
one independent coupling constant $g$ if we fix the coupling constant in
front of the Gaussian term.
Eq. \rf{5.15} gives the relation between the bare cosmological coupling
constant $\m$ and the coupling constant $g$ of the matrix models. We
have seen that there is a critical $\m_c$ such that the continuum
limit should be taken for $\m \to \m_c$. Corresponding to $\m_c$ there
will be a $g_c$. If we now introduce the lattice spacing $a$ the
area of our universe for a given triangulation is given by $N_T a^2$,
$N_T$ being the number of triangles in $T$. It is natural to introduce the
{\it renormalized} cosmological constant $\L$ by:
\beq
\m-\m_c = \L a^2,~~~~~{\rm i.e.}~~~g_c-g \sim \L a^2.
\label{5.44}
\eeq
In this way the renormalization follows the renormalization of
one-dimensional gravity and is in accordance with the general
additive renormalization of dimensionful coupling constants.
In case we extend our model and consider the gluing of arbitrary
polygons (but with positive weight) we expect nothing new, except
that the critical point $g_c$ will now be a ($n$-$1$) dimensional
hyper-surface if we have $n$ coupling constants.
The identification of this hyper-surface
is easy in the present formalism. Let us for simplicity consider the
complex matrix model. The density of eigenvalues $\rho(\l)$
is given by \rf{*4.46} and \rf{5.35}:
\beq
\rho(\l) \sim M(\l)\sqrt{x^2-\l^2},~~~~M(\l) = \sum_k M_k(x^2-\l^2)^{k-1}.
\label{5.45}
\eeq
The only non-analytic behaviour is associated with the endpoints
of the distribution $|\l| \to x$. If $M_1[g_i]$ is positive the
behavior will be identical to that of the Gaussian model.
$M_1$ involves a relation between the positive coupling constant
in front of the Gaussian term and the negative coupling constants
of the polygons. It can be fine tuned to zero on a $n-1$ dimensional
hyper-surface. All higher $M_k$ will be negative if 
the coupling constants $g_k$ are negative. We conclude that
the critical hyper-surface is determined by:
\beq
M_1[x(g_i),g_i] = 0.
\label{5.46}
\eeq
A glance on $W_0 (z_1,...,z_s)$ corroborates this observation since
it will be singular precisely when $M_1=0$. Let us denote
a critical point on the hyper-surface by $g_i^c$ and the
corresponding endpoint of the eigenvalue distribution by $x_c$.
In statistical mechanics the masses and running coupling constants
are defined {\it by the approach to the critical point}. We have already
seen examples of this in the case of random walks and non-critical
strings. Let us therefore move slightly away from the critical hyper-surface
by scaling the coupling constants according to \rf{5.44}:
\beq
g_i=g^c_i(1-\L a^2)=g_i^2+\del g_i.
\label{5.47}
\eeq
Corresponding to this change there will
be a change $x_c^2 \to x_c^2-\del x^2$.
It can be calculated directly from \rf{5.35a} using
$dM_k/dx^2=(k+1/2)M_{k+1}$ and $M_1[x_c,g_i^c] =0$. Let us introduce
the notation:
\beqn
M^c_i = M_i[x_c,g_i^c],~~~~~{\rm i.e}~~M_0^c=2,~M_1^c=0,~M^2_c \neq 0.
\eeqn
Expanding $M_0[x,g_i]$ leads to:
\beq
2=M_0[x,g_i]=M_0^c(1-\L a^2) +
\frac{3}{8} M^c_2\;(\del x^2)^2 + O(a^3),
\label{5.47a}
\eeq
or
\beq
(\del x^2) =\frac{16 }{3M_2^c} \;\L a^2, ~~~~{\rm i.e.}~~~
(\del x^2)^2 \sim \del g_i.
\label{5.48}
\eeq
In addition $M_1[x(g_i),g_i]$ will now be different from zero:
\beq
M_1[x,g_i] = \frac{3}{2 M_2^c}\; \del x^2 + O(a^2).
\label{5.48a}
\eeq
By a trivial rescaling of the cosmological constant we can there
{\it define} it by
\beq
x^2= x^2_c -a \sqrt{\L}.
\label{5.49}
\eeq

{\it Let us now calculate $\g_{s}$}. As discussed in detail in the
last section $\g_{s}$ will
be related to the  $s$-point function by\footnote{In the string case we had
to integrate the $s$-point function over all space, but here  is
no target space and the integration simply drops out.}:
\beq\label{5.61}
W_0(i_1,\ldots,i_s) \sim \frac{1}{(\m-\m_c)^{\g_2+s-2}} \sim
\frac{1}{ (\L a^2)^{\g_{s}+s-2}},
\eeq
where \rf{5.44} is used. We get the contribution to the $s$-point
function in the scaling limit from \rf{5.41} by first performing
the contour integral with some finite powers of $z_i$. This
produce a non-singular function of $x$. Using \rf{5.48} and \rf{5.48a}
we can take $a \to 0$ and we get:
\beq\label{5.62}
W_0(i_1,\ldots,i_s) \sim (\L a^2)^{5/2-s}.
\eeq
{\it We conclude that $\g_{str} = -1/2$}

Let us now turn to objects which are not readily calculated
in Liouville theory, {\it correlation functions between genuine macroscopic
loops} \cite{ajm}. These are object of fundamental interest in a quantum gravity
theory.

From the explicit formulas for the multi-loop correlators it follows
that the complex variable $z$ appears in the combination $z^2-x^2$ and
it is natural to introduce a scaling of $z^2$:
\beq\label{5.50}
z^2 = x_c^2+a\pi.
\eeq
With this notation we can write in the scaling limit :
\beq\label{5.50a}
M(z)= M_1 + M_2 (z^2-x^2) +\cdots = a M_2^c (\pi-\oh \sqrt{\L}) + O(a^{2})
\eeq
$\pi$ serves the same role in the scaling limit as $z$ at the
discretized level. Knowing $W_0(z_1,...z_s)$ allow us to
reconstruct the multi-loop correlators consisting of discretized
boundary loops of length $n_1,...,n_s$ by multiple contour integration.
In the scaling limit the physical length of these loops will
be $l_i=n_ia$, i.e. they will go to zero. If we want {\it genuine
macroscopic loops}
we have to scale $n_i$ to $\infty$ at the same time as $a \to 0$
such that $l_i$ is constant.  By substituting \rf{5.49} and \rf{5.50}
in the expressions for $W_0(z_1,...z_s)$ we get an expression
$W_0(\pi_1,...\pi_s;\L)$ and  we can reconstruct the corresponding
multi-loop amplitude $W_0(l_1,....,l_s;\L)$ by an inverse Laplace
transformation. The reason why the contour integration is changed into
an inverse Laplace transformation is that the cut $[-x,x]$
(or $[y,x]$ in the Hermitean matrix model) by the substitution
\rf{5.49} and \rf{5.50} is changed into a cut  $] -\infty,\sqrt{\L}]$.
The contour integration around the cut can now be deformed:
\bea\label{5.51}
\lefteqn{\oint_{C_i}\prod_{i=1}^s\d z_i \,z^{2n_i}\;W_0(z_1,\ldots,z_s) \to}\\
&& x_c^{2(n_1+\cdots+n_s)}
\int^{c+i\infty}_{c-i\infty} \prod_{i=1}^s \d\pi_i\,\e^{l_i\pi_i}\;
W_0(\pi_1,\ldots,\pi_s;\L)\, ,~~~~~~c > \sqrt{\L},   \nn
\eea
since we have
\beq\label{5.52}
z^{2n} = x_c^{2n}(1+a\pi)^{l/a} \sim x_c^{2n}\e^{\pi l},~~~l\equiv n/ax_c^2.
\eeq
The highly divergent factor $x_c^{2(l_1+\cdots+l_s)/ax_c^2}$
is a wave function renormalization of the macroscopic boundaries.
It is to be expected  since it is possible in the continuum to  add a term
\beq\label{5.53}
S_{\prt\cM} = \l \int_{\prt \cM} \d l = \l \int \d\xi \;g^{\frac{1}{4}}(\xi)
\eeq
to the action. This is just the induced one-dimension gravity on the
boundary of the manifold. Since $\l$ has the dimension of mass
we expect in the discretized version that the bare coupling constant
will undergo an additive renormalization (like the cosmological
constant itself), i.e. we will have to cancel a term like $x_c^{2n}$.

From eq. \rf{5.41} we immediately get the generating
functional for macroscopic multi-loop amplitudes:
\beq
W_0(z_1,\ldots,z_s)\sim a^{5-7s/2} w_0(\pi_1,\ldots,\pi_s;\L),\label{5.54}
\eeq
\beq
w_0(\pi_1,\ldots,\pi_s;\L) = \frac{\d^{s-3}}{\d\L^{s-3}}\,\frac{1}{\sqrt{\L}}
\prod_{k=1}^s \frac{1}{(\pi_k+ \sqrt{\L})^{3/2}},~~~~s\geq 3. \label{5.55}
\eeq
The expressions for $w_0(\pi,\L)$ and $w_0(\pi_1,\pi_2;\L)$ are slightly
more complicated since $W_0(z)$ contains a non-universal part ($V'(z)/2$).
and we will not give them here.

We can calculate the inverse Laplace transform \rf{5.51}
\bea
w_0 (l_1,\ldots,l_s;\L) &=& \int_{c-i\infty}^{c+i\infty}
\prod_{i=1}^{s} \d\pi_i\, \e^{\pi_il_i} \; w_0(\pi_1,\ldots,\pi_s;\L)
\nonumber \\
& =&\frac{\d^{s-3}}{\d\L^{s-3}}\,\frac{\sqrt{l_1\cdots l_s}}{\sqrt{\L}}
\;\e^{-\sqrt{\L}(l_1+\cdots+l_s)}.  \label{5.60}
\eea

\subsection{Generalizations}

In the last subsection we considered the spherical limit
of ordinary discretized gravity.
All graphs appeared with positive weight
and the scaling limit was independent of the precise class of
graphs used: Any polynomial
\beq\label{5.63}
V(\pdp) = \sum_{n=1}^{n_0} \frac{g_n}{n} (\pdp)^n,~~~g_1 >0,~~g_n < 0,~n >2
\eeq
leads to the same scaling limit.

Let us lift the
constraint in eq. \rf{5.63}. In this way we clearly              
move away from pure gravity since some of the polygons are glued
together with negative weight. However, as already mentioned, formulas
like \rf{5.39}-\rf{5.41} are valid for {\it any} potential where
$g_1 >0$. Without the constraint it is possible to get a new critical
behavior in the scaling limit \cite{kazakov} since one can fine tune
\beq\label{5.64}
M_k [x(g_i),x_i]=0,~~~k < m,~~~~~M_m[x(g_i),g_i] \neq 0.
\eeq
The hyper-surface in coupling constant space satisfying \rf{5.64} is called
{\it the $m^{th}$ multicritical hyper-surface} and pure gravity
corresponds to $m=2$. On this surface the eigenvalue
density $\rho(\l) \sim M(\l) \sqrt{x^2-\l^2}$ changes from the generic
Wiener form to :
\beq\label{5.65}
\rho(\l) \sim (x^2-\l^2)^{m+1/2}.
\eeq
Let us move away from the critical surface in a way similar to \rf{5.47}:
\beq\label{5.66}
g_i =g_i^c (1 - (\L a^2)^{m/2})= g_i^c+\del g_i.
\eeq
We want to maintain \rf{5.49} and \rf{5.50} which allow us to identify
$a$ with  the link-length. This being the case, \rf{5.64}
forces us to replace \rf{5.48} and \rf{5.48a} with
\beq\label{5.67}
(\del x^2)^m \sim \del g_i,~~~~~~~M_1[x_i(g_i),g_i]\sim (\del x^2)^{m-1},
\eeq
and one is finally led to the power $a^m\L^{m/2}$ used in \rf{5.66}.
We can repeat the arguments used for pure gravity and find corresponding to
\rf{5.62}:
\beq\label{5.68}
W_0 (i_1,\ldots,i_s) \sim (\L a^2)^{m+\oh} \left[(\L a^2)^{m/2}\right]^{-s}
\eeq
If we compare this formula for $s=0$ to the free energy
$F \sim (\L a^2)^{2-\g_{s}}$ we conclude:
\beq\label{5.69}
\g_{s} = -m +\frac{3}{2}.
\eeq
Let us tentatively compare this to a $(p,q)$ minimal conformal field theory
coupled to gravity. The central charge is $c = 1-6(p-q)^2/pq$ and 
from \rf{5.0} we have:
\beq\label{5.70}
\g_{s} = 1-\frac{q}{p},~~~q>p~~{\rm and~co-primes}.
\eeq
\rf{5.69} leads to {\it the identification of the $m^{th}$ multicritical
model with a (2,2m-1) minimal conformal theory coupled to quantum gravity}.
It might be confusing that the s-point function $W_0(i_1,...,i_s)$
does not behave like \rf{5.61}. But the explanation
is perfectly in accordance with the identification suggested. $(p,q)$-models
have operators of negative scaling dimension, the most negative being
$\Delta_0 = (1-(p-q)^2)/4pq$. After coupling to gravity they might produce a
more singular behavior than the cosmological term.
The change in potential \rf{5.66} excite all operators, including
possible negative dimensional ones. These will dominate over the
cosmological term  for $m >2$ and the scaling dimension of the
most negative one integrated over the whole surface
produces the observed behavior $(\L a^2)^{m/2}$. 
This implies that each time one inserts ``an arbitrary puncture''
on the surface, i.e an excitation which involves the negative 
dimensional operator, it will multiply the partition function by 
$\tilde{\L}^{-1}$, where $\tilde{\L}= \L^{m/2}$. From this 
point of view $\tilde{\L}$ acts like the cosmological term in 
the unitary theories and we can rewrite eq. \rf{5.28} as:
\beqn
W_0 (i_1,\ldots,i_s) \sim (\tilde{\L} a^m)^{2+\frac{1}{m}} 
\left[(\tilde{\L} a^m)\right]^{-s} 
\eeqn
from which one would be tempted to define a 
\beqn
\tilde{\g}_s = -\frac{1}{m}.
\eeqn 
It is of course a matter of definition whether one 
uses \rf{5.69} or $\tilde{\g}_s$.

\vspace{12pt}
\noindent
The second point concerns the expansion beyond the spherical limit.
Using the loop equation it can be performed without any problem.
Let us here just mention the result in the double scaling limit:
Approaching the $m^{th}$ multicritical hyper-surface \rf{5.64}
according to $\del g_i \sim O(a^m)$ leads to the following
behavior of the moments $M_k$:
\beq\label{5.71}
M_k \sim \m_ka^{m-k},~~~1 \leq k\leq m,~~~~~M_k = O(1),~~~k \geq m.
\eeq
The free energy has a genus expansion \cite{iz,ackm}:
\beq\label{5.72}
F = \sum_{g=0}^\infty \frac{1}{N^2 a^{2m+1}} F_g,~~~~
F_g = \sum_{\a_i >1} \bra \a_1\cdots\a_s\ket_g
\frac{\m_{\a_1} \cdots \m_{\a_s}}{\m_1^{2g-2+s} d^{g-1}}
\eeq
where $\sum_j \a_j = 3g-3+s$ and $d=x-y$ is the length of the cut.
The second equation is valid only for $g>0$. 
For $g=0$ there are additional non-universal parts.
The numbers $\bra\a_1\cdots\a_s\ket_g$ are independent
of the multicritical point and can be identified with so-called
{\it intersection indices} in the moduli space $\cM_{g,s}$ of Riemann surfaces
of genus $g$ with $s$ punctures.

This expansion points towards two apects of the theory which we have no
space to discuss further 
\begin{itemize}
\item[(1):] {\it The double scaling limit} where
$N^2 a^{2m+1}$ is kept fixed
should be identified with the renormalized gravitational constant via
\rf{5.15d}. In fact we get $\tilde{c}=(2m+1)/4$.
\item[(2):] The strong indication of topological nature of the
theory due to the surprising and beautiful
appearance of the intersection indices of moduli spaces of Riemann surfaces.
(see \cite{iz} for a review, and \cite{ak} for the connection
to the so-called Kontsevich model.)
\end{itemize}

What we would rather like to stress here is the perfect analogy to
standard statistical mechanics. The ``masses'' of the theory
($\m_k$) are fixed by the specific approach to the critical surfaces.
These are hyper-surfaces in an (in principle) infinite dimensional
coupling constant space and they are all of finite co-dimension,
precisely as we expect in the  general analysis of critical phenomena.

\newsection{The mystery of $c >1$}\label{largec}

\subsection{The Ising model}

The coupling of any matter fields to two-dimensional quantum
gravity at the discretized level is in principle simple.
First consider the theory in ordinary two-dimensional space.
Most two-dimensional field theories
can be ``latticized'', i.e. they can be formulated as
statistical models on a regular two-dimensional lattice in such
a way that the continuum limit is recovered as the scaling
limit where the ``bare'' coupling constants are scaled to
a critical point. At this critical point a correlation length
 is divergent and one can forget the
underlying lattice. In order to couple the theory to gravity
we formulate the model on  random lattices and take
the {\it annealed average over a suitable ensemble of
random lattices} which can be identified with surfaces
of a certain topology. Usually we have in mind triangulations,
but as we saw in the last section, it is possible to
glue together  a large variety of polygons without changing
the critical behavior of pure gravity. To the extend that  it is
possible to formulate the matter theory on such polygons, we
expect the same universality after coupling to matter.
After taking the annealed average there might still be a
phase transition and at the transition point it might
be possible to define a continuum limit of the theory.
This continuum limit will then qualify as an explicit realization
of the original theory coupled to quantum gravity.
In general the critical properties of the theory defined by
taking the annealed average over the class of random lattices
will differ from the corresponding critical properties
of the theory defined on a regular lattice. This change
is interpreted as the influence of quantum gravity on the matter fields.
However, the weight attributed to the random surfaces is influenced
by the presence of the matter fields and {\it at} the critical point
the non-analytic part of the partition function can
change. Critical exponents like $\g_s$ change too, and this
has the interpretation as a {\it back reaction of the matter on gravity}.

In sec.\,\ref{smallc} this program was realized, although
the identification with the $(2,2m-1)$ minimal conformal
field theories were made in a rather indirect way for $m >2$.
They all corresponded to $c < 0$  for $m > 2$,
i.e. to non-unitary field theories. Conceptually it is more
interesting to consider unitary theories, i.e. theories with
$c > 0$. The Ising model was the first model with $c > 0$
where it was possible to take the annealed average and in this
way calculate the critical exponents of the theory after
coupling to gravity \cite{kazakov1,bk}. 
Later the results were confirmed by the use of
continuum methods.

Following the strategy just outlined we
define the model on a regular lattice. This is just
the ordinary Ising model\footnote{On a regular lattice the
spins are usually placed at the sites of the lattice. Alternatively
one could place the spins in the centers of the simplexes or hyper-cubes
etc. depending on the structure of the lattice. By connecting the
centers we can view the latter assignment as a spin system on the dual
lattice. When it comes to critical properties of the system
one would not expect any difference since we will usually be interested
in long distance behavior where we {\it want} to forget
about the underlying lattice. Historically the spins on dynamical
triangulations were put at the centers of the triangles and we
will follow this convention.}.
Next we define the Ising model on a random triangulation as follows:
\beq\label{6.1}
Z(\m,\b) = \sum_{T \in \cT}\frac{1}{S_T} \e^{-\m N_T}{\sum_{\{\sg_i\}}}
\exp \left(\frac{\b}{2} \sum_{(ij)} (\sg_i \sg_j -1)\right)
\eeq
where $i,j$ refers to the triangles in $T$ and $\sum_{(ij)}$ is 
over pairs of neighboring triangles.  
We assume that the triangulation has spherical topology.
The explicit solution of this model is made possible by mapping it
on a two-matrix model. To each triangle $i$ is associated a spin $\sg_i$
variable which can take two values ($\pm 1$) and the model \rf{6.1}
has the following representation in terms of triangulations: we have
to glue together two kind of triangles (with labels $\pm$)
 in all possible ways compatible a given topology (here taken
to be spherical). The weight of the gluing along  links will be 1
if the triangles are identical and $e^{-\b}$ if they are different.
In addition we have the usual weight given by the total number of triangles.
This generalized gluing process can be realized by a two matrix model
where the two kind of triangles are represented by two different matrix
potentials $\phi^3$ and $\psi^3$, $\phi$ and $\psi$ being Hermitean matrices
as in the one-matrix model. In order to perform the gluing we use a
Gaussian part:
\beq\label{6.2}
\frac{1}{2(1-\e^{-2\b})}\Tr
\left(  \phi^2 + \psi^2 -2\e^{-\b} \phi \psi \right).
\eeq
The inverse of this quadratic form will be given by
\beq\label{6.3}
\pmatrix{1 & \e^{-\b} \cr \e^{-\b} & 1}\; \del_{\a_1\b_1}\del_{\a_2,\b_2}
\eeq
where $\a,\b$ refers to the matrix indices, while
the $2\times 2$ matrix refers the indices $\pm$. It glues triangles with
different spin with weight $e^{-\b}$ as desired.

This two-matrix model can be solved explicitly, as was the case
for the one-matrix model. We will not discuss the solution here,
only mention the result. For large $\b$ (low temperature)
the system is magnetized. The spins are aligned. All triangulations
have the same magnetic energy for this spin configuration
and this means that gravity can fluctuate as if there where
no spins at all: The fractal geometry will be as in the
pure gravity case and $\g_s = -1/2$. For very low $\b$ (high
temperature) the spins will fluctuate wildly and not care
about any underlying lattice structure either. The magnetization
will be zero, the geometry will still be independent of the
spin system and $\g_s=-1/2$. As $\b$ is increased
from zero the spin system
will start to interact with geometry. This is possible since the
magnetic energy is proportional to the length of boundaries
between spin clusters.
On a regular lattice a large spin cluster of area $A$ will have
a length $L \geq \sqrt{A}$. This is not necessarily  so on a
dynamical lattice. A glance at fig.\,\ref{6_1} shows that we
can have very large spin clusters separated by boundaries of only
a few links. In this way it is clear that the matter
system will have a tendency to deform the triangulations
towards geometry with small ``bottle necks'' like the
ones shown in fig.\,\ref{6_1}. {\it This gives us a direct visualization
of the back-reaction of matter on geometry}. For the Ising model this
back-reaction is not sufficiently strong to the change $\g_s$ before
we reach  a $\b$ so high that it is favorable for the spins
to align: we have a transition. At the transition $\g_s$ changes
to $-1/3$. Above the transition it jumps back to $-1/2$.
Not only does the spin system affect the fractal structure of the geometry.
The fluctuating geometry will change the critical properties of the
spin system. Intuitively one would expect it to soften the transition
and this is what happens. The second order transition for the
Ising model on a regular lattice is changed to a third order
transition and the specific heat exponent $\a$ is changed
from 0 to $-1$. The critical exponents $\b$ for the
magnetization and the spin susceptibility exponent $\g$
(not to be confused
with $\g_s$) are also changed. Needless to say, the exponents calculated
this way agree with the KPZ exponents of a $c=1/2$ conformal field
theory coupled to two-dimensional quantum gravity.

\subsection{Multiple Ising spins}

It is now trivial to couple many Ising spins to quantum gravity.
We simply put several independent species of Ising spins on
each triangle. On a regular lattice this does not lead to any
new physics since they are non-interacting. The only change is that
the central charge will be $c=n/2$, $n$ being the number of Ising copies.
On dynamical triangulations they will interact in a non-trivial
way via the back-reaction on geometry as described above.
It is clear how one can formulate the models as {\it multi-matrix models}.
However, we cannot solve these. If the number of copies $n$
of Ising models is larger than 2 we enter into the region
of strong interaction between geometry and matter in the sense that
the KPZ formula \rf{5.0} breaks down.  In this region it is
still not fully understood what happens. This is the same region where
the non-critical string is defined.
However, for $n \to \infty$ it is possible to analyze what
happens using mean field theory\footnote{The discussion here follows 
\cite{adj1}, since the interpretation in terms of random surfaces
is very simple. However, the same mean field arguments have appeared
in slightly different contexts in a number of papers 
\cite{indians,korchemsky,alvarez,wexler,wheater}.}.

As a starting point we use some very important  inspiration 
from the extensive numerical simulations
of multiple Ising models on dynamical triangulations
\cite{bj,adjt,at,ckr}. The simulations indicate that:
\begin{itemize}
\item[1)] There is still a critical
point $\b_c$ below which there is no magnetization and above
which the system is magnetized.
\item[2)] Above $\b_c$ the geometry seems to be that of pure
gravity. Below $\b_c$ there is a region where the situation is not clear,
and where the fractal structure of the surface is very pronounced.
For sufficiently small $\b$ the geometry again is that of pure
$2d$ gravity.
\item[3)] For a large number of spins it seems as if the system
increasingly fast will be totally magnetized when $\b > \b_c$.
I.e. for $\b > \b_c$, but quite close to $\b_c$,
the system will essentially be identical to the one at $\b = \infty$.
\item[4)] For a sufficiently large number of Ising spins $\g_s(\b_c)$
will be positive.
\end{itemize}

\begin{figure}
\input{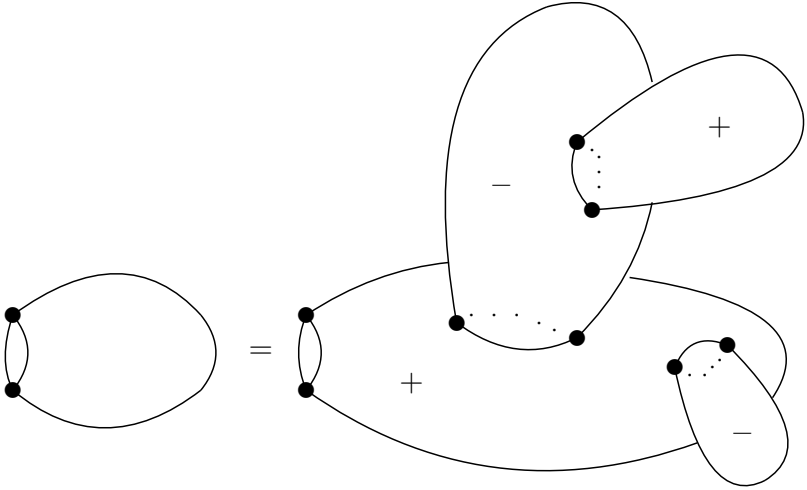}
\caption[6_1]{A graphical representation of eq. \rf{6.9}. The complete
one-loop function allows a recursive decomposition
into baby universes with definite spin assignment.}
\label{6_1}
\end{figure}

In view of point 2) and 3) it seems reasonable to attempt a description
of the model (for a large number of Ising copies) in a region around
$\b_c$ and for large $\b$ in terms of an effective model which
has only spin excitations with  minimal boundaries separating
$\pm$ regions since these excitations have a minimum energy.
Let us for convenience of the following arguments specify the
class of dynamical triangulations $\cT_2$ used in the annealed average to
include all gluings of triangles such that the corresponding surfaces
are spherical and the minimal length of a closed loop of links
is two. Such minimal loops are shown in
fig.\,\ref{6_1}. For this  class of triangulations a minimal
boundary  between spin clusters will have length two and
we can view the surfaces as glued together of $\pm$ decorated baby
universes, i.e. parts of the surface connected to the rest by a
small loop of length two. In other words
we consider the summation over all triangulations in the class $\cT_2$
and on a given triangulation we consider self-consistent iterations
of the minimal spin-energy excitation. This is obviously identical
to the first term in a low temperature expansion on a regular lattice.
Here it will interact non-trivially with the geometry.

Let us in this mean field model consider the one-point (or
more precisely one-loop)  function $G(\m,\b)$,
where the boundary just consist of two (marked) links. The
boundary eliminates the symmetry factor $1/S_T$ in eq. \rf{6.1} and
we have
\beq\label{6.4}
G (\m,\b) = \sum_{T \in \cT'_2} \e^{-\m N_T}{\sum_{\{\sg_i\}}}'
\exp \left(\frac{\b}{2} \sum_{(ij)} (\sg_i \sg_j -1)\right)
\eeq
where $\cT'_2$ denotes the class of triangulations where the boundary is
a loop consisting of two (marked) links and
${\sum_{\{\sg_i\}}}'$ denotes the
summation over the restricted class of spin configurations.
 Similarly, one defines
$n$-point functions that are essentially derivatives w.r.t. $\m$ of
the one-point function. The susceptibility $\chi(\m,\b)$ is defined as
\beq\label{6.7}
\chi(\m,\b) = - \frac{\partial G}{\partial \m}
\eeq
and the string susceptibility exponent $\g_s (\b)$ is given by
\beq\label{6.8}
\chi(\m,\b) = \frac{c(\b)}{(\m -\m_c(\b))^{\g_s(\b)}} +
{\rm less~~singular~~terms},
\eeq
for $\m\to \m_c(\b)$, where $\m_c(\b)$ denotes the critical cosmological
constant as a function of $\b$.

The corresponding quantities in  pure 2d gravity will be denoted by
$\chi_0(\m)$, $\g_s^{(0)}$ and $\m_0$.

Recall that the reason why the Ising model on the dynamical triangulations
could be solved was the ability to map it to a random surface
model (gluing two kind of triangles). For our mean field model
we have this mapping explicitly given to us by the recursive
decomposition in baby universes of alternating spin orientation
as illustrated in fig.\,\ref{6_1}.

Using this decomposition procedure and summing first over baby
universes with different spin configurations we obtain
\beq\label{6.9}
G(\m,\b) = \sum_{\bar{T} \in \cT'_2} \e^{-\m N_{\bar{T}}}
\left( 1+ \e^{-2\b} G(\m,\b)\right)^{L_{\bar{T}}}
\eeq
where $L_{\bar{T}}$ denotes the number of links in $\bar{T}$,
i.e. $L_{\bar{T}} = 1+\frac{3}{2}N_{\bar{T}}$. In eq. \rf{6.9}
the factor $\e^{-2\b}$ represents the coupling of a baby
universe across the phase boundary to the rest of the surface and
the factor 1 in the parentheses originates from the empty baby universe.
We can now write
\beq\label{6.10}
G(\m,\b) = \sum_{T\in \cT'_2} \e^{-\bar{\m} N_T} = G_0(\bar{\m}),~~~~~~~~
\bar{\m}=\m-\frac{3}{2} \log \left(1+ \e^{-2\b}G(\m,\b)\right).
\eeq
Note that the last equation can be written as
\beq\label{6.11}
\m = \bar{\m} + \frac{3}{2} \log \left( 1 + \e^{-2\b} G_0(\bar{\m})\right)
\eeq
which expresses $\m$ in terms of known functions of $\bar{\m}$ and $\b$
since pure gravity can be solved.

From eqs. \rf{6.10} and \rf{6.11}  we get
\beq
\chi(\m,\b) = \chi_0 (\bar{\m}) 
\frac{\partial \bar{\m}}{\partial \m},\label{6.12}
\eeq
\beq
\frac{\partial \bar{\m}}{\partial \m} =
\frac{\e^{2\b} + G_0 (\bar{\m})}{\e^{2\b}-
(\frac{3}{2} \chi_0 (\bar{\m}) -G_0(\bar{\m}))}. \label{6.13}
\eeq
It is clear that the derivation and the equations are very similar to
the ones used to prove $\g_s=1/2$ for the non-critical strings.
Indeed, we will find the same results here, but the additional
coupling constant $\b$ allows a new critical behavior as will
be clear.

\begin{figure}
\input{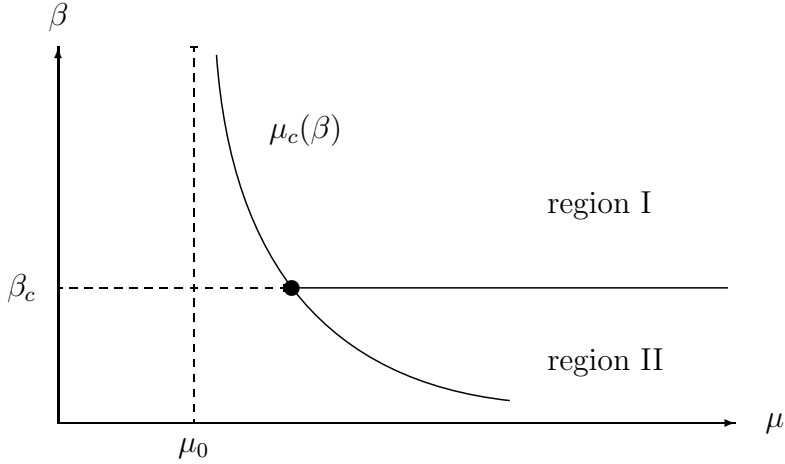}
\caption[6_2]{The phase diagram. The partition function is defined
(convergent) to the right of the critical line $\m_c(\b)$. The infinite
volume limit is obtained by keeping $\b$ fixed and approaching
$\m(\b)$ from the right. For $\b \to \infty$ $\m(\b)$ approaches
$\m_0$ for pure gravity. $\b_c$ denotes the phase transition point.}
\label{6_2}
\end{figure}

Since the string susceptibility exponent $\g_s^{(0)} = -1/2 < 0$
in the case of pure
gravity both $G_0(\m_0)$ and $\chi_0(\m_0)$ are finite.
This implies that a $\b_c$ exists such that
the denominator in \rf{6.13} is different from
zero for all $\m \geq  \m_c(\b)$ provided $\b >\b_c$ .
$\b_c$ will be the phase transition point.
Fig.\,\ref{6_2} shows the phase diagram in the $(\m,\b)$-plane.

Let us first consider region I in fig.\,\ref{6_2}.
The only chance for a singular
behavior is that $\bar{\m} (\m_c(\b) = \m_0$ and in this case the
leading singularity has to come from $\chi_0(\bar{\m})$. We conclude:
\beq\label{6.14}
\bar{\m}(\m_c(\b)) = \m_0 ~~~{\rm and}~~~~
\g_s(\b) = \g_s^{(0)}~~~~{\rm for}~~~~\b > \b_c.
\eeq
This is the phase where the model is magnetized, where
the spin fluctuations are small and where the geometry of the
surfaces is not  affected by the spins.

The number $\b_c$ is characterized by being the largest $\b$  for which
$\partial \m/\partial \bar{\m}$ equals zero for some $\bar{\m}$,
i.e. (by \rf{6.13}) the largest $\b$ for which the equation
\beq\label{6.15}
\e^{2\b} = \frac{3}{2} \chi_0(\bar{\m}) - G_0 (\bar{\m}) ~~
(=\sum_{T \in \cT'_2} (3N_T/2-1)\,\e^{-\bar{\m} N_T})
\eeq
has a solution. If we define $\bar{\m}_c (\b)$ by
\beq\label{6.16}
\left. \frac{\partial \m}{\partial \bar{\m}} \right|_{\bar{\m}_c(\b)} =
0~~~~{\rm for}~~~\b \leq \b_c ,
\eeq
then $\bar{\m}_c(\b)$ obviously  solves eq. \rf{6.15} and we have
\beq\label{6.17}
\bar{\m}_c(\b_c) = \m_0,~~~~~~\bar{\m}_c(\b) > \m_0 ~~~{\rm for}~~~\b < \b_c.
\eeq

Let us now consider region II in fig.\,\ref{6_2} where $\b < \b_c$.
If we use eq. \rf{6.12} and \rf{6.13}
this implies that the lhs will be singular
for  $\bar{\m}\to \bar{\m}_c(\b)$.
In fact, since $\bar{\m}_c(\b) < \m_0$ both $\chi_0 (\bar{\m})$
and $G_0(\bar{\m})$ will be regular around
$\bar{\m}_c(\b)$ and we can Taylor expand
the lhs of eq. \rf{6.12}:
\beq\label{6.18}
\chi(\m,\b) \sim \frac{c}{\bar{\m} -\bar{\m}_c(\b)} \sim
\frac{ \tilde{c}}{\sqrt{\m -\m_c(\b)}}
\eeq
To derive the last equation we have used \rf{6.13} and
\rf{6.16} which tell us that
\beq\label{6.19}
\m= \m_c(\b) + {\rm const.} (\bar{\m}-\bar{\m}_c(\b))^2 +\cdots.
\eeq
We conclude that $\g_s(\b) = 1/2$ for $\b < \b_c$. In this phase
baby universes are  dominant. Effectively we have branched
polymers  and the total magnetization of the system is zero.

Let us finally consider the system at the critical point $\b_c$.
This point is characterized by the fact that $\bar{\m}_c(\b_c)$ coincides
with $\m_0$. Although the singularity of $\chi(\m,\b_c)$ for
$\m \to \m_c(\b_c)$ is still dominated by the zero of
$\partial \m/\partial \bar{\m}$ we can no longer Taylor expand $\m (\bar{\m},\b_c)$
around $\bar{\m}_c(\b_c)$ ($=\m_0$) since the functions in \rf{6.13} are singular
in $\m_0$. On the other hand we can use the known singular behavior of
the pure gravity functions $G_0$ and $\chi_0$
at $\m_0$ to deduce from \rf{6.13}, remembering that $\g_s^{(0)} < 0$,
\beq\label{6.20}
\frac{\partial \m}{\partial \bar{\m}} \sim (\bar{\m}-\bar{\m}_c(\b_c))^{-\g_s^{(0)}},
~~~~~{\rm i.e.}~~~\m-\m_c(\b_c) \sim (\bar{\m}-\bar{\m}_c(\b_c))^{-\g_s^{(0)}+1}.
\eeq
From \rf{6.12} we finally get, using \rf{6.20}
\beq\label{6.21}
\chi(\m,\b_c) \sim \frac{c}{(\bar{\m}-\bar{\m}_c(\b_c))^{-\g_s^{(0)}}} \sim
\frac{\tilde{c}}{(\m-\m_c(\b_c))^{-\g_s^{(0)}/(-\g_s^{(0)}+1)}},
\eeq
and we have derived the remarkable relation:
\beq\label{6.22}
\g_s(\b_c) = \frac{-\g_s^{(0)}}{-\g_s^{(0)}+1} =\frac{1}{3}.
\eeq

\vspace{12pt}

The model can easily be extended to include the coupling to a magnetic
field and one can explicitly verify that the system is magnetized for
$\b > \b_c$ and has zero total magnetization for $\b < \b_c$.
Further the transition is a third order transition (like in the
full Ising model on dynamical triangulations). The model captures
the essential features observed numerically for a large number of
Ising models on dynamical triangulations. It also strongly suggests
the existence of a region below $\b_c$ where the surfaces
consist of numerous totally magnetized baby universes with
different spin orientations such that the total magnetization
is zero. The spin transition observed is just the final
alignment of the spins of the baby universes.

\vspace{12pt}

\noindent The following should be noticed:\\
\begin{itemize}
\item[(1):] For a finite number of Ising models the mean field approximation
above clearly fails for $\b \to 0$. In the infinite temperature limit
$\g_s$ has to return to the value $-1/2$. In the mean field approximation
$\g_s$ stays equal $1/2$. For high temperatures there
will be important spin configurations which are different
from the ones considered in the mean field approximation.
If we for sufficiently high central charge $c$ 
{\it assume} the existence of an interval
below $\b_c$ where $\b_s =1/2$, it is an interesting and unsolved
question how the transition to $\g_s=-1/2$ takes place.
Is there a single transition, a continuous change or  a
cascade of transitions, and how are these transitions characterized?\\
\item[(2):] The transition at $\b_c$ had $0 < \g_s < 1/2$ and it was
related to $\g_s^{(0)}$ for pure gravity by eq. \rf{6.22}.
The formula reflects that the individual baby universes are totally
magnetized, i.e. have the fractal structure of pure gravity. This can
be generalized \cite{durhuus}:\\

\vspace{12pt}\noindent
{\bf Theorem}: given a multiple spin model with
$ 0<\g_s<1/2$ it follows that
\beq\label{6.23}
\g_s = \frac{-\g_n}{-\g_n+1}= \frac{1}{n+1},~~~~~\g_n = -\frac{1}{n}.
\eeq

\vspace{12pt}\noindent
The interpretation of eq. \rf{6.23} is that the individual baby universes
have a fractal structure corresponding to a
unitary conformal field theories {\it with $c <1$ coupled
gravity}. Such theories are characterized by a $c_n= 1-6/n(n+1)$ and
a $\g_s =-1/n$. Formula \rf{6.22} represented the simplest example,
$c=0$, but it can be shown, using renormalization group arguments,
that \rf{6.23} is the only possible critical behavior if
$\g_s >0$. It rules out the possibility that $\g_s$ can
change continuously while larger than zero. It also tells us the
possible solution to the $c >1$ disaster in the continuum approach:
{\it The interaction between matter and geometry becomes so strong that
the surfaces disintegrate dynamically into baby universes which
individually have $c <1$}. What is missing in our understanding
is how to predict a specific $c_n \leq 0$ from $c \geq 1$.
The only thing we know is that for $c \to \infty$ it follows that $c_n = 0$.
\end{itemize}

\subsection{Random surfaces with extrinsic curvature (II)}

Let us return to the random surface theory with extrinsic curvature.
Recall that numerical simulations indicated that $\g_s \approx 1/4$.
Can we understand the theory in the general setting outlined
in the beginning of this section?
To investigate this possibility
consider the model {\it on a fixed regular triangulation T}:
\beq\label{6.24}
Z_T(\l) = \int \prod_{i\in T/\{i_0\}}\d x_i \;\e^{-\sum_{(ij)}(x_i-x_j)^2
-\l \sum_{\triangle_i,\triangle_j} (n_{\triangle_i} -n_{\triangle_j})^2}.
\eeq
In \rf{6.24} $i$ denotes a vertex in $T$ which is mapped
into  $x_i \in R^3$ (but it is easy to formulate
the model in $R^D$) and  $n_{\triangle}$ is a normal to the oriented
triangle $(x_ix_jx_k)$ where $\triangle =(ijk)\in T$.

Eq. \rf{6.24} defines the  model of so-called 
{\it crystalline surfaces} \cite{crystalline}
(see the lectures of Peliti and Nelson for details). Since $T$ is
regular we can view it as a lattice theory, either in terms
of the variables $x_i$, in which case it is non-polynomial, or
in terms of the normals $n_{\triangle}$ after integrating out some
of the degrees of freedom of the $x_i$'s. If we go to the limit $\l=0$
it is a purely Gaussian model and it can be solved trivially.
Viewed as a lattice model in terms of the normals it is a weird 
theory \cite{adj2}.
If we use a continuum notation and denote the lattice coordinates by
$\xi$ the normal-normal correlator is:
\beq\label{6.25}
\bra \vec{n}(0) \vec{n}(\xi)\ket \sim
\frac{\del_{\L}^{(2)}(\xi)}{\xi^2} -\frac{1-\e^{-\L^2\xi^2}}{2\pi \xi^4}
\sim \frac{-1}{2\pi \xi^4}
\eeq
for $\xi \neq 0$. In this formula $\L$ is just a convenient
continuum proper-time cut-off instead of the lattice spacing\footnote{
To be more precise the continuum Gaussian action is used and, if $\L$ is an
ultraviolet cut-off and $L$ an infrared cut-off,
the propagator  is $$\la x^\m(0) x^\n(\xi)\ra = \del_{\m\n}
\int^{\L^2}_{L^{-2}} d\a e^{-\a\xi^2}/4\pi\a.$$}.
Except at coinciding points the normals are anti-correlated.
This truly remarkable situation can only appear because the surface
is totally crumpled in $R^3$. It is easy to calculate the
average radius of the surface in $R^3$:
\beq\label{6.26}
\bra r \ket_{N_T} \sim \log N_T,
\eeq
i.e. the Hausdorff dimension $d_H =\infty$.

\begin{figure}
\input{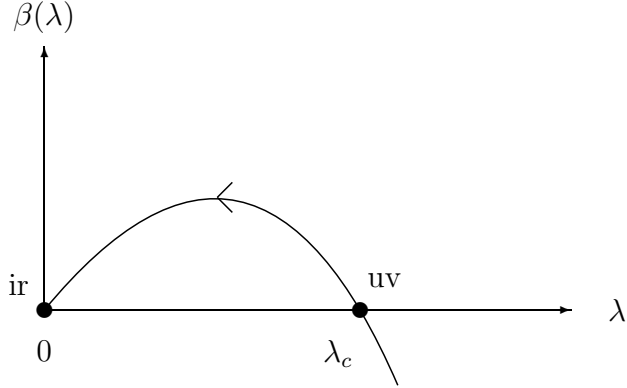}
\caption[6_3]{The $\b$-function for the crystalline surface model
\rf{6.24}.}
\label{6_3}
\end{figure}

With increasing  $\l$  the short distance correlation
between the normals becomes positive, but the long distance tail
remains negative, which implies that $d_H = \infty$.
However, for a finite value $\l=\l_c$ the system undergoes
a phase transition, the so-called {\it crumpling transition},
after which the normal-normal correlation becomes positive
everywhere and the Hausdorff dimension finite. {\it The transition
seems to be second order}. Large scale computer simulations
have verified the existence of this transition and
strongly suggest that it is second order \cite{hw,fw,espriu}.
Calculations of the $\b$-function $\b(\l)$ for continuum
toy models related to the model defined by eq. \rf{6.24} lead to \cite{dg}
\beq\label{6.27}
\b(\l) = c_1 \l -c_2 \l^2,~~~~~c_i > 0,
\eeq
i.e. an ultraviolet fixed point and applying the
renormalization group we would have the flow shown in fig.\,\ref{6_3}.
There should be a central charge
$c$ associated with this transition. Its value is not known,
but it would be most interesting if $c >1$. If $\l=0$ we have
3 Gaussian fields and $c=3$. Since $\l=0$ is an infrared fixed point
(see fig.\,\ref{6_3}) one would naively expect from the c-theorem that
$c_{crump} \geq 3$.

The summarize: Although weird, the model fulfills the requirement
for a spin model we can couple to quantum gravity. It is done
by taking the anneal average of \rf{6.24} over random
triangulations (of the sphere), i.e. we arrive
precisely at the random surface model studied earlier.
Numerical simulations showed that this model  had
a phase transition to smooth surfaces. {\it It is natural
to conjecture that it is the quantum gravity version of
the crumpling transition.} If $c_{crump} > 1$ we will be in
same situation as for the multiple Ising models. The
numerical measurement of $\g_s$ gives some support to this
idea. It would be most interesting to have a better analytic
understanding of the transition since it is relevant both
from the viewpoint of string theory, as described above,
and from the viewpoint of membrane physics: It is a model of a fluid membrane,
not entirely physical since it is self-intersecting, but still
of considerable interest.
   
\newsection{Euclidean quantum gravity in $d >2$}\label{highd}

\subsection{Basic questions in Euclidean quantum gravity}
Until now we have considered various aspects of two-dimensional
quantum gravity. As long as we restrict topology we deal
with a well defined theory. There might even be some hope that
one could define the summation over all two-dimensional
topologies although this question is not yet settled.
At the moment we address Euclidean quantum gravity in
dimensions higher than two a frightening number of basic questions
appear. Let us just list some of them:
\begin{itemize}
\item[(1):] How do we cure the unboundedness of the Einstein-Hilbert action
in $d >2$?
\item[(2):] Does the non-renormalizability of the gravitational
coupling constant not spoil any hope of making sense of the theory?
\item[(3):] What is the relation between Euclidean and Lorentzian
signature in the absence of any Osterwalder-Schrader axioms to
ensure that we can rotate from Euclidean space to Lorentzian space-time?
\item[(4):] What is the role of topology, keeping in mind that
four-dimensional topologies cannot be classified?
\end{itemize}

It is possible to take  these questions and our inability to
answer them in a fully satisfactory way
as an indication that there exists no theory of Euclidean
quantum gravity for $d=4$.
Let us briefly discuss the problems. While (1) and (2) in
principle are unrelated the solution of one of the problems often
solves the other problem too. The unboundedness of the
Euclidean Einstein action is due to the conformal mode.
An obvious way to cure this problem is to assume that the Einstein-Hilbert
action is only an infrared approximation to the complete theory.
A simple minded way to make the action bounded is to add higher derivative
terms like $R^2$ or $R_{\m\n}R^{\m\n}$. Such terms also tend to cure the
problem of renormalizability since the propagators will contain
higher inverse powers of the momentum. In fact adding suitable
combinations of $R^2$ terms one can get a theory which is renormalizable
and where the Euclidean action is bounded from below \cite{stelle}. It is not
known whether the theory qualifies as a unitary field theory if
we apply the standard rotation back to Lorentzian signature.
A more general discussion along the lines of viewing the Einstein-Hilbert
term as the first (infrared leading term) term of the full action
goes back to Weinberg who introduced to concept of asymptotic 
safety \cite{weinberg}.
The idea is that we should be able, by means of the renormalization
group, to work our way back from the long distance region where
the Einstein-Hilbert action is a good effective description to some
non-trivial ultraviolet fixed point. In addition the associated critical
surface is assumed to be of finite co-dimension, which means that only a
finite number of parameters need to be fine tuned to reach the critical
surface and from this point of view the theory will not differ
in spirit from ordinary renormalizable theories. The effective Lagrangian
description of the field theory by means of fields suitable for
the infrared fixed point might then be an infinite series
\beq\label{7.1}
\cL =\sqrt{g} \left[\Lambda -
\frac{1}{16\pi G} R + f_2 R^2 + f_2'R_{\m\n}R^{\m\n}+\cdots \right]
\eeq
which might even be non-polynomial, but which {\it might} make sense
both with Riemannian and Lorentzian signatures.

A weakness in such a scenario is that the existence of the ultraviolet
fixed point is hypothetical up to now and we have not exactly been flooded
with examples of non-trivial fixed points
in four-dimensional field theory. However,
the interesting results obtained by  use of the expansion in $2+\ep$
dimensions give some support to this idea \cite{kn}.

Finally one could hope that once the correct non-perturbative
formulation of the quantum theory has been found these problems
will resolve themselves and it will become clear why one should
not necessarily think in terms of perturbative expansions
around fixed background metrics.

Not much is known about the question of summing over topologies.
As long as we think in terms of continuum physics and write
down the path integral it offers the possibility of
summing over different manifold structures as well as integrating
over inequivalent Riemannian structures of  a given manifold:
\beq\label{7.2}
Z = \sum_{\cM \in {\rm Top}} \int_{\cM}
\frac{\cD g_{\m\n}}{{\rm Vol(diff)}} \e^{-S[g]}.
\eeq
In two and three dimensions we do not have to worry about the
meaning of {\it Top}, since there is equivalence between smooth
manifolds and topological manifolds. In two dimensions the manifolds
are uniquely characterized by their Euler characteristic $\chi$ and the
summation over {\it Top} is simply a summation over $\chi$ or the genus
$g =1-\chi/2$ of the surfaces. In spite of this simple prescription
surprisingly little progress has been made in defining the sum in
\rf{7.2} using continuum methods. Matrix models gave  non-perturbative
definitions, as described in sec.\,\ref{smallc}. However, so far the
results have been ambiguous. If we move to three dimensions we
encounter a slight classification problem in the sense that
no simple parametrization like $\chi$ of the various
topologies exists. However, the problem seems to get completely out of
control when we move to four dimensions. For four-dimensional
manifolds there is not equivalence between smooth and topological
structures. Topological manifolds exist which do not
admit smooth structures and some topological manifolds admit
infinitely many inequivalent smooth structures. If we insist on summing
over all smooth structures $\sum_{Top}$ will be rather unwieldy.
To complicate the operational meaning of $\sum_{Top}$ it should
be added that four-dimensional manifolds are not algorithmic
classifiable, i.e. no finite algorithm {\it in the sense
of Turing}  exists which allows us to decide if two arbitrary
four dimensional manifolds are equivalent. On the other hand
arguments (not known to me) might exist which dictate a restriction
of the allowed class of manifolds. Since there seems to be
fermions in the world one could argue that the manifold should
be a spin manifold. If one  makes the additional (rather arbitrary)
restriction that it should be simply connected it is possible
to show that such a (smooth) manifold is characterized by its Euler
number and its signature where these in four dimensions are given
by:
\bea
\chi(M) &=& \frac{1}{128\pi^2} \int_M \d^4\xi\sqrt{g}\; R_{abcd}R_{a'b'c'd'}
\ep^{aba'b'}\ep^{cdc'd'}   \label{7x.1}\\
\tau(M) &=& \frac{1}{96 \pi^2} \int_M \d^4\xi\sqrt{g}\; R_{abcd}
R^{ab}_{~~c'd'}\ep^{cdc'd'} \label{7x.2}
\eea
For simply-connected  spin manifolds the signature $\tau$ is a
multiple of 16, while $\chi$ is an integer $\geq 2$. Thus eq.
\rf{7x.2} seems a minor extension compared to two dimensions where
the summation is over $\chi$, but the restriction to
simply connected spin manifolds is not natural at this stage
of a quantum theory of space-time.

While all these problems seem to discourage any attempt to
make sense of the path integral of Euclidean quantum gravity,
it is still our obligation to try to investigate if it is
possible. Below I will argue that the use of dynamical triangulations,
which works so well in two dimensions,
allows us to discuss several of the above issues in some detail and
might be a candidate for non-perturbative definition of quantum
gravity even in higher dimensions.

\subsection{Definition of simplicial quantum gravity for $d >2$}

Let us define simplicial quantum gravity as a generalization
of the construction in sec.\,\ref{surface} and sec.\,\ref{smallc}
for two dimensions: In $d$ dimensions (where $d=3$ or $d=4$)
we construct all closed (abstract) simplicial manifolds
from $K$ $d$-dimensional simplexes. As in two dimensions we imagine that
the length of the links in the simplexes are $a$ (which we
take as 1 unless explicitly stated). For such a combinatorial
or, equivalently, piecewise linear manifold we can apply Regge calculus
and in this way assign a Riemannian metric to the manifold.
By such an assignment we see that the discretization is able, for
finite $K$, to assign a meaning to  the sum and 
integral in \rf{7.2}\cite{weingarten1,adj3}
\beq\label{7.3}
\sum_{\cM \in {\rm Top}} \int_{\cM}
\frac{\cD g_{\m\n}}{{\rm Vol(diff)}} \to \sum_{T}.
\eeq
{\it The discretization has the same virtue as in two dimensions: 
In principle it allows
a unified treatment of the summation over topologies
and Riemannian structures.}

\begin{figure}
\input{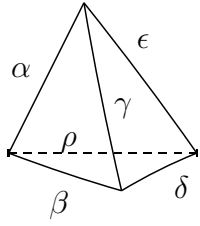}
\caption[7_2]{Labels of a tetrahedron}
\label{7_2}
\end{figure}

A few remarks should be said about the formula \rf{7.3}. First
one could try as in two dimensions to make the gluing automatic,
and in this way arrive at a generalized matrix 
model \cite{adj3,sakura,gross}. In $d$ dimensions
the task is to glue together ($d$-$1$)-dimensional sub-simplexes.
Let us discuss the situation in $d=3$ (the generalization to
higher dimensions is trivial). In fig.\,\ref{7_2} we have
shown a building block for three-dimensional simplicial
complexes: a tetrahedron (the generalization of the equilateral
triangle used up to now) where the edges has labels $\a,\b,\g,\del,
\ep,\rho$, which takes values $1,...,N$, where $N$ is an
abstract label which we in the end want to take to infinity like
in the two-dimensional case.
To each oriented face of the tetrahedron, $\a,\b,\g$, say,
we assign a complex variable $\phi_{\a\b\g}$ whose value
is invariant under even permutations of the indices but conjugated
under odd permutations (which correspond to a reversal of
orientation).  To the tetrahedron we associate the generalization
of $\Tr \phi^3$ for the triangles:
\beq\label{7x.10}
A(\phi) = \sum_{\a\b\g\del\ep\rho} \phi_{\a\b\g}\phi_{\g\del\ep}
\phi_{\ep\rho\a}\phi_{\b\rho\del}
\eeq
Consider the following integral:
\beq\label{7x.11}
Z(g) = \int \d\phi \; \exp\left(-\frac{1}{6} \sum_{\a\b\g}
|\phi_{\a\b\g}|^2 +g A(\phi) \right).
\eeq
If the exponent is expanded in a power series in $g$, and
all possible Wick contractions are performed on the powers of $A(\phi)$
it corresponds to all possible gluings of the tetrahedra along 
triangles. The expansion leads  to
\beq\label{7.x12}
\log Z(g) \sim \sum_{T} \frac{1}{S_T} N^{N_1(T)} g^{N_3(T)},
\eeq
where the sum is over all possible gluings $T$. $N_1(T)$ denotes the
number of links in $T$ and $N_3(T)$ the number of tetrahedra.
Let us postpone the discussion of the terms $N^{N_1(T)} g^{N_3(T)}$
and concentrate on $\sum_{T}$.
This kind of  ``blind'' gluing
of $d$-dimensional simplexes along their $d-1$ dimensional
sub-simplexes will not in general create combinatorial manifolds,
but only so-called {\it combinatorial pseudo-manifolds}: i.e simplicial
complexes where  $d-1$-dimensional sub-simplexes which
are not boundary simplexes are contained in precisely
two $d$-simplexes and where it is possible to connect any two $d$-simplexes
by a sequence of $d$-simplexes, each intersecting along some
$d-1$ simplex. For such pseudo-manifolds
the neighborhood of a vertex will not necessarily  be equivalent with
a $d$-dimensional ball. Whether or not this is important
for quantum gravity is not clear. 
In the two-dimensional case it did not course any problems. 
One could create one-loops and two-loops which cannot be 
present on a combinatorial manifold, but the pseudo-manifold
still had a transparent surface interpretation and the 
Euler number was perfectly well defined. 
In the three dimensional
case the modification is more severe in the sense that
the Euler characteristic can be
different from zero, while $\chi =0$ for any odd dimensional
{\it manifold}.  In the following we will (to be conservative)
restrict ourself to so-called combinatorial manifolds,
i.e. combinatorial pseudo manifolds where the neighborhood
of any sub-simplex is homeomorphic to a $d$-simplex. For closed
manifolds this enforces certain relations between the number
of $i$-simplexes $N_i$ in the triangulation, known as the Dehn-Sommerville
relations:
\beq\label{7x.13}
N_i = \sum_{k=i}^d (-1)^{k+d} \pmatrix{k+1\cr i+1} N_k.
\eeq
If we define $N_{-1}=\chi$, the Euler characteristic of the manifold,
\rf{7x.13} is valid for $i=-1,...,d-1$.

Secondly, in light of the complicated relation between
topology and diffeomorphism  for four-dimensional manifolds
one could be worried that similar problems arise when
we compare combinatorial, i.e. piecewise linear, manifolds and
smooth manifolds. However, for dimensions $d < 7$ we have
equivalence between piecewise linear and smooth structures.
Whatever subtleties might be involved in defining the sum on the
lhs of eq. \rf{7.3} it should be captured at the rhs of eq. \rf{7.3}.
Of course eq. \rf{7.3} itself is rather formal as it stands.
Even in the two-dimensional case the precise meaning of
the lhs was not clear from a mathematical point of view and in
sec.\,\ref{smallc} we tried to use the rhs to {\it define} what is
meant by the summation over all two-dimensional manifolds.
The problem was non-trivial in the sense that the
number of triangulations  grows as fast as $K !$, $K$ being the number
of triangles. No reasonable discretized action 
could kill this entropy factor \cite{weingarten1,adj3} and
\beq\label{7.4}
Z = \sum_T \e^{-S_T}
\eeq
will be divergent. This problem will of course be present also
when we move to higher dimensions. The naive expression \rf{7.4}
makes no sense. In two dimensions the way out was to fix  topology.
This bounds the number of triangulations  exponentially and
eq. \rf{7.4} will be well defined. Only afterwards the topology is allowed
to fluctuate  and the {\it double scaling limit taken}. In higher
dimensions it is sensible, as a minimum, to try to define eq. \rf{7.4}
for a fixed topology. We need the following conjecture :

\vspace{12pt}
\noindent{\bf Conjecture:}
The number of combinatorial equivalent $d$-dimensional manifolds
is an exponentially bounded function of the number of $d$ dimensional
simplexes.

\vspace{12pt} \noindent
We call two simplicial complexes combinatorial equivalent if they
have a common subdivision and when we talk about equivalence classes
of piecewise linear manifolds we have in mind combinatorial equivalence.
While the conjecture is true in two dimensions, there is no general
proof in higher dimensions\footnote{Very
recently there has appeared a proof \cite{xxxx}.}.
 For $d\geq 4$ manifolds exist
which are not algorithmic recognizable, the reason
being that any finitely presented group can appear
as the fundamental group for some four-dimensional manifold.
However, finitely presented groups cannot be algorithmically 
classified due to the famous ``word problem''. As a curiosity we can mention
that for such manifolds their number as a function of $N_4$, the 
number of 4-simplexes,  is not algorithmic
calculable. This does not mean that the number cannot be
exponentially bounded, only that there is no finite algorithm
which allows us to calculate the exact number. In fact numerical simulations
seem to work well for the simplest topology ($S^4$) and the give support to the 
conjecture for $S^4$ \cite{av1,ckr1,bm1}\footnote{Recently 
some controversy arose \cite{ckr2}, 
but my opinion the question is now settled.}.

After these general remarks it is natural as a first explorative step
to fix the topology of our four-dimensional manifold to be
the simplest possible, that of $S^4$. The Einstein-Hilbert action
for a simplicial manifold can be calculated by Regge calculus. However,
we do not need the full machinery in our case where the simplexes
are identical and all link length equal. The Regge version
of the Einstein action is the sum over deficit angles of the
$d-2$ dimensional sub-simplexes times their $d-2$ dimensional
volume. In our case the deficit angle associated with a $d-2$ dimensional
sub-simplex $n_{d-2}$ is $2\pi -{\rm const.} o(n_{d-2})$, where
$o(n_{d-2})$ is the {\it order} of $n_{d-2}$, i.e. the number of $d$
dimensional simplexes of which $n_{d-2}$ is a sub-simplex. It follows
that
\beq\label{7.5}
\left\{\int \d^d\xi \sqrt{g}R\right\}_{Regge} \propto \sum_{n_{d-2}}
(2\pi - {\rm const.} o(n_{d-2})).
\eeq
If we note that the number of $d-2$-dimensional  sub-simplexes in
a $d$-dimensional simplex is $d(d+1)/2$ we can write:
\beqn
S[g;\l,G]=\int \d^d\xi \sqrt{g}\left(\l
- \frac{1}{16\pi G} R\right)  \to 
\eeqn
\beq
S_T [k_d,k_{d-2}] = k_d N_d (T) -k_{d-2} N_{d-2} (T), \label{7.6}
\eeq
where the piecewise linear manifold is given defined by the triangulation
$T$ and $N_d(T)$ and $N_{d-2}(T)$ denote the number of $d$- and
$d-2$-dimensional simplexes in the triangulation $T$.
We can view $1/k_{d-2}$ as a bare gravitational coupling constant.
At first sight the action might  seem much too
simple to have anything to do with gravity. Our point of view will be
the opposite: {\it The fact that the action is so simple reflect
the beauty and simplicity of quantum gravity, and
hopefully this simplicity will be reflected in the
solution of the theory}.

Our final prescription will be:
\beqn
Z[\l,G]=\int_{S^d} \frac{\cD g_{\m\n}}{{\rm Vol(diff)}} \e^{-S[g;\l,G]} \to
\eeqn
\beq
Z[k_{d-2},k_d] = \sum_{T \in S^d} \e^{-S_T[k_{d-2},k_d]}.\label{7.7}
\eeq
We will be interested in $d=4$ and sometimes in $d=3$, in which cases
we get
\bea
Z[k_1,k_3]& = &\sum_{T \in S^3} \e^{-k_3N_3(T)+k_1 N_1}. \label{7z.10}\\
Z[k_2,k_4]& =& \sum_{T \in S^4} \e^{-k_4N_4(T)+k_2 N_2}.   \label{7z.11}
\eea
For $d=3,4$ there are only two independent coupling
constants as long as we only want an action which depends on
global quantities like $N_i$. It follows from the Dehn-Sommerville relations.
If we include higher derivative terms in the action
we will certainly loose the simplicity of  eqs. \rf{7z.10} and \rf{7z.11}.
The higher derivative terms will contain explicit the order of the
sub-simplexes which carry the curvature. We will not discuss the lattice
implementation of these.

\begin{figure}
\input{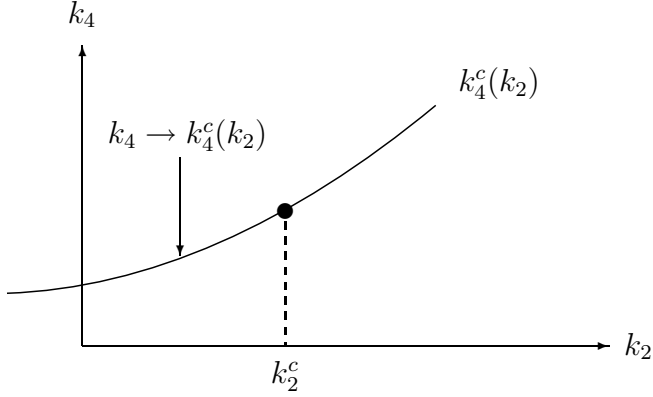}
\caption[7_3]{A hypothetical  phase diagram for four-dimensional
gravity.}
\label{7_3}
\end{figure}

Let us now discuss the phase diagram. Assume $d=4$.
Since it is easy to prove that $N_2 (T) \leq {\rm const.} N_4 (T)$,
the conjecture above implies that for each $k_2$,
a $k_4^c(k_2)$ exists such that the lhs of \rf{7.7} for a given $k_2$
is well-defined for $k_4 >k_4^c(k_2)$ and divergent for $k_4 <k_4^c(k_2)$.
A potential continuum limit should be taken as $k_4 \to k_4^c$ from
above. The corresponding phase diagram is shown in fig.\,\ref{7_3}.
If we for a fixed $k_2$ approach $k_4^c$ we will be probing the infinite
volume limit of the discretized system. It does not imply that there
necessarily will be a continuum limit. Rather we should view the
system as a lattice system where the infinite volume limit is
taken. For some specific values of the bare couplings
critical points might exist where a correlation length diverges
and where a continuum limit exists. Such a point is tentatively
indicated at the figure. Approaching this point in a
specific way will then define the renormalized cosmological
constant and the renormalized gravitational constant.
Since we are in unchartered territory one should be open-minded
for other possibilities, e.g. the possibility that a whole
range of $k_2$'s might correspond to a topological gravity
where the metric, and correspondingly concepts like divergent
distances, play no role.

Let us for a moment return to the three-dimensional partition function
given by eq. \rf{7z.11}.
Comparing with eq. \rf{7.x12} derived from the tensor model
one has the identification:
\beq\label{7.7b}
k_3 = -\log g,~~~~~k_1 = \log N.
\eeq
This relation is the same as we encountered in two dimensions and
the relation will be valid for the matrix model in any dimension:
\beq\label{7.7c}
k_d = -\log g,~~~~~~k_{d-2} = \log N.
\eeq
From a formal point of view the $N \to \infty$ limit corresponds
to taking the ``bare'' gravitational coupling constant $G \to 0$.
In two dimensions this limit played a very important role:
since the Einstein-Hilbert action in two dimensions  is  topological, an
expansion in $1/G$ automatically becomes an expansion in topology.
However, in three and higher dimensions we have not reason to
expect the large $N$ limit to classify topologies for us,
and it does not!. It has not yet been possible to construct anything
like the double scaling limit in higher dimensions than two.

\subsection{Observables}

It is possible to define the same critical exponents for
higher dimensional simplicial quantum gravity as we
have already defined for the various two-dimensional
theories we have considered.

First we can define an entropy or susceptibility
exponent $\g_s$. In the following it will always be assumed
that the topology is spherical, i.e. the combinatorial manifolds
are combinatorial equivalent to the boundary of a 5-simplex.
According to the conjecture the number $\cN (N_4)$ of triangulations
which can be constructed from $N_4$ 4-simplexes is exponentially
bounded. Let us now fix $k_2$. According to the remarks
above there will be a critical point $k_4^c(k_2)$. This implies that
\beq\label{7x.20a}
\cN(k_2,N_4) \equiv \sum_{T \in S^4(N_4)} \e^{k_2 N_2}
\sim \e^{k_4^c(k_2) N_4}f_{k_2}(N_4)
\eeq
where $f_{k_2}(N_4)$  stands for subleading corrections.
{\it If} the subleading correction is power-like we define $\g_s(k_2)$ by
\beq\label{7x.20}
\cN(N_4) \sim N_4^{\g_s(k_2) -3} \e^{\k_4^c N_4}(1+O(1/N)).
\eeq
However, it is not clear that we have such a behavior.
It is possible to imagine an asymptotic behavior like:
\beq\label{7x.21}
\cN(N_4) \sim \e^{\k_4^c N_4-c(k_2)N_4^\a(k_2)-
\cdots}(1+O(1/N))
\eeq
where $0 < \a < 1$. In this case the exponential  correction
given by $\a$ will always dominate over the power-like
correction determined by $\g_s(k_2)$. There are strong indications
that there are several regions with different asymptotic
behavior, depending on the value of $k_2$. This will be discussed below.

Apart from the entropy- or susceptibility exponent $\g_{s}$ we can
introduce the critical exponents $\n$ and $\eta$ already discussed
numerous times in the context of strings and random walks
where they were determined from the
properties of the two-point function in target space.
However, here these quantities
will refer to {\it intrinsic} geometry. Let us define
the two-point function as follows: Consider the ensemble of combinatorial
manifolds with two marked vertices $i_1$ and $i_2$.
For each manifold, build of regular 4-simplexes, we have by Regge's
prescription a metric. This means that we can define the geodesic distance
between the vertices $i_1$ and $i_2$. As a rough definition one can
consider paths along the links in the triangulation and define the geodesic
distance as the shortest link-path\footnote{Alternatively
one could talk about the distance between the 4-simplexes, defined as
by moving between successive 4-simplexes via their common 3-simplex boundary.
Again this is not a true geodesic distance, but should not differ drastically
from the genuine geodesic distance. For a given combinatorial manifold
there might be a significant deviation between link distances and
4-simplex distances, but we expect concepts like Hausdorff dimensions
etc. to be the same.} between $i_1$ and $i_2$. Let us consider the
sub-ensemble of manifolds where the geodesic distance between $i_1$ and $i_2$
is fixed to be $r$. If we use the link-paths in the definition of geodesic
length $r$ will be an integer. Let us denote this ensemble of manifolds
by $\cT(2,r)$. We define the two-point function by:
\beq\label{7y.17}
G_{k_4}^{(k_2)}(r) = \sum_{T\in \cT(2,r)} e^{-S_T[k_4,k_2]}
\eeq
{\it Provided that we have the asymptotic behavior \rf{7x.20} }
we expect a generic behavior of this spherical two-point function
(we suppress the explicit dependence on $k_2$):
\beq
G_{k_4} (r) \sim r^{-\a} e^{-{m}(k_4) r},~~~~~~~~~~~~~~~~
r\gg 1/{m}(k_4);\label{7y.18}
\eeq
\beq
G_{k_4} (r) \sim  r^{1-\eta},~~~~~~~~~~~~~~~~~~~~~~~~
1 \ll r \ll 1/{m}(k_4); \label{7y.19}
\eeq
\beq
\chi(k_4) \equiv \sum_r G_{k_4}(r) \sim \frac{1}{(k_4-k_4^c)^{\g_{s}}}.
\label{7y.20}
\eeq
The short distance behavior of $G_{k_4}(r)$ arises from the generic
behavior since there is an angular average involved in the 
definition.

{\it If ${m}(k_4)$  scales to zero at the critical point $k_4^c$}:
\beq\label{7y.21}
{m}(k_4) \sim (k_4-k_4^c)^{\n}
\eeq
we can define in a simple way the continuum limit: Introduce the
continuum parameters $R = r\cdot a$ with the dimension of length and
${m}_{{\rm ph}}={m}(k_4)/a$ and
keep them fixed as $k_4\to k_4^c$. This fixes $a(k_4) \sim (k_4-k_4^c)^\n$.
In addition one can readily prove Fischer's scaling relation:
$\g_s= \n(2-\eta)$.

As usual one can prove that
the exponent $\n$ is related to the Hausdorff dimension of
the ensemble of manifolds. However, in this case we have in mind
an {\it intrinsic Hausdorff dimension} $d_h$. For a manifold
with $N_4$ 4-simplexes the volume $\sim N_4$ as long as we take $a=1$.
The average volume is therefore $ \sim~\bra N_4 \ket_r$, where
$\bra \cdot \ket_r$ refers to the ensemble $\cT(2,r)$.
We can now define the internal Hausdorff dimension $d_h$ by
\beq\label{7y.22}
\bra N \ket_r \sim r^{d_h}~~~~~{\rm for}~~r\to\infty,~~
{m} (k_4) r={\rm const.}
\eeq
From \rf{7y.17} we have
\beq\label{7y.23}
\bra N \ket_r = -\frac{\d \log G_{k_4} (r)}{\d k_4} =
\frac{\d {m}(k_4)}{\d k_4}\; r,
\eeq
and using \rf{7y.21} we get
\beq\label{7y.24}
\bra N \ket_r \sim r^{1/\n},~~~{\rm i.e.}~~d_h=1/\n.
\eeq

As is seen from the  above consideration one can apply without
any problems general scaling considerations known from
critical phenomena provided that we have the asymptotic behavior \rf{7x.20}.
It can be formulated even stronger: It is hard to imagine
that any continuum limit of a conventional kind can be taken
unless that mass $m(k_4) \to 0$ for $k_4 \to k_4^c(k_2)$.

A good test case could be to return to $2d$ gravity,
but surprisingly it turns out that $m(\m)$  ($\m \equiv k_4$ in $d=2$)
and the associated critical exponents have only recently been
calculated even if they are the most fundamental scaling variables
in two dimensional quantum gravity. Unfortunately there is no space
for a detailed discussion of the calculation,
which uses the so-called transform matrix formulation \cite{wak}.
Details can be found in \cite{aw} and the surprising result is
that the two-point function $G_\m(r)$ can be calculated even at the
discretized level:
\beq\label{tax15}
G_\m(r) = {\rm const.} \; (\triangle \m)^{3/4}
\frac{\cosh \left[(\triangle \m)^{\oq} \b r\right]}{\sinh^3
\left[(\triangle \m)^{\oq} \b r\right]},
\eeq
where {\it const.} and $\b$ are positive constants
of $O(1)$ and $\triangle \m = \m-\m_c$.

We conclude the following:
\begin{enumerate}
\item $G_\m(r)$  falls of like
$\e^{-2(\triangle \m)^{\oq} \b r}$ for $r \to \infty$,
i.e. the critical exponent $\n = \oq$ and the Hausdorff dimension
$d_H =4$.
\item $G_\m(r)$ behaves like $r^{-3}$ for
$1 \ll r \ll \triangle \m^{-\oq}$, i.e.
the scaling exponent $\eta = 4$.
\item The pre-exponential factor to $\e^{-2(\triangle \m)^{\oq} \b r}$
is $m( \m)^{\eta-1}$, and this factor is precisely the one
needed if we should be able to define a continuum two-point
function up to an allover scaling factor by the assignment
\rf{7y.22}, which states that $r\, m(\m)$ is constant
when $r \to \infty$ and $\m \to \m_c$, since the short distance
behavior is given by $r^{1-\eta}$.
\item The exponent $\a$ giving the power correction to the
exponential decay is zero. This can be understood intuitively from
from the fact universes for $r \gg 1/m(\m)$ have to be thin tubes,
and it is not difficult to show that these have no correction
to the exponential decay.
\item From Fisher's scaling relation we get $\g_s = \n (2-\eta) = -1/2$.
This well known result can of course also be derived directly from
\beq\label{tax17}
\chi(\m) =  \sum_{r=1}^\infty G_\m(r) =
{\rm const.} -c^2 (\triangle \m)^{\oh} + \cdots
\eeq
by use of \rf{tax15}, but it should be clear that the explicit calculation
in \rf{tax17} is nothing but a specific example of the general calculation
used in proving Fisher's scaling relation. What is somewhat
unusual compared to ordinary statistical systems is that the
anomalous scaling dimension $\eta >2$. $\eta =0$ is the ordinary
free field result, while $\eta =2$ is the infinite temperature limit,
and for statistical systems we expect $\eta < 2$.
\item The continuum two-point function can now be defined as
\beq\label{tax19}
G(R;M) = \lim_{a\to 0} (\sqrt{a})^{\eta-1} G_\m (r) \sim
M^{3}\frac{\cosh [ M\, R]}{\sinh^3 [ M\, R]}.
\eeq
where $R$ is a continuum distance and $M$ a continuum mass
proportional to $\L^{1/4}$, the unusual power of the ``mass'' being
due to $d_h=4$. The factor in front of $G_\m(r)$ is the usual
``wave function renormalization'' present in the path integral
representation of the propagator.
\end{enumerate}

It is interesting to give a direct physical interpretation
of the short distance behavior of $G_{k_4} (r)$ as defined
by \rf{7y.17}\footnote{The remarks to follows are of course equally valid for
$G_\m(r)$ in two-dimensional gravity.}
(in the following we suppress the index $k_2$).
In order to do so let us change from the  grand canonical
ensemble given by \rf{7y.17} to the  canonical ensemble defined by
$\cT(2,r,N_4)$, the class of triangulations with $N_4$ 4-simplexes where
two of the 4-simplexes are marked and separated by a distance $r$.
On this ensemble we can define the discretized analogue of $G_{k_4}(r)$:
\beq\label{hx1}
G(r,N_4) = \sum_{T \in \cT(2,r,N_4)}  1,
\eeq
For  $r=0$ we have the following
$N_4$ dependence (for the $r$ dependence see \rf{hx6} below)
\beq\label{hx2}
G(0,N_4) \sim N_4^{\g_s-2}\, \e^{\m_c N_4},
\eeq
assuming that $\g_s$ exists. This remark is important since it is by
no means obvious that that $\g_s$ exists in four-dimensional gravity,
as discussed above.
The partition function$Z(N_4) \sim N_4^{\g_s-3}\,\e^{\k_4^{c} N_4}$
and the one-point function for large $N_4$ is proportional to
$N_4 Z(N_4)$ since it counts the triangulations with one marked simplex
For $r=0$ (or just small) there is essentially no difference
between the one-point function and $G(0,N_4)$.

$G(r,N_4)$ is related to $G_{k_4}(r)$ by a (discrete) Laplace transform:
\beq\label{hx3}
G_{k_4}(r) = \sum_{N_4} G(r,N_4)\, \e^{-k_4^{c} N_4}.
\eeq
The {\it long distance behavior} of $G(r,N_4)$ is determined by the
long distance behavior of $G_{k_4}(r)$. Close to the scaling
limit it follows by direct calculation (e.g. a saddle point
calculation\footnote{In addition to the exponentially decaying
part of $G(r,N_4)$ there is also a power correction coming from the
quadratic integration in the saddle point approximation. We shall
not consider the explicit form of the power correction here.}) that
\bea
G_{k_4}(r) &\sim &\e^{-r\,(k_4-k_4^c)^{1/d_h} }~~~ \Rightarrow   \nonumber\\
G(r,N_4) &\sim & \e^{-c \left(r^{d_h}/N_4\right)^{\frac{1}{d_h-1}}}\;
\e^{k_4^c N_4}
~~~~{\rm for}~~~~ r^{d_h} > N_4,    \label{hx4}
\eea
where $c=(d_h-1)/d_h^{d_h/(d_h-1)}$.

On the other hand the {\it short distance behavior} of $G_{k_4} (r)$
is determined by the short distance behavior of $G(r,N_4)$ which is
simple. Eqs. \rf{7y.22}-\rf{7y.24} define the concept of
Hausdorff dimension in the grand canonical ensemble. A
definition in the canonical ensemble would
be: Take $N_4^{1/d_h} \gg r$ and simply count the  volume
(here number of 4-simplexes) of a ``spherical shell'' of thickness 1 and
radius $r$  from a marked simplex,
sum over all triangulations $T_{N_4}$ with one marked 4-simplex and
divide by the total number of such triangulations.
Call this number $\la n(r)\ra_{N_4}$. The Hausdorff dimension is then defined by
\beq\label{hx5}
\la n(r) \ra_{N_4} \sim r^{d_h-1}~~~{\rm for}~~~1 \ll r \ll N_4^{1/d_h}.
\eeq
It follows from the definitions that we can write
\bea
\la n(r) \ra_{N_4} &\sim & \frac{G(r,N_4)}{G(0,N_4)}, ~~~~{\rm i.e}~~~
\nonumber\\
G(r,N_4)& \sim& r^{d_h-1} N_4^{\g-2} \e^{k_4^c N_4}~~~{\rm for}~~~
1 \ll r \ll N_4^{1/d_h}.                            \label{hx6}
\eea
We can finally calculate the short distance behavior of $G_\m(r)$
from eq. \rf{hx3}. For $k_4 \to k_4^c$,  we get:
\beq\label{hx7}
G_{k_4}(r) \sim r^{d_h-1} \sum_{N_4} N_4^{\g-2}
\,\e^{-c \left(r^{d_h}/N_4\right)^{\frac{1}{d_h-1}}}
 \sim r^{\g_s d_h -1}
\eeq
This shows that
\beq\label{hx8}
\eta = 2-\g_s d_h,~~~~{\rm i.e.}~~~ \g_s = \n(2-\eta),
\eeq
which is Fishers scaling relation.
The above arguments highlight that the
anomalous scaling dimension $\eta$ is a function of the two
kinds of fractal structures we can define on  the ensemble
of piecewise linear manifolds: the Hausdorff dimension and
the baby universe proliferation probability.  In addition the arguments
show  that the canonical and grand canonical definitions of
Hausdorff dimension agree.

\subsection{Branched polymers}

The model of branched polymers ($BP$)
provides us with a simple, but non-trivial
example of the above scenario \cite{adj2} and will play an
important role in the following. In a certain way it can be viewed
as the lowest dimensional fractal structure and it will
appear as the limiting case of higher dimensional gravity theories.

Let us define branched polymers as the sum over all
tree graphs (no loops in the graphs) with certain weights
given to the graphs according to the following definition of the
partition function:
\beq\label{k2}
Z(\m) = \sum_{BP} \frac{1}{C_{BP}}\,\rho(BP) \; \e^{-\m |BP|},
\eeq
where $|BP|$ is the number of links in a $BP$ and $\m$ is a chemical
potential for the number of links, while
\beq\label{k3}
\rho(BP)= \prod_{i \in BP} f(n_i),
\eeq
where $i$ denotes a vertex, $n_i$ the number of links joining at
vertex $i$ and $f(n_i)$ is non-negative. $f(n_i)$ can be viewed
as the unnormalized branching weight for one link branching into $n_i-1$
links at vertex $i$. Finally $C_{BP}$ is a symmetry factor such
that rooted branched polymers, i.e. polymers with the first link marked,
is counted only once.

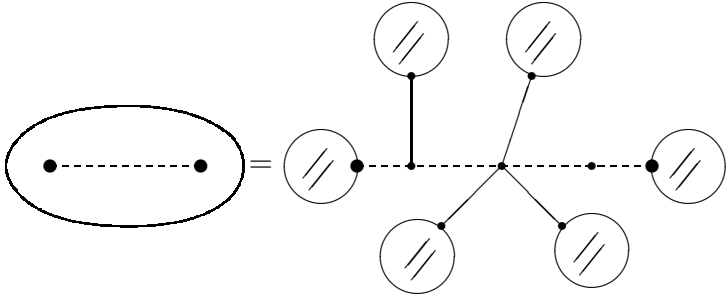
\begin{figure}
\unitlength=0.80mm
\linethickness{0.6pt}
\begin{picture}(127.00,52.00)
\put(15.00,25.00){\circle*{2.00}}
\put(40.00,25.00){\circle*{2.00}}
\put(15.00,25.00){\dashbox{1.00}(25.00,0.00)[cc]{}}
\put(60.00,25.00){\circle{12.00}}
\put(57.00,23.00){\line(4,5){4.00}}
\put(58.00,21.00){\line(4,5){4.00}}
\put(66.00,25.00){\circle*{2.00}}
\put(66.00,25.00){\dashbox{1.00}(49.00,0.00)[cc]{}}
\put(115.00,25.00){\circle*{2.00}}
\put(75.00,25.00){\line(0,1){15.00}}
\put(75.00,40.00){\circle*{1.50}}
\put(75.00,25.00){\circle*{1.50}}
\put(90.00,25.00){\circle*{1.50}}
\put(90.00,25.00){\line(-1,-1){10.00}}
\put(80.00,15.00){\circle*{1.50}}
\put(90.00,25.00){\line(1,-1){10.00}}
\put(100.00,15.00){\circle*{1.50}}
\put(90.00,25.00){\line(1,3){5.00}}
\put(95.00,40.00){\circle*{1.50}}
\put(105.00,25.00){\circle*{1.50}}
\put(75.00,46.00){\circle{12.00}}
\put(72.00,44.00){\line(4,5){4.00}}
\put(73.00,42.00){\line(4,5){4.00}}
\put(97.00,46.00){\circle{12.00}}
\put(94.00,44.00){\line(4,5){4.00}}
\put(95.00,42.00){\line(4,5){4.00}}
\put(76.00,10.00){\circle{12.00}}
\put(74.00,8.00){\line(4,5){4.00}}
\put(75.00,6.00){\line(4,5){4.00}}
\put(105.00,11.00){\circle{12.00}}
\put(102.00,9.00){\line(4,5){4.00}}
\put(103.00,7.00){\line(4,5){4.00}}
\put(121.00,25.00){\circle{12.00}}
\put(118.00,23.00){\line(4,5){4.00}}
\put(119.00,21.00){\line(4,5){4.00}}
\put(50.00,25.00){\makebox(0,0)[cc]{{\large $=$}}}
\bezier{56}(10.00,30.00)(5.00,25.00)(10.00,20.00)
\bezier{52}(45.00,30.00)(49.00,25.00)(45.00,20.00)
\bezier{76}(28.00,35.00)(39.00,35.00)(45.00,30.00)
\bezier{80}(10.00,30.00)(15.00,35.00)(28.00,35.00)
\bezier{80}(10.00,20.00)(15.00,15.00)(28.00,15.00)
\bezier{76}(28.00,15.00)(40.00,15.00)(45.00,20.00)
\end{picture}
\caption[figbranch]{The graphical representation of the two-point
function for branched polymers. The dashed line represents the
unique shortest path between the two marked vertices. The ``blobs''
represent the contribution from all rooted polymers branching
out from a vertex.}
\label{figbranch}
\end{figure}

This model can be solved \cite{adf}.
It has a critical point $\m_c$ (depending on $f$)
and close to the critical point we have:
\beq\label{k4}
Z''(\m) \sim (\m-\m_c)^{-1/2},
\eeq
i.e. $\g_s =1/2$ for branched polymers.
On the branched polymers we define the ``geodesic distance'' between
two vertices as the shortest link path, which is unique since we
consider tree-graphs. The graphic representation of the
two-point function is shown in fig.\,\ref{figbranch}.
Had it not been for the ability to branch, the two-point function
would simply be
\beq\label{k5}
G_\m(r) = \e^{-\m r}.
\eeq
However, the insertion of one-point functions at any vertex leads
to a non-analytic coupling constant renormalization and the
result is changed to \cite{adj2}
\beq\label{k6}
G_\m (r) = {\rm const.}\; \e^{-\kappa\, r\sqrt{\m-\m_c}}~~~
{\rm for}~~~\m-\m_c \to 0,
\eeq
where $\kappa$ is a positive constant depending on $f$.
We can now find $G(r,N)$ by an inverse Laplace transform:
\beq\label{k7}
G(r,N) = {\rm const.} \, N^{-3/2} r \,\e^{-\kappa^2r^2/4N}.
\eeq
We confirm from this explicit expression
that the (internal) Hausdorff dimension of
branched polymers is 2 (like a smooth surface !)
and that $\g_s = 1/2$ since the
prefactor of $G(r,N)$ for small $r$ should be $N^{\g_s-2} r^{d_H-1}$.

\subsection{Numerical simulations}

Presently it has not been possible to make much progress in analyzing
\rf{7.7} by analytical methods except for $d=2$.
However, the action is well suited
for the use of Monte Carlo simulations and the results
to be discussed later for $d=3$ and $d=4$ have been obtained
by such simulations. Let me shortly discuss the principles
involved in such simulations since there are a number of
interesting aspects involved compared to ordinary lattice 
simulations.

Monte Carlo simulations usually operate by means of a
stochastic process, usually taken to be a Markov chain.
The chain is selected such that it has a stationary
distribution equal to the desired probability distribution,
in this case the probability distribution given by \rf{7.7}.
The Markov chain is a prescription for moving around in
the configuration space, each step being independent of the
former steps and chosen with a certain probability among a
set of possible  steps. Two conditions exist
which, if fulfilled, are sufficient to ensure that
the chain converges to the desired probability
distribution. The first condition is {\it ergodicity},
i.e. by applying the available steps successively one should
be able to move between any two configurations. The other
condition is {\it detailed balance}. Let the steps be
defined by a probability $P(A\to B)$ for going
from state $A$ to state $B$ and let the desired
probability distribution be given by $\e^{-S(A)}$. The equation of
detailed balance state:
\beq\label{7z.12}
\e^{-S(A)}P(A \to B) = \e^{-S(B)}P(B \to A).
\eeq

\begin{figure}
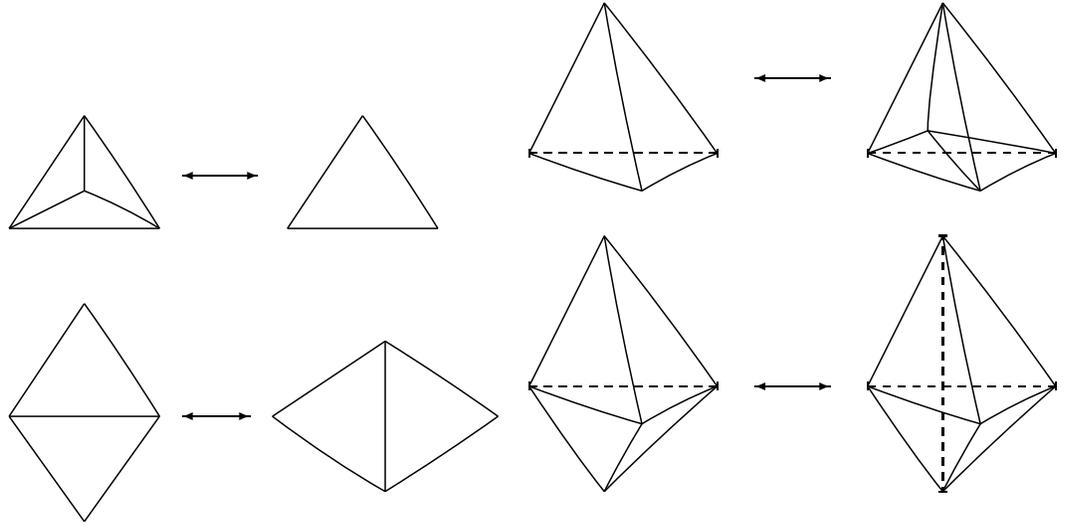

\input{7_4.tex}
\input{7_6.tex}
\caption[7_4]{The moves in $d=2$ and $d=3$.}
\label{7_4}
\end{figure}

Let us first discuss the question of ergodicity. Given the class
combinatorial triangulations of a manifold we want to find a sequence of
so-called moves which are ergodic. Such moves have been known
for a long time and they are known as the Alexander moves \cite{alex}. They can
be simplified somewhat and in $d$ dimensions there are $d+1$ of
them \cite{gv}. They can be described as follows: Given a  sub-simplex $i$
of order $d$+1--$i$  the $d$+1--$i$ $d$-simplexes which
share $i$ form a $d$-ball $B(i)$. Remove $i$ and all the higher dimensional
simplexes to which $i$ belongs. Instead insert
an ``orthogonal'' $d-i$ dimensional sub-simplex,
i.e. the sub-simplex constructed from the $d-i$ vertices of $\prt B(i)$
which did not belong to $i$, together with all the higher dimensional
sub-simplexes. In this way  $B(i)$ has been exchanged with a new
$d$-ball $B'(d-i)$ with the same boundary: $\prt B(i)=\prt B'(d-i)$.
This is illustrated for $d=2$ and $d=3$ in fig.\,\ref{7_4}.

This set of moves is ergodic for a given combinatorial manifold.
In two dimensions it is seen that one of the moves preserves
$N_2$, the number of triangles in the triangulation. In fact it can
be shown that this move alone is ergodic on the set of
triangulations which a fixed topology {\it and} and fixed volume
(i.e. $N_2$). From a computational point of view it is
very convenient to be able to keep the volume fixed.
Can it be done in higher dimensions with a suitable
set of moves of the same local nature as the ones mentioned above?
The answer is no, at least in four- and higher dimensions,
the reason being that it is known that in these dimensions
manifolds exist which are not algorithmic recognizable.
More precisely this means that manifolds  $\cM_0$ exist such that
no finite algorithm  in the sense of {\it Turing} allows
us to decide if an arbitrarily chosen manifold is
combinatorial equivalent to the given $\cM_0$. {\it If}
there were ergodic moves which kept the volume fixed one
can argue that it takes only a finite number of operations,
computable as a function of the volume,
to list all combinatorial manifolds equivalent with $\cM_0$, and
for a given combinatorial manifold one can now by inspection
check if it is in the list. Again it can be argued that this only
takes a finite number of operations, computable as a function
of the volume \cite{benav}.

It is seen that there are deep reasons which makes it
impossible to have  a local set of moves  which preserves
the volume and are ergodic for all manifolds. The situation
is even more drastic: Let $N$ be the volume and let us ask
the question: what is the smallest number of moves
needed in order to connect any two configuration with
volume $N$. The answer is that for manifolds like $\cM_0$ this
number cannot be bounded by any recursively definable
function. This result is only possible if we on the way from
one configuration to another with volume $N$ has to make large
detours to configurations with a volume $N' >> N$ \cite{benav}.

\vspace{12pt}
Let us finally discuss the implementation of detailed balance:
The transition from a configuration $A$ to a neighboring one $B$
can be realized in two steps. First we pick an $i$-simplex shared by
$d+1-i$ $d$-simplexes. The probability of doing this is $1/n_i (A)$,
$n_i(A)$ being the number of $i$-simplexes in $A$. Afterwards we
exchange $i$ with an $d-i$ simplex with some probability $P_i(A\to B)$.
The equation for detailed balance reads:
\beq\label{7z.14}
\e^{-S(A)} \frac{1}{n_i (A)} P_i(A\to B) =\e^{-S(B)}\frac{1}{n_{d-i}(B)}
P_{d-4} (B\to A),
\eeq
where $i=0,1,...,d$.
One can now choose $P_i$'s according to some standard Metropolis
algorithm, but it is worth to emphasize the appearance of the
combinatorial prefactors $1/n_i$ which are usually absent
in ordinary lattice simulations in the sense that they cancel out.
They appear here because  we have a dynamical lattice which change
during the updating.

\subsection{Results}

Let me now discuss the results of the numerical simulations.
They can be summarized as follows \cite{ajk,am4,bm,ckr3,bs,ajjk,abjk}

\begin{itemize}
\item[(1):] There seem to be two
different regions, as a function of the bare {\it inverse} gravitational
coupling constant $k_2$: For small or negative values of
$k_2$ the typical quantum universe will be very crumpled,
with almost no extension and a very large, if not infinite
Hausdorff dimension, while the universes for large
values of $k_2$ will be elongated with a Hausdorff dimension as
small as two. In fig.\,\ref{7_1} we have shown the
average radius for universes of size  9000, 16000, 32000 and 64000
4-simplexes as a function of $k_2$.
\begin{figure}
\centerline{\epsfysize5.5cm\epsfbox{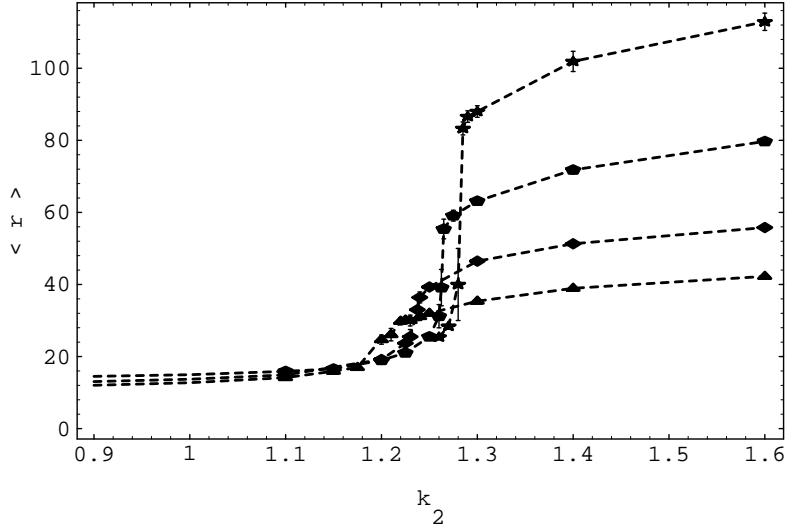}}
\vspace{24pt}
\caption[7_1]{The average radius of the universes of sizes 9000 (triangles),
16000 (squares), 32000 (pentagons) and 64.000 (stars) versus $k_2$.}
\label{7_1}
\end{figure}
The two phases are separated by a phase transition which is
of order two or higher. {\it At} the transition point, $k_2^c$, the Hausdorff
dimension might be finite (the precise value is not well determined,
but it could be close to four).
\item[(2):] The same results are valid for three dimensional simplicial
quantum gravity except that the phase transition seems to be of
{\it first order}, rather than of higher order \cite{am2,bk,av2,abkv}.
\end{itemize}

\begin{figure}
\centerline{\epsfysize5.5cm\epsfbox{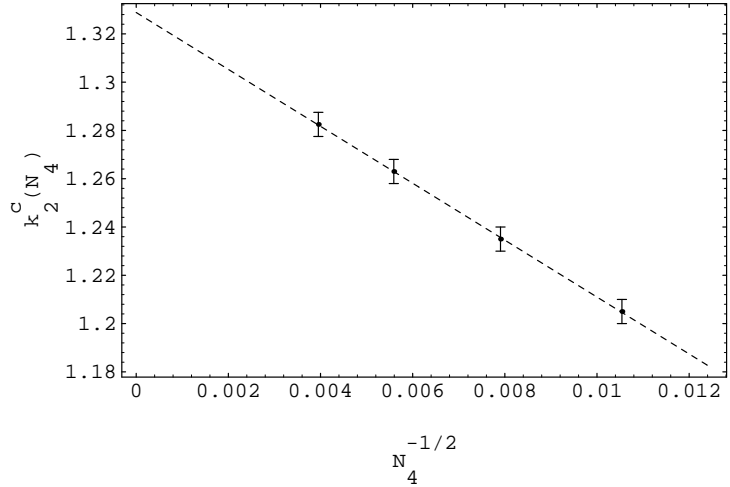}}
\vspace{24pt}
\caption[fig7_2]{The pseudo critical  point $k_2(N_4)$.}
\label{fig7_2}
\end{figure}

From fig.\,\ref{7_1} it is seen that the change between the
elongated region and the crumpled region becomes increasing
visible as the size of the system increases. In addition the
critical point seems to move to higher values of  $k_2$.
We observe a so called pseudo critical point $k_2^c(N_4)$.
The indication of convergence to a limiting value $k_2^c$
as the volume $N_4 \to \infty$ is shown in fig.\,\ref{fig7_2}
and we conclude that we have a genuine phase transition.
From the general theory of finite size scaling  the pseudo
critical point $k_2^c(N_4)$ of a first order
phase transition scales to the limiting value $k_2^c$ as $1/N_4$. Since we
observe a scaling like $1/\sqrt{N_4}$ we conclude that the transition
most likely is a higher order transition. In three dimensional
simplicial quantum gravity it is difficult to perform
this kind of measurement since we observe
pronounce hysteresis around the transition point. This is a
typical sign of a {\it first order} transition.

\begin{figure}
\centerline{\epsfysize5.5cm\epsfbox{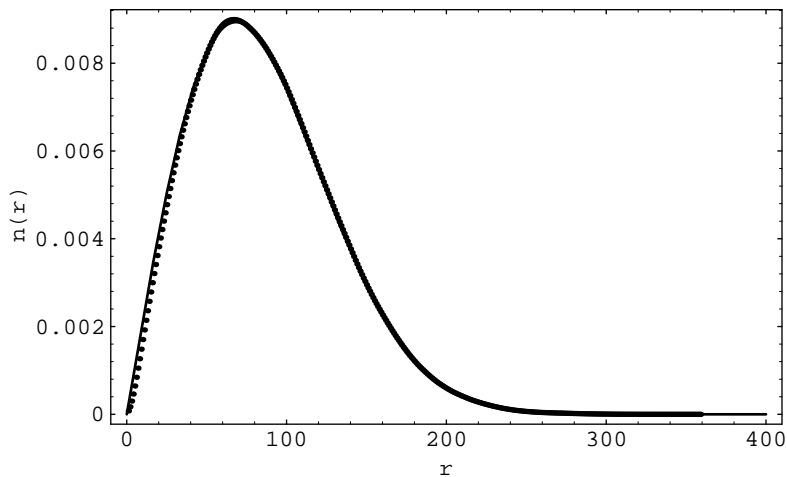}}
\vspace{12pt}
\caption[fig7_3]{The measured distribution (dots, the error bars smaller
than the dots), and the fit (the curve) using the functional form
$r\exp(-r^2/N_4)$.}
\label{fig7_3}
\end{figure}

\begin{figure}
\centerline{\epsfysize5.5cm\epsfbox{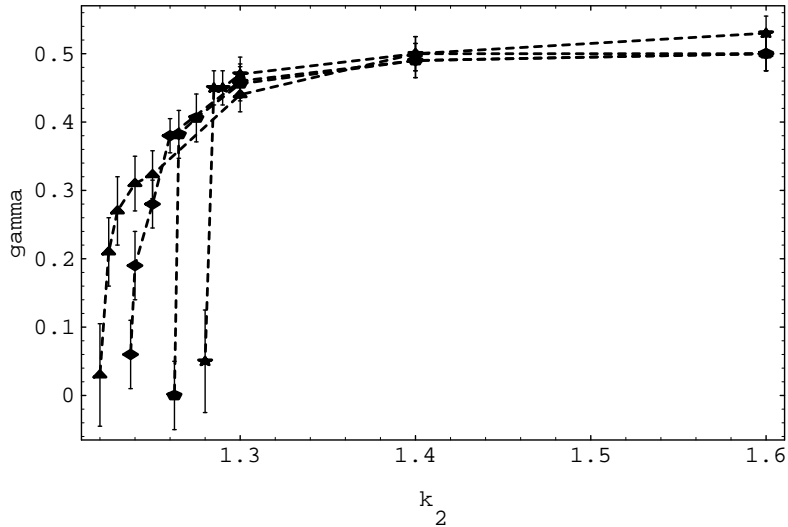}}
\vspace{24pt}
\caption{The measure $\g$ in the elongated phase for various size
lattices ($N_4$ = 9000, 16000, 32000 and 64000).}
\label{fig7_4}
\end{figure}

Let us first discuss the measurements in the elongated phase,
i.e. for $k_2 > k_2^c$.
An ideal quantity to measure in the computer simulations is the
number of four-simplexes $\la n(r) \ra_{N_4}$ within a
spherical shell of thickness 1, a geodesic
distance $r$ from a marked four simplex, the average taken over all
spherical triangulations with one marked four-simplex. It will depend
on the coupling constant $k_2$ and it is related to the two-point
function precisely as discussed above:
\beq\label{7.30}
G(r,N_4) \sim N_4^{\g(k_2)-2} \e^{k_4^c(k_2) N_4}\; \la n(r) \ra_{N_4},
\eeq
provided the subleading corrections are power-like.
In fig.\,\ref{fig7_3} we show the measured distribution
$\la\ n(r)/N_4\ra_{N_4}$ just above the transition to the
elongated phase and a corresponding fit to $r\exp(-r^2/N_4)$.
We see a very good agreement and it becomes even better if
we move further into the elongated phase. Now recall the general scaling
relations
\bea
\la n(r) \ra_{N_4} &\sim& r^{d_h-1} ~~~~{\rm for}~~~1\ll r \ll N_4^{1/d_h},
\label{7.31}\\
\la n(r) \ra_{N_4} &\sim& r^\a \e^{-c (r^{d_h}/N_4)^{1/(d_h-1)}}
~~~~{\rm for}~~~N_4^{1/d_h} < r < N_4, \label{7.32}
\eea
we conclude that $d_h=2$ (and $\a=1$). In addition we can measure
the critical exponent $\g_s$ very conveniently by baby universe
counting. Again the arguments are identical to ones
presented in the two-dimensional case.
The result is shown in fig.\,\ref{fig7_4}
and it is natural from the figure to conjecture that
$\g=1/2$ as long as we are in the elongated phase.
From \rf{7.30}  and $\la n(r) \ra_{N_4}$ we know $G(r,N_4)$
and we can construct the two-point function $G_{k_4} (r;k_2))$  by Laplace
transform (as in the two-dimensional case):
\beq\label{7.33}
G_{k_4}(r;k_2) = \sum_{N_4=1}^\infty N^{-3/2}
e^{-\Delta(k_4) N_4}\;r \e^{-c r^2/N_4} \sim
\e^{-\tilde{c}\,r \sqrt{\Delta k_4}},
\eeq
where $\Delta (k_4) = k_4-k_4^c(k_2)$ is assumed to be small.
This function is precisely the two-point function of the so called
branched polymers, which are known to have internal Hausdorff dimension
$d_h =2$.
We conclude that the numerical simulations provide convincing
evidence that the elongated phase of simplicial quantum
gravity corresponds to a well known statistical theory,
the one of branched polymers. {\it The tendency  to create
baby universes is so pronounced in this phase that the
geometry degenerates to the generic lowest dimensional
fractal structure possible, i.e. that of branched polymers.}

\vspace{12pt}

\noindent
When we lower the value of $k_2$ and move below the
critical point $k^c_2$ the fractal structure of our
ensemble of piecewise linear manifolds changes drastically.
A glance on fig.\,\ref{7_1} shows that the average radius
hardly changes with the volume. This is an indication that the
Hausdorff dimension is large or maybe infinite. If
we move deep into this phase the average curvature is negative and
in addition there are only few baby universes and they are small.
This could lead to the
idea that we entered a phase with ``smooth'' manifolds of negative
curvature. For such manifolds one would expect that the volume
of geodesic balls of radius $r$ would grow exponentially with
the radius, which is what we observe. Clearly this is a ``fake''
infinite Hausdorff dimension and indeed we should observe
the dimension $d_h=4$ in the sense that $\la n(r) \ra_{N_4} \sim r^3$
for small geodesic distances. A closer look at ``typical'' members of
the computer generated manifolds indicates that they cannot
be considered as ``smooth''. Rather they have a few
vertices of very high orders which connect to almost any
other vertex in the manifold and in such a situation
it is not surprising that the linear extension will be small.
In addition we have not been able to fit to any sensible power dependence
$r^{d_H-1}$ for small $r$.
A plot of $\log \la n(r) \ra_{N_4}$ shows indeed a linearly
growing function of $r$ up to some $r_0 (N_4)$ which is not much
different from the average value $\la r \ra_{N_4}$. A
reasonable fit to $\la r \ra_{N_4}$ is
\beq\label{3.4.7}
\la r \ra_{N_4} = a(k_2) + b(k_2) \log N_4.
\eeq
This again gives some support to the idea that the Hausdorff dimension
is infinite in this phase since it appears as a limit of
 $\la r \ra_{N_4} \sim N^{1/d_H}$ for $d_H \to \infty$.
Finally the extrapolation of \rf{7.32} to infinite Hausdorff
dimension indicates that in such a case one should observe
\beq\label{3.4gg}
\la n(r)\ra_{N_4} \sim c_1\,\e^{-m_1(k_2)\,r}
\eeq
down to quite small distances $r \gg N_4^{1/d_h}$, $d_h \to \infty$,
which naturally is replaced by $r > a_1 \log N_4 +a_2$. This is
indeed what we measure.

The observation that $\la n(r) \ra_{N_4}$ grows exponentially
from $r \approx 6$ out to $r \approx r_0$  and then falls
off exponentially indicates that we deal with an infinite
Hausdorff dimension at all distances and it is easy to
get a quite good ``phenomenological'' fit to $\la n(r) \ra_{N_4}$
which incorporates both these features by choosing e.g.:
\beq\label{3.4.7a}
\la n(r) \ra_{N_4} \sim \exp\left(- m_1(k_2) r -c_2\e^{-m_2(k_2) r} \right).
\eeq
It will grow like $\e^{(c_2m_2 -m_1)r}$ for small distances
and fall off like
\beq\label{3.4.6}
\la n(r)\ra_{N_4} \sim c_1\,\e^{-m_1(k_2)\,r}-
c_2\,\e^{-(m_1(k_2)+m_2(k_2)\,r}+\cdots
\eeq
for large distances, while a  $N_4$ dependence
in the coefficient $c_2$ would explain the observed $N_4$ dependence
of $r_0$.
\begin{figure}
\centerline{\epsfysize5.5cm\epsfbox{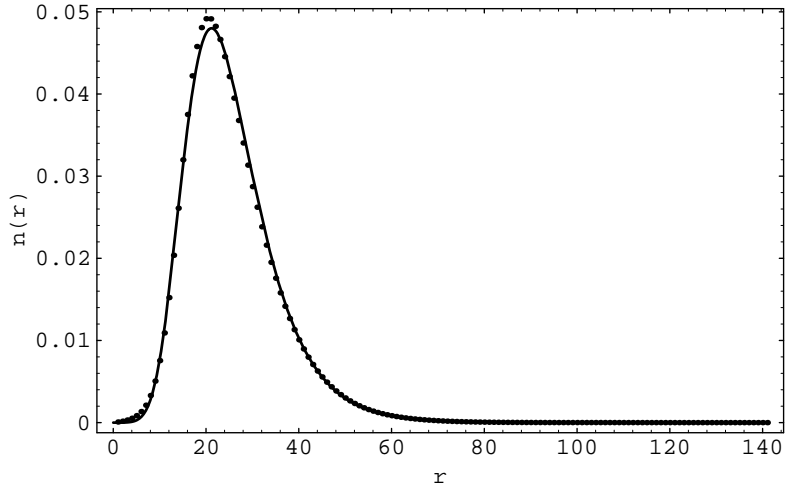}}
\vspace{12pt}
\caption[fig7]{exponential fit (curve) of the form \rf{3.4.7a}
to the measured $n(r;N_4)$ (dots, error bars less than  dot-size)
in the crumpled phase ($N_4 =$ 64000, $k_2=1.26$).}
\label{fig7}
\end{figure}
The data and a fit of the form \rf{3.4.7a} are shown in fig.\,\ref{fig7}
for $k_2 = 1.26$ and $N_4 =64000$, i.e. right below the transition
to the crumpled phase, where the fit is worst. But even so
close to  critical point \rf{3.4.7a} works quite well over the
whole range of $r > 6$. It should be mentioned
that the coefficient in front of the second exponential in eq.
\rf{3.4.6} is negative. This implies that the term cannot be
given the interpretation as an additional heavier mass excitation.
However, just looking at the long distance tail the distribution
$\la n(r) \ra_{N_4}$ allows us to determine $m_1$, $m_2$ and $c_2$
from \rf{3.4.6}. On the other hand we can determine $c_2 m_2 -m_1$
from the short distance exponential growth alone and find
good agreement. This indicates that long and short distance
behavior are intertwined  in the case of  infinite Hausdorff
dimension, as they are in the case of finite Hausdorff dimension,
where $d_H$ appears both in the short distance and long
distance expression for $\la n(r) \ra_{N_4}$ (see e.g. \rf{7.31}-\rf{7.32}).

\vspace{12pt}

\noindent
{\it We conjecture that the internal Hausdorff dimension is
infinite for $k_2 < k_2^c$}.

\vspace{12pt}

\begin{figure}
\centerline{\epsfysize5.5cm\epsfbox{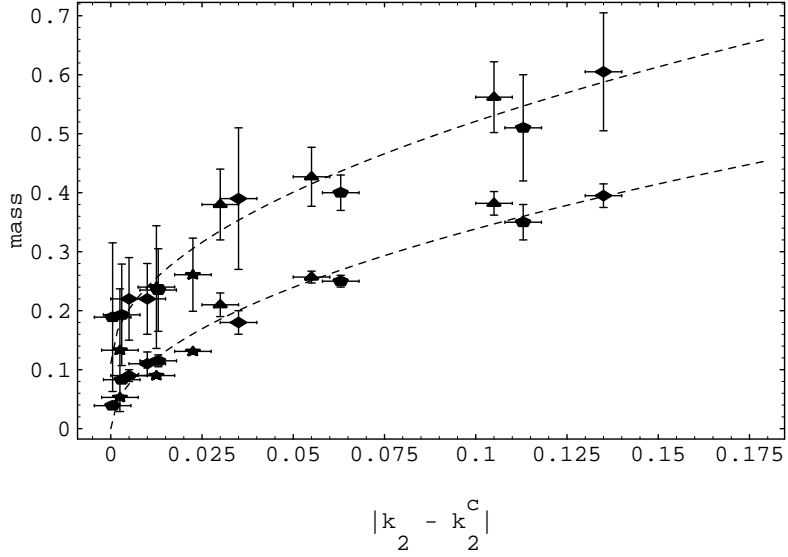}}
\vspace{32pt}
\caption[fig8]{The behavior of the masses $m_1(k_2)$ and
$m_1(k_2)+m_2(k_2)$ from \rf{3.4.7a} in the crumpled phase as
a function of $k_2^c-k_2$.}
\label{fig8}
\end{figure}

\noindent
In accordance which this conjecture we have in the elongated
phase no ``mass'' term $m(k_4) = (k_4-k_4^c(k_2))^\n$ which scales to
zero as we approach $k_4^c(k_2)$. The exponential coefficient $m_1(k_2)$
is finite in the infinite volume limit $k_4 \to k_4^c(k_2)$. However,
it is most interesting that $m_1(k_2)$ scales to zero as $k_2 \to k_2^c$,
the phase transition point between the crumpled and the elongated phase.
{\it This gives a strong indication that the system {\it at} the transition
might have a finite Hausdorff dimension, which could very well be
larger that the generic $d_h=2$ found in the elongated phase.}
In fig.\,\ref{fig8} we have shown the scaling of the mass $m_1(k_2)$
as a function of $k_2$.

\vspace{12pt}

\noindent
The above mentioned numerical ``experiments''
suggest the following scenario:
The typical quantum universe, determined without any Einstein action
(i.e. $k_2=0$) has (almost) no extension. Its Hausdorff dimension
might be infinite and internal distances between points are always
``at the Planck scale''. By that we simply mean that no consistent
scaling can be found which will be compatible with  finite continuum
volume and finite Hausdorff dimension.
For a finite value of the bare gravitational coupling
constant there is a phase transition (second or higher order)
to a phase with a completely different geometry with pronounced
fractal structures. It is tempting to view the transition between the two
kind of geometries as a transition where excitations related to the
conformal mode are liberated, since large $k_2$ is a region which
formally corresponds to small values of the gravitational coupling constant.
Right at the
transition it seems as if we have the chance to encounter genuine extended
structures with a finite Hausdorff dimension. Maybe the fact that the
transition between the two types of geometry is of second (or higher)
order can be  used as the starting point for a non-perturbative
definition of quantum gravity.

Of course it is crucial to be able to perform high statistics simulations
{\it at} the critical point in order to investigate this possibility
in greater detail. It would be even better if we could perform
some analytical calculations. The branched polymer picture indicates
that mean field methods should be available in some regions of the
coupling constant space.
\newsection{Discussion}

We have seen in some detail that it is possible to discretize
reparametrization invariant theories and apply with success the methods
known from the theory of critical phenomena.
In this way we deal with the theories of fluctuating geometries.
However, we did not really answer the most interesting
question: How wildly should we allow the geometries to fluctuate?
For a fixed topology it was possible to formulate a regularized
Euclidean quantum theory. In two dimensions one can take the
scaling limit of the regularized theory and the corresponding
quantum theory correctly describes the interaction between
matter and gravity. In higher dimensions it is not yet known if
it leads to any interesting theory, but at least there is a well
defined non-perturbative procedure for how to investigate this
question. In addition the discretized action in this approach is
remarkable simple and one could hope that it will be possible
to solve the theory in the same detail as in two dimensions.
At the moment we let loose topology we have also lost control
of the theory. In two dimensions the double scaling limit
gave us a hint that it might be possible to perform the summation over
topologies. At least there are some
prescriptions for how it should be done. They are not yet unambiguous
and this  reflects our lack of understanding of the
physics which goes beyond a simple perturbative expansion in topology.
Nevertheless I view it as very encouraging that one at least
seems to have by now the tools which allow such questions to be asked.
These tools have their origin in the discretized approach and even
in three and four dimensions the discretized approach has the
advantage that it in a natural way is able to  combine the summation
over Riemannian structures and topology with the volume of the
universe acting as the cut-off. It remains to be seen if the approach  will
result in an interesting theory of gravity, but it offers at least
a playground for a fascinating interplay between  pure mathematics,
theoretical physics and computational physics.

\vspace{24pt}

\noindent
{\bf Acknowledgment}  It is a pleasure to thank the organizers and the
participants of  the school for creating a  most simulating
environment during the lectures. The presentation has be
heavily flavored by discussions and collaborations with Bergfinnur Durhuus,
Thordur Jonsson, Jerzy Jurkiewicz, Charlotte Kristjansen, Yuri Makeenko
and Gudmar Thorleifsson. They are responsible for most of the interesting
observations presented, while any mistakes can only be blamed on me.

\end{document}